\documentclass[12pt, letterpaper]{article}
\usepackage[titletoc,title]{appendix}
\usepackage{color}
\usepackage{booktabs}
\usepackage[usenames,dvipsnames,svgnames,table]{xcolor}
\definecolor{dark-red}{rgb}{0.75,0.10,0.10}
\definecolor{bluish}{rgb}{0.05,0.05,0.85}
\PassOptionsToPackage{unicode}{hyperref}
\PassOptionsToPackage{naturalnames}{hyperref}
\usepackage[margin=1in]{geometry}

\usepackage{natbib}

\usepackage{float}
\usepackage{placeins}

\usepackage{geometry}  % see geometry.pdf on how to lay out the page. There's lots.
\usepackage{graphicx}  % Handles inclusion of major graphics formats and allows use of
\usepackage{amsfonts,amssymb,amsbsy}
\usepackage{amsxtra}
\usepackage{setspace} % Permits line spacing control. Options are:
\usepackage{pdflscape}
\usepackage{fancyhdr}   % Permits header customization. See header section below.
\usepackage{xurl}        % Correctly formats URLs with the \url{} tag
\usepackage{fullpage}   %1-inch margins
\usepackage{multirow}
\usepackage{verbatim}
\usepackage{rotating}
\usepackage{chngcntr}
\usepackage{longtable}
\usepackage{adjustbox}
\usepackage{dcolumn}

\ExplSyntaxOn
\cs_new:Npn \my_expandableinput:n #1
{ \use:c { @@input } { \file_full_name:n {#1} } }
\AddToHook{env/tabular/begin}
{ \cs_set_eq:NN \input \my_expandableinput:n }
\ExplSyntaxOff

\makeatother

\usepackage{footmisc}
\usepackage{color}

\usepackage[
    skip            =0pt,
    labelfont       =bf, 
    font            =small,
    textfont        =small,
    figurename      =Figure,
    justification   =justified,
    singlelinecheck =false,
    labelsep        =period]
{caption}
\usepackage{subcaption}

\usepackage[linkcolor=blue,
			colorlinks=true,
			urlcolor=blue,
			pdfstartview={XYZ null null 1.00},
			pdfpagemode=UseNone,
			citecolor={bluish},
			pdftitle={social proof}]{hyperref}
\usepackage[nameinlink, capitalize, noabbrev]{cleveref}

\usepackage{tocloft}

\usepackage{lscape}

\makeatletter
\providecommand\phantomcaption{\caption@refstepcounter\@captype}
\makeatother

\title{Social Proof is in the Pudding: The (Non)-Impact of Social Proof on Software Downloads\thanks{You can download the replication materials from \href{http://github.com/themains/social_proof_stars}{https://github.com/themains/social\_proof\_stars/}.}
}

\author{
Lucas Shen\thanks{Institute for Human Development and Potential, Agency for Science, Technology and Research,  \href{mailto:lucas@lucasshen.com}{\footnotesize{\texttt{lucas@lucasshen.com}}}}
\and Gaurav Sood\thanks{\href{mailto:gsood07@gmail.com}{\footnotesize{\texttt{gsood07@gmail.com}}}} 
}
\date{\today}

\begin{document}
\maketitle

\thispagestyle{empty}
\begin{abstract}
\noindent
Open-source software is widely used in commercial applications. Pair that with the fact that when choosing open-source software for a new problem, developers often use social proof as a cue. These two facts raise concerns that bad actors can game social proof metrics to induce the use of malign software. We study the question using two field experiments. On the largest developer platform, GitHub, we buy `stars' for a random set of GitHub repositories of new Python packages and estimate their impact on package downloads and broader repository activity. We find no discernible impact on downloads, nor on forks, pull requests, issues, or other measures of developer engagement. In another field experiment, we manipulate the number of human downloads for Python packages. Again, we find no detectable effect on subsequent downloads or on any measure of repository activity.\\

\noindent
\textit{Keywords:} Field Experiment; RCT; GitHub; Security; Malware; Open-Source Software; Social Proof; Social Computing\\

\noindent
% \textit{Word count:} 4100 words in text
\end{abstract}

%TC:ignore
% \quickwordcount{social_proof}

%TC:endignore

\clearpage
\setcounter{page}{1}
\doublespace
\section{Introduction}
Most commercial software relies on open-source software \citep{Mojica2014, eghbal2016roads, Vargas2020, Wermke}. But vetting the quality of software with respect to security is hard \citep{Munaiah2017}. Hence, most software developers use cheap heuristics like social proof to choose between multiple open-source software that purport to solve the same problem \citep{simtech, Pickerill_2020}. This raises the concern that bad actors can game social proof metrics to induce developers to use malicious software \citep{Ohm2020, cao2022fork, Ladisa2023, Wermke, CheckPoint2024, Forbes2024, he2024, Sonatype2024}. In this paper, we explore the possibility that faked social proof of software can induce popularity and usage using two field experiments.

Our first field experiment was conducted on GitHub, the world’s largest developer platform. We manipulated the perceived popularity of a random set of repositories associated with new Python packages by purchasing `stars' for them. For a random subset of the treated repositories, we further increased the `stars' by asking people in our network to `star' the repositories. In all, our manipulation raises the median number of stars from 0 to 20--65 stars, depending on the treatment group. This meaningful but modest increase in the number of stars, however, has little impact on the number of package downloads or contributions to the repository. We cannot reject the null hypothesis that the manipulation had no effect three months after the treatment.

In a second field experiment, we manipulate social proof of Python packages by increasing the download count in the official downloads registry. We use a script to download a random set of packages multiple times, nearly quintupling the median download count from about 20 to 100. (In absolute terms, the treatment manipulation is still modest, and deliberately so.) But there is little impact of the treatment on the number of downloads or contributions to the associated GitHub repository, three months later.%
\footnote{In the Appendix, we report results from an analysis of whether bot and human downloads Granger-cause subsequent human downloads. The results suggest that human downloads predict future human downloads. However, because this pattern is inconsistent with the experimental findings, we are cautious about placing too much weight on it.}

Our study allays but does not dispel concerns about the possibility of spreading malware by gaming social proof. For one, even a single organization adopting a malign piece of software may constitute a practically important effect. But our experiments are underpowered to detect anything so subtle. (In fact, it is nearly impossible to design experiments that can.) For another, it is entirely plausible that a more intense treatment would persuade developers.

\section{Social Proof}
\label{sec:theory}
Others' choices, especially those of similar others or those we admire, can be informative \citep{rao2001fool, cialdini2003influence, salganik2006experimental, simtech, amblee2011harnessing, Dabbish2012, messing2014selective, venema2020doubt}. They can tell us what is useful or desirable. For instance, if someone is in the market for a car, they may pay attention to the cars in the office parking lot to infer what a good car is for them. Companies understand the value of social proof and regularly use it to try to influence customers. For example, many companies prominently show lists of customers who have bought their products on their websites. Retailers like Amazon allow customers to sort by best sellers and show how many customers bought a product from a particular seller over the last month or year. News media companies show lists like `most read articles' to aid readers \citep{messing2014selective}.

The Elaboration Likelihood Model (ELM) provides a framework for understanding when social proof is likely to influence decisions \citep{ELMppr, ELMbook}. The model distinguishes two routes of persuasion: a central route, in which decision makers carefully evaluate the merits of the object itself, and a peripheral route, in which they rely on heuristic cues such as popularity or endorsement counts. The peripheral route dominates when decision makers lack either the motivation or the ability to engage in careful evaluation. Social proof metrics---star counts, download tallies, best-seller rankings---function as peripheral cues, and their influence is predicted to be strongest when the decision maker cannot easily assess quality directly.

Signaling Theory offers a complementary perspective \citep{Spence1973}. A signal is informative to the extent that it is costly or difficult to produce without possessing the underlying quality it purports to indicate. When a signal becomes cheap to produce regardless of quality, its correlation with quality degrades, and rational receivers who recognize this degradation discount the signal accordingly \citep{Campbell1979}. Critically, receivers need not know the exact cost of producing a fake signal; it is sufficient that they observe or infer that the signal is noisy---for instance, by encountering cases where high signal values do not correspond to high quality.

Software adoption is an instructive case for both frameworks. Inspecting code for safety and quality takes time and skill, and most developers cannot afford to do so for every dependency they adopt. Social proof can therefore be especially influential when choosing between packages that purport to provide the same functionality \citep{Dabbish2012, simtech, he2024}. At the same time, the ELM predicts that developers may be less susceptible to peripheral cues than users of other platforms: developers face tangible consequences from poor choices (broken builds, security vulnerabilities, maintenance burden), giving them strong motivation to evaluate quality through the central route. They can also draw on richer signals than raw popularity, including code documentation, commit frequency, contributor activity, and project responsiveness \citep{Dabbish2012, Tsay2014, BORGES2018112}. From the perspective of Signaling Theory, the informational value of GitHub stars has been eroded by commercial markets that sell stars at scale for trivial cost \citep{he2024}. To the extent that developers have learned---whether through direct awareness of these markets or through experience with highly starred but low-quality repositories---that star counts are a noisy signal, they have reason to discount them.

These predictions of attenuation are not unambiguous, however. Even when decision makers are aware that a cue is noisy, anchoring effects can cause high values to inflate quality judgments \citep{TverskyKahneman1974}. A developer who is skeptical of stars may still perceive 500 stars differently from 5, because the number sets a reference point that is difficult to fully adjust away from. Moreover, the ability to engage in central-route evaluation varies: assessing package safety and fitness for purpose is difficult without downloading and inspecting the code, and in time-pressured contexts, developers may default to peripheral heuristics even when they would prefer not to. Whether social proof in the form of manipulated platform metrics actually shifts developer behavior is therefore an empirical question that the competing predictions above cannot resolve on their own.

To shed light on how manipulable developers' choices are, we conduct two field experiments. In the first, we manipulate the number of stars for a random set of new Python packages on GitHub and assess the impact on downloads of the associated package and activity on the repository. In the second, we manipulate the number of downloads of Python packages to assess its impact on future downloads and contributions to associated GitHub repositories.

\section{GitHub Experiment}\label{sec:github}
GitHub is the most popular platform for creating, storing, managing, and sharing code. Over 100 million developers use \href{https://github.com/}{GitHub}, and the platform hosts over 420 million repositories, of which over 28 million are public.\footnote{Figures as of January 2023; see GitHub Octoverse 2023.}

GitHub prominently displays four social proof signals for each repository: the number of stars, watchers, forks, and open issues. These signals differ in what they convey and how readily they function as quality or popularity indicators. Forks indicate derivative work rather than endorsement, and open issues can signal either active use or neglect. Watching provides a clearer signal of interest, but relatively few users watch repositories because doing so subscribes them to notifications about repository activity. That leaves stars, an analog of `likes,' as the most widely used metric of popularity and social proof on GitHub \citep{Munaiah2017, BORGES2018112, Qiu2019, Pickerill_2020, Vargas2020, Wermke, koch2023fault, ghwfh}. Unlike other platforms, GitHub does not readily show visit counts or similar engagement metrics, making stars all the more salient.

Users star repositories for various reasons. Some use stars as a bookmarking tool; GitHub lets users browse, search, and organize their starred repositories into lists. Others star repositories to signal approval to their followers, since starred repositories appear in followers' news feeds. Still others star a repository simply to show support for a project or its maintainer. GitHub also uses stars to customize a user's news feed, recommending related repositories and surfacing changes to starred projects. Whatever the motivation, each star adds to a repository's visible social proof.

Repository owners try hard to get people to star their repositories to increase visibility and hence usage. Starring can increase visibility through three channels. First, starred repositories appear in the news feeds of the starring user's followers, exposing the repository to a wider audience. Second, stars contribute to a repository's chances of appearing on GitHub's \href{https://github.com/trending}{trending page}, which attracts additional attention, including from the media.\footnote{While GitHub uses a combination of factors to determine trending repositories, stars are generally suspected as a key factor. See, for example, \url{https://github.com/orgs/community/discussions/3083}.} Third, a high star count serves as social proof: a developer evaluating the repository may be persuaded by the visible endorsement of others to adopt the software.
The number of stars a repository has is widely considered the primary signal of its popularity, which is why it is the metric we intervene on. We believe our experiment operates primarily through the social proof channel rather than the news feed or trending channels, for two reasons. First, our manipulation is modest enough that it does not cause treated repositories to trend. Second, the accounts we use to add stars have follower networks whose interests are unrelated to the treated packages, limiting the informational value of the stars for those followers.

\subsection{Sample and Randomization}\label{sec:gh-randomization}
\label{sec:gh-design}
Our population of interest is repositories associated with new Python packages. We focus on new Python packages because we conjecture that quality is the least certain when a repository is new. Hence, for a new repository, social proof provides the strongest signal. Our choice of studying repositories associated with Python packages stems from the fact that we can get reliable data on the number of downloads for a Python package. Our sample includes new packages listed between 24 and 30 April 2023. (We identified new PyPI packages by taking a set difference of the PyPI index on April 30 and April 24.) Of these packages, we only keep Python packages with a public GitHub repository. (We used the GitHub source URL from the package's setup configuration file to link the package to a GitHub repository.) In all, we identified 622 new packages with a public GitHub repository. Of the 622 public repositories, we assigned 100 to the treatment group.

\cref{tab:baltest-repo-treated-01} (column (1)) reports (pre-treatment) descriptive statistics of the packages. Most new packages were created in 2022 but were released in 2023.  96\% of the packages had an open issue on GitHub. On average, a repository had been forked 20 times, and the user under whom the repository was listed had, on average, four subscribers (\cref{sec:baltests}).

\subsection{Treatment Conditions}\label{sec:gh-treatment-conditions}
Our treatment includes stars from two sources: market-bought and network-based clicks. We asked users in our network (not tabulated for anonymized submission) to `star' the 100 repositories. This group serves as our `low dosage' treatment group. The median number of followers of users in our network who starred the treatment repositories was 9 (the mean was 64, \cref{tab:organic_starrers_profile_summary}). We expect the benefits of these stars to also come primarily from greater social proof because the repositories that the users starred did not focus on the interests and specializations of the users. We triggered our network to `star' the repositories on May 12, 2023. They took about 10 days to `star' all the 100 repositories (see \cref{fig:timeseries-stars-treated-012-medians}).

Of the 100 packages, we randomly selected 25 packages and bought 50 stars for each from \href{https://web.archive.org/web/20240619193418/https://baddhi.shop/product/buy-github-followers/}{Baddhi Shop} on May 12, 2023. 50 stars is equivalent to doubling the stars of control repositories (mean of 53.2 stars).%SD = 389.2
The stars we bought from the vendor were assigned over a few days, plausibly to avoid triggering GitHub's anomaly detection algorithms. One interesting feature of the stars bought on the market is that they were all from users whose accounts were created around April 20, 2023. We expect some of these stars to be taken down by GitHub integrity teams, so these stars likely only have a short-run impact. For the stars we bought, we have no reason to expect the users who starred the repository or their followers to be authentic. So, we only expect the benefits of these stars to accrue from people who look at the GitHub repository before downloading a package.

\subsection{Attrition and Analytic Strategy}\label{sec:attrition}
About one week after the start of the intervention, on May 20, 2023, we took a snapshot of the GitHub repository. By that time, we could only retrieve 582 repositories (\cref{sec:baltests}). Thirty-seven repositories in the control group and three from the treatment group were lost because the GitHub repository was deleted or moved to private status. It is plausible that the somewhat lower attrition in the treatment group is because the repository owners were buoyed by seeing more stars. 

%We see one fewer user (from 538 to 537). (Of the 582 repositories, eight users contributed to two repositories.) 

Our primary estimand is Intent-To-Treat (ITT) effects on Python package downloads. We use the entire dataset of 622 Python packages for that. In the SI, we also estimate the Local Average Treatment Effect (LATE) that subsets on the compliers: GitHub repositories for which we see a net increase of at least 20 stars during the treatment period. (Note that some of the stars in the treatment period could be organic.)

\subsection{Balance Tests}
\label{sec:baltests}
To confirm that the randomization was done correctly, we compare the means of various attributes across the treatment and control groups. Looking at the primary outcome variable, downloads, in the pre-treatment period, we find little difference between treatment and control packages (see \cref{tab:baltest-readme-treated-01}).

We supplemented this check with other balance tests that rely on GitHub data. We did not take a snapshot of GitHub attributes before applying the treatment. We only took a snapshot of GitHub user characteristics linked to the PyPI packages on May 20 (\cref{sec:attrition}). So our balance tests that rely on GitHub data use post-treatment data, though one right at the end of the treatment period. As we noted above, there was attrition between the start and end of treatment.

% https://github.com/soodoku/social_proof_stars/blob/main/github_exp/get_baseline_profile/src/get_participant_repo_profile.ipynb
\cref{tab:baltest-repo-treated-01} reports statistical summaries of various attributes of the Python packages associated with treated and control repositories. Columns (2) and (3) report the means and standard deviations (in parentheses) for packages in the control and treatment groups, respectively. The last column reports the standardized mean difference. We compare (a) repository size (in megabytes), (b) whether the repository is forked from a previously existing repository, (c) the year of repository creation, (d) the number of subscribers, (e) whether issues exist (or is enabled), (f) the number of forks into other repositories, (g) the number of open issues, (h) the number of topics listed,\footnote{Examples of topics from \href{https://github.com/tensorflow/tensorflow}{TensorFlow} include ``machine-learning,'' ``deep-neural-networks,'' ``deep-learning,'' ``neural-network,''  and ``distributed.''} and (i) whether the primary detected language is Python (not all Python packages have Python as the primary source code language). As the table shows, except for two characteristics (the number of subscribers and the number of forks), the treatment and control are statistically indistinguishable.

\crefrange{tab:baltest-repo-treated-012}{tab:baltest-readme-treated-012} repeat the balance tests, making an additional split by dosage, and reach similar conclusions.\footnote{High-dosage repositories and high-dosage users are different from the control group on two aspects: high-dosage users have more gists (blogs/short code snippets) and are likelier to list the company they work for at the 5 percent level (\cref{tab:baltest-user-treated-012}). Given that our groups are randomly assigned, we ascribe these two statistically significant but substantively small differences to chance.}
% https://github.com/soodoku/social_proof_stars/blob/main/github_exp/get_baseline_profile/src/get_participant_user_profile.ipynb

Another way to look at balance is in terms of the user characteristics of repository owners. There are 545 users behind the 582 repositories. Eight users appear in more than one group. \cref{tab:baltest-user-treated-01} reports whether users releasing treatment and control packages differ in how they decorate their user profile by listing their company, email, personal webpage, and a brief biography of themselves. In each instance, we detect no difference between the publishers of treatment and control packages. 

% https://github.com/soodoku/social_proof_stars/blob/main/github_exp/get_baseline_profile/src/get_pypi_desc_readme.ipynb

On May 20, we also took a snapshot of the PyPI Python package metadata: the number of dependencies and the package description length. The former was captured through setup configuration files, and the latter from the ``readme'' documentation or equivalent. While these attributes were collected post-intervention, these constitute balance tests to the extent that these characteristics are slow-moving. \crefrange{tab:baltest-repo-treated-01}{tab:baltest-readme-treated-01} report differences between treated and control groups.

Lastly, to further supplement the balance tests, we also constructed pre-treatment balance tables using historical GitHub event data from the GitHub Archive Project (\cref{sec:balance-gharchive}).
Overall, the pre-treatment activity metrics (stars, forks, pushes, pull requests, opened issues, closed issues, and releases) confirm balance between the treatment groups.

\subsection{Manipulation Check}
\begin{figure}[ht]
\centering
    \includegraphics[width=\textwidth]{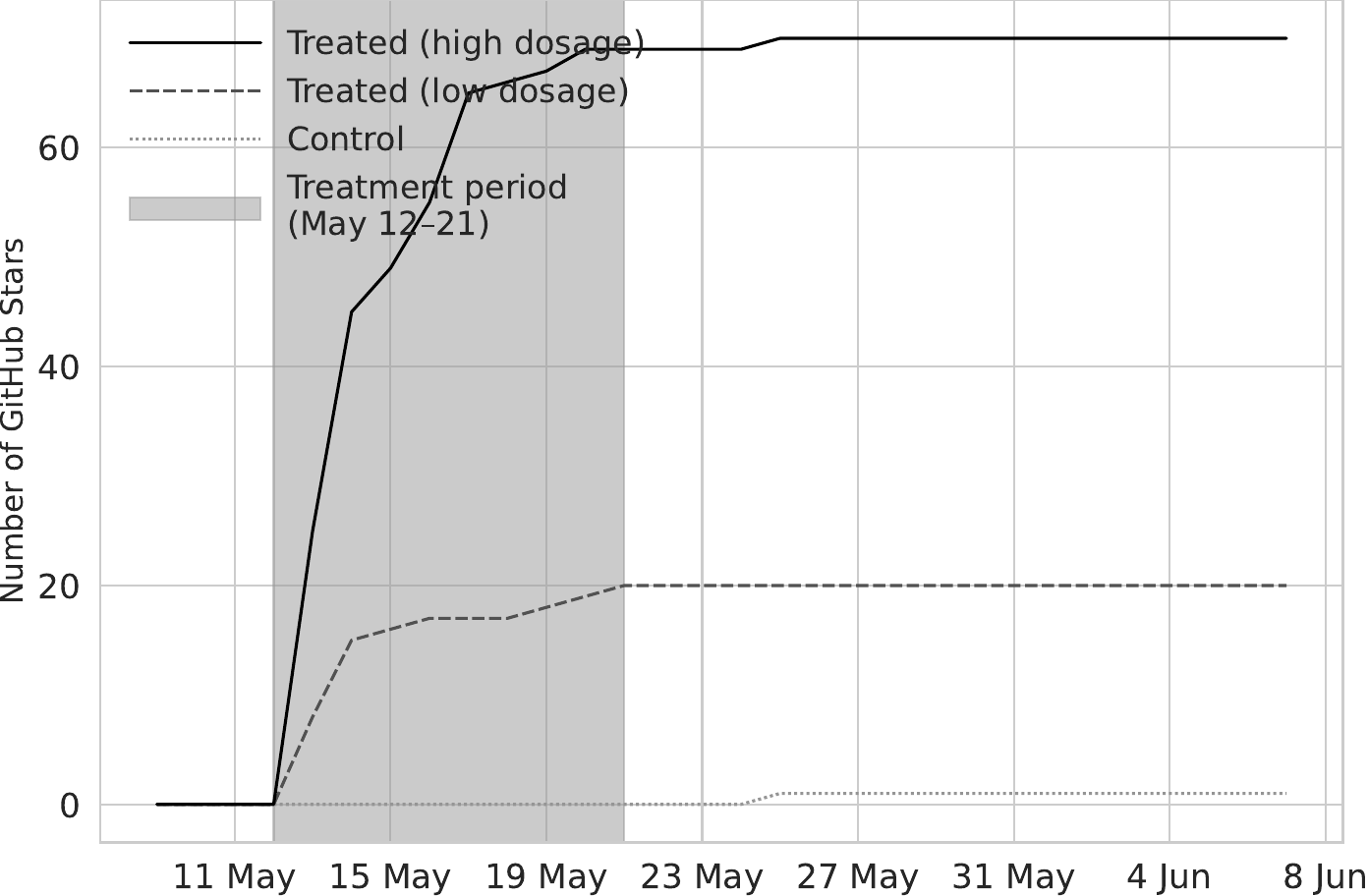}
\caption{
\textbf{Manipulation Check: GitHub Stars for Treated vs. Control}. The figure plots the median number of stars on a particular day for the three groups: high-dosage treatment (market and network stars), low-dosage treatment (network stars), and control (no stars). 
% for numbers, see https://github.com/soodoku/social_proof_stars/blob/main/github_exp/metrics-timeseries/src/plot_timeseries_stars.ipynb
In all, we plot data for 585 packages and 17,550 package days. The shaded vertical bar indicates the period during which the treatment was applied. \cref{tab:github_exp_stars_regtable} presents formal estimates of this manipulation check. See \cref{fig:timeseries-stars-treated-012} for the figure that shows the means.}
\label{fig:timeseries-stars-treated-012-medians}
\end{figure}

While historical download logs are immutable, users can rescind stars. Anticipating that bought stars might be flagged by GitHub and removed shortly after being added, we took a snapshot of the stars of each package at the end of every day. This allows us to more accurately capture the change in stars during our experiment period.\footnote{Historical stars with timestamps are available from the API, but do not retain records of removed stars.}

\cref{fig:timeseries-stars-treated-012-medians} shows the time trend of the median number of stars across the three groups: high-dosage treatment (market and network stars; $n = 25$), low-dosage treatment (network stars only; $n = 75$), and control (no stars; $n = 485$). By the end of the intervention (May 21), relative to the control group, the median number of additional stars in the low-dosage group was 19 ($p < .001$), while the median number of additional stars in the high-dosage group was 69 ($p < .001$, \cref{tab:github_exp_stars_regtable}).

\subsection{Outcome Measure: PyPI Downloads}
\label{sec:measures-pypi-downloads}

The primary outcome for both the field experiments, the GitHub experiment, and the PyPI experiment (\cref{sec:pypi-experiment}) is Python package downloads. These Python package download metrics come from the centralized Python Package Index (PyPI) repository. It hosts open-source Python packages uploaded to it by package developers. Users download packages as needed from PyPI. 

To log package downloads from PyPI, the \href{https://github.com/pypi/linehaul-cloud-function}{Linehaul project} implements a daemon that listens for download events and logs them. Specifically, it tracks details like the package name, date and time the package was downloaded, package version, and the type of installer software. Linehaul then feeds the download logs to the publicly available Google BigQuery. Some installers are known bots (\cref{tab:human_bot_download_classification}). These bots tend to be caching mirrors for distributional and security purposes. For instance, Bandersnatch is the official mirroring service of PyPI to help improve accessibility and download speed for users of various geographical regions. We restrict the analysis to human downloads (\cref{tab:human_bot_download_classification}).

The main analyses focus on differences in medians, given the huge volatility in means induced by a few extreme outliers (\crefrange{fig:timeseries-downloads-treated-012}{fig:timeseries-downloads-treated-012-without-top2-extreme-outliers}).

\subsection{Results}
\label{sec:results}

\begin{figure}[ht]
\centering
    \includegraphics[width=\textwidth]{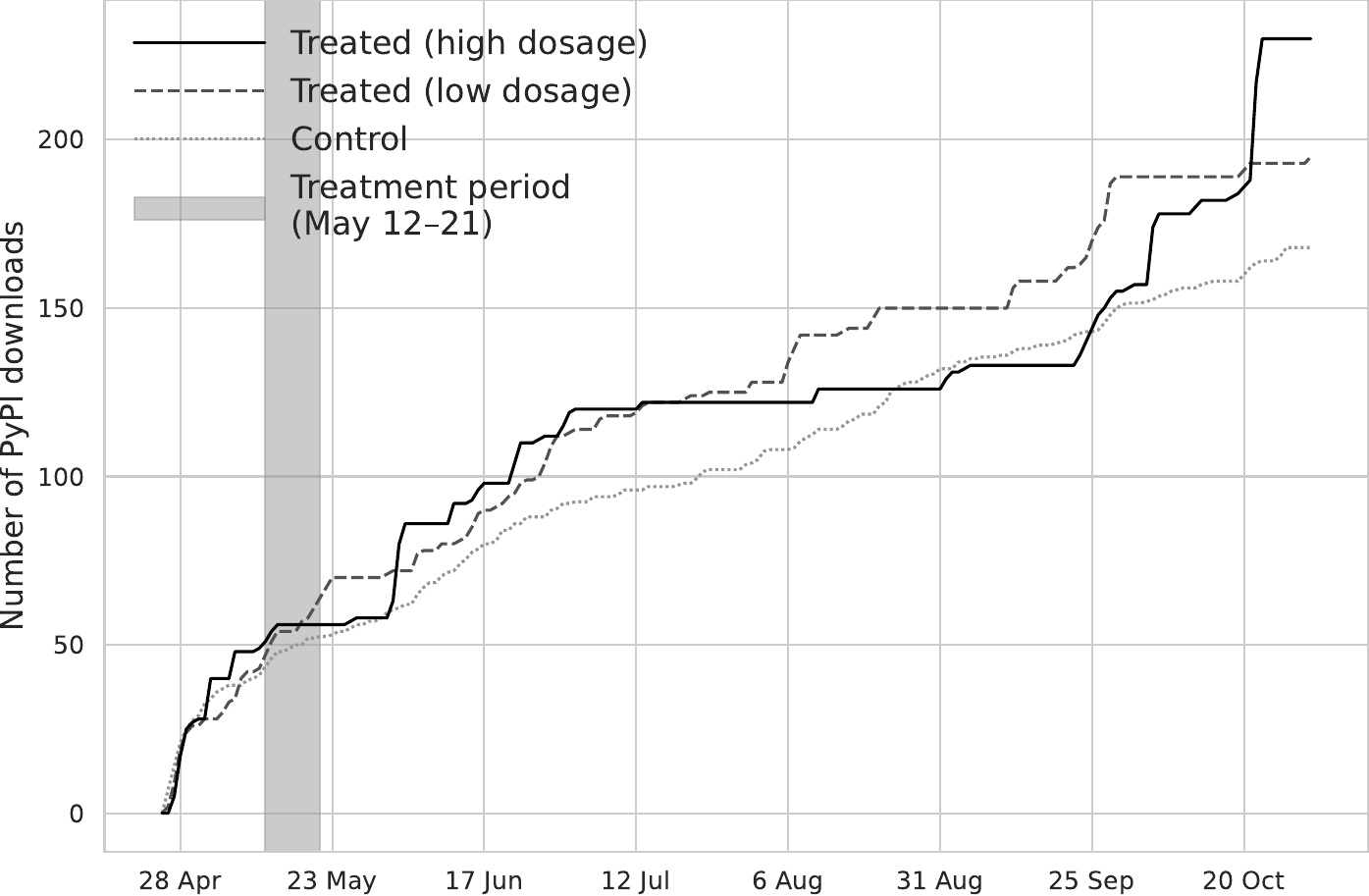}
\caption{
\textbf{Median PyPI Downloads for Treatment vs. Control in GitHub Experiment.}  The figure plots the median of cumulative downloads for each of the three groups. Each point is a day averaged within the group for 622 packages and 118,180 package days. 
% for numbers, can look at https://github.com/soodoku/social_proof_stars/blob/main/github_exp/get_metrics/src/consolidate_pypi_downloads_allinstallers.ipynb
Treatment is distinguished by low and high dosage (see \cref{sec:gh-design}). The shaded vertical bar indicates the treatment period.  Downloads include only human downloads (\cref{tab:human_bot_download_classification}). See also \crefrange{fig:timeseries-downloads-treated-012-individual}{fig:timeseries-downloads-treated-012-without-top2-extreme-outliers} for the time series of individual packages. \cref{tab:github_exp_medians_regtable} reports estimates of differences in medians. \cref{fig:timeseries-downloads-treated-012} plots the means.}
\label{fig:timeseries-downloads-treated-012-medians}
\end{figure}

\cref{fig:timeseries-downloads-treated-012-medians} traces the median number of PyPI downloads over time. If the treatment shifted adoption behavior, we would expect the download trajectories of treated and control packages to diverge after intervention. No such divergence is apparent. The ITT estimates in \cref{tab:github_exp_medians_regtable} confirm this. One month after intervention (on June 21), the median number of downloads in the low-dosage and high-dosage groups was statistically indistinguishable from the control group. We estimated differences in medians at four additional time points over the following four months and found no significant difference between any treatment group and the control at any snapshot.
We also estimated a model allowing treatment effects to vary linearly over time. Neither the low-dosage nor the high-dosage group exhibits a trend distinguishable from that of the control group (see \cref{tab:github_exp_medians_regtable}). In \cref{sec:difference-means}, we examine differences in means and reach similar conclusions. We also estimate the LATE using treatment assignment as an instrument for compliance (receiving at least 20 stars; see \cref{fig:timeseries-stars-treated-012-medians}). The mean differences in downloads for compliers are larger than the ITT estimates, as expected, but remain non-significant (\cref{tab:github_exp_regtable_allhumaninstallers_late}).\footnote{Subsetting downloads to those installed by \emph{pip} (\cref{tab:human_bot_download_classification}), the most common human installer, yields similar results.}\footnote{We also tested whether the manufactured stars in the treatment groups attracted additional organic stars beyond those we added and found no evidence of this (untabulated).}

Additionally, to test whether packages with greater technical complexity benefit more from social proof, we tested if treatment effects varied by package complexity, proxied by repository size (total size of all files in MB) and \textit{README} documentation length (\cref{sec:het-analyses}). We found no significant interactions at any post-treatment snapshot date; however, the confidence intervals on the interaction terms are sufficiently wide that we cannot rule out meaningfully large differential effects. We return to the power limitations of these tests in the discussion.

Finally, we tested whether the increase in stars spilled over into broader GitHub activity on treated repositories. Using post-treatment data from GitHub Archive, we estimated treatment effects on six additional metrics: forks, pushes, pull requests, issues opened, issues closed, and releases. We found no significant differences between treated and control groups on any metric (\cref{sec:gharchive-outcomes}).

\section{PyPI Experiment}
\label{sec:pypi-experiment}
We experimentally manipulate human downloads and test the hypotheses that higher human downloads lead to yet higher future human downloads.\footnote{We also analyzed if human downloads Granger-cause human downloads. We find that they do. However, we are cautious about overinterpreting the results, given that they conflict with the experimental findings.}

\subsection{Design}
We randomly sampled 50,000 packages from the PyPI repository and filtered to those with at least five human downloads (\cref{sec:measures-pypi-downloads}). 
% for numbers, see https://github.com/soodoku/social_proof_stars/blob/main/pydownloads/scripts/consolidate_pypi_downloads.ipynb
This left us with 23,916 packages. We then randomly assigned 20\% of the packages to the treatment group ($n = 4,814$) and the remaining 80\% to the control group ($n = 19,102$). As with the GitHub experiment, we confirmed pre-treatment balance across various GitHub activity metrics (\cref{sec:balance-gharchive}). For packages in the treatment group, we wrote a script to download the same package 100 times. We downloaded the package in a way that each download shows up in the official Linehaul numbers (\cref{sec:measures-pypi-downloads}).\footnote{When installing packages, most user systems usually cache the source download files (e.g., \texttt{.tar.gz}, etc.) on the local drive to save on bandwidth resources. Cognizant of this, we wrote the script so that it installs packages without using the cached download directory. This option forces the process to always download the package from PyPI instead of using the cached source files.}. We treated the packages between June 3, 2023, and June 8, 2023. \cref{fig:median_downloads_pypi_experiment} serves as our manipulation check, with the gray region indicating the sharp increase in downloads within the treatment period.

\subsection{Results}

\begin{figure}[ht]
\centering
    \includegraphics[width=\textwidth]{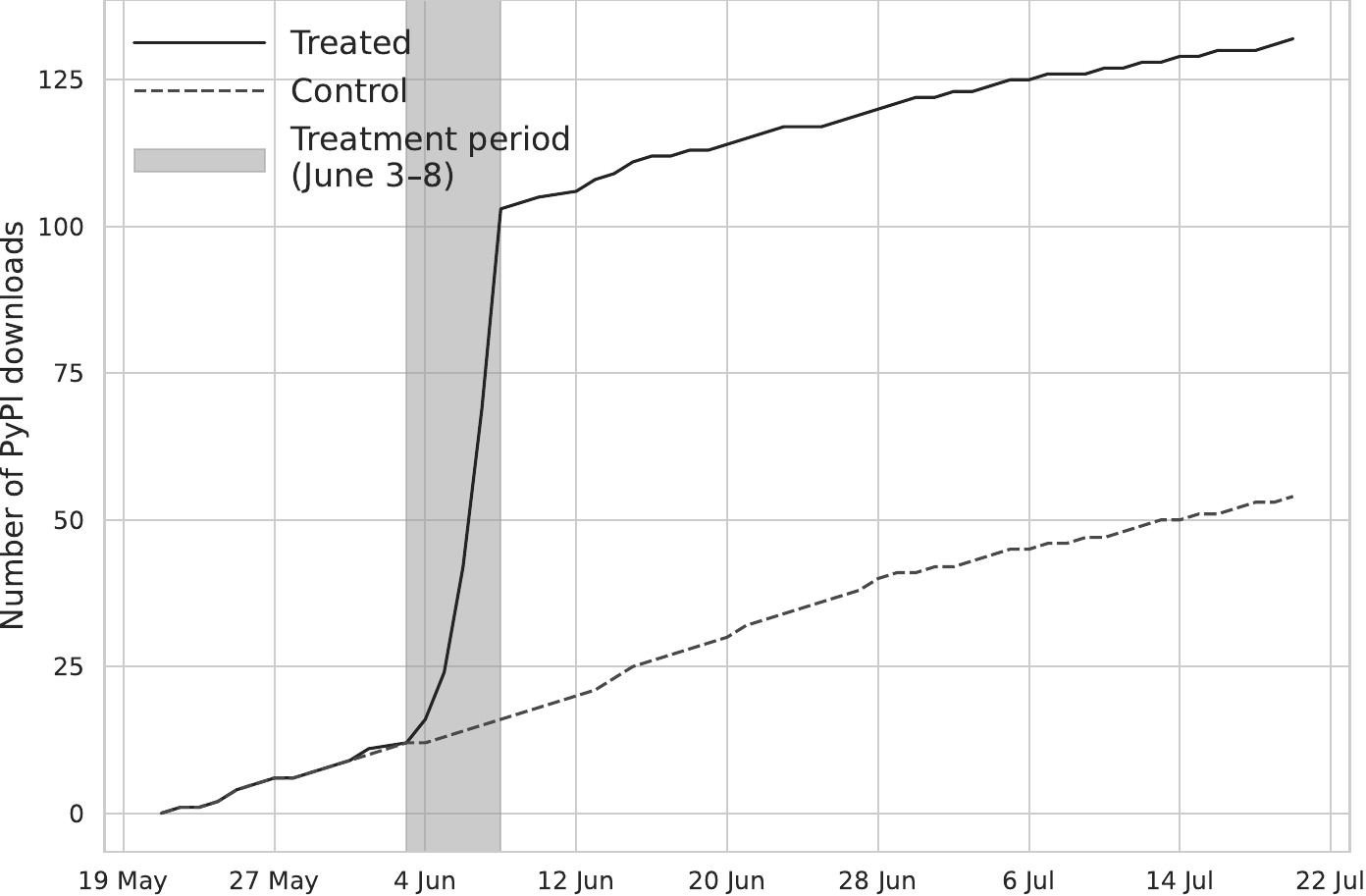}
\caption{
\textbf{Median PyPI downloads for Treated vs. Control in the PyPI Experiment.} The figure shows trends in median daily downloads for the treated packages (n = 4,814) and control group packages (n = 19,102) for 1,458,876 package-day observations. 
% for numbers, see https://github.com/soodoku/social_proof_stars/blob/main/pydownloads/scripts/consolidate_pypi_downloads.ipynb
The shaded vertical bar indicates the treatment period. Downloads include only human downloads (\cref{tab:human_bot_download_classification}). See \cref{fig:mean_downloads_pypi_experiment} for the same figure of mean downloads. \cref{tab:pypi_exp_regtable} reports the estimates for the differences in medians.
}
\label{fig:median_downloads_pypi_experiment}
\end{figure}

\cref{fig:median_downloads_pypi_experiment} visualizes our results for the PyPI experiment. Downloads are extremely volatile. To mute the effects of extreme outliers, we focus on the medians (as with \cref{sec:github}). \cref{fig:median_downloads_pypi_experiment} plots medians for the treatment and control groups. Pre-treatment, little separates the treatment and control groups. The treatment causes the daily median series for the treatment and control group to diverge over the treatment period (June 3--8). On June 8, the median treatment package had 83 more downloads relative to the control group ($p < .001$, column (1) of \cref{tab:pypi_exp_regtable}). However, after the treatment application period, the difference between the series is roughly constant. Treated packages, if anything, have a less sharp slope than the control group; the median difference is .2 fewer downloads per day ($p < .001$, column (2) of \cref{tab:pypi_exp_regtable}). The LATE estimates for compliers are no different (\cref{tab:pypi_exp_regtable_late}). Overall, providing social proof doesn't appear to increase downloads of the treated packages.

Finally, we tested whether the increase in downloads spilled over into GitHub activity. We find that manipulated download counts on PyPI did not lead to higher numbers of stars, forks, pushes, pull requests, issues, or releases in the corresponding GitHub repositories (\cref{sec:gharchive-outcomes}).

\subsection{Detectable Effect Bounds}
\label{sec:bounds}
To help interpret the absence of observed effects, we estimate the minimum effect sizes that our data could have reliably detected, given the observed variance. 
Based on the standard error of the estimated treatment effect (\cref{tab:pypi_exp_regtable}), our experiment had 80\% power to detect treatment effects exceeding a minimal detectable effect size of approximately 0.03 standard deviations.
This corresponds to an effect size of over 86k downloads, roughly 83\% of the average number of downloads in our sample.
Thus, we can confidently rule out large effects of this magnitude or greater.%
\footnote{
    Based on the estimated standard error of the mean treatment effect ($\hat{SE} \approx 30.8k$ downloads, column (4) of \cref{tab:pypi_exp_regtable}) and using the two-sided test with $\alpha = .05$ and power of 80\%, the minimal detectable effect size is $\approx (1.96 + 0.84) \times \hat{SE} \approx 86.2k$ downloads.
    %  3,011,376 = SD of control group in post-treatment period
    % cohen's D = 86.2k / 3,011,376 ~= 0.0286
    Using the observed post-treatment standard deviation of $3,011.4k$ downloads in the control group, this corresponds to approximately $0.029$ standard deviations.
    % 86.2k / 104.1k = 0.828
}
The 95\% confidence interval yields a similar conclusion.
% 73% = 76.4k / 104.1k
Expressed in relative terms, the true effect is very unlikely to exceed 73\% of the mean outcome.

Given the skewed distribution of downloads, we also assess differences in medians.
Based on the estimated standard errors (column (2) of \cref{tab:pypi_exp_regtable}), the minimum detectable median effect is 101.12 downloads (i.e., 100 mechanical from our intervention + 1.12 behavioral effect), corresponding to a 0.4\% increase relative to the median baseline (23.9k downloads).
In practical terms, this means our experiment was powered to rule out behavioral amplification effects at the median in the range of 10–30\%.
Ultimately, as we discuss further in \cref{sec:discussion}, our design prioritized ethical constraints and ecological realism, which limits our ability to detect subtle behavioral shifts.
As such, while we can rule out large effects, smaller or targeted effects remain within statistical uncertainty.

\section{Discussion}
\label{sec:discussion}
% CONTRIBUTION AND NULL RESULTS
Many existing studies examine how people use cheap decision-making heuristics in technology adoption \citep{simtech, Pickerill_2020}. Separately, a growing body of research has documented the rise in malware delivered through dependencies within the open-source software supply chain \citep{Ohm2020, cao2022fork, Ladisa2023, Wermke, CheckPoint2024, Forbes2024, he2024, Sonatype2024}. \cite{he2024} find that most faked stars on GitHub are used to promote disguised malware. We contribute to this literature by conducting two field experiments to evaluate whether artificially inflated social proof influences software adoption. In both experiments, we find no detectable effect. Manipulating star counts on GitHub did not increase downstream package downloads, and manipulating download counts on PyPI did not increase subsequent organic downloads. These null results obtain at the treatment intensities our ethical constraints permitted; we cannot rule out effects at the scale that commercial astroturfing campaigns deploy.

% COMPARISON TO PRIOR WORK
Our findings contrast with those of prior studies \citep{salganik2006experimental, Fang2022}, including field experiments \citep{Muchnik2013, Rijt2014}. In an artificial music market, \cite{salganik2006experimental} find that displaying download counts amplifies the success of higher-ranked songs. \cite{Rijt2014} conducted field experiments across four platforms (Kickstarter, Epinions, Wikipedia, and Change.org), finding that small early endorsements increase the likelihood of subsequent success. \cite{Muchnik2013} find that manipulated upvotes on a social news aggregation site increased subsequent positive ratings, demonstrating herding effects, although effects were topic-dependent and critically absent for IT-related posts. In a non-randomized natural experiment on GitHub, \cite{Fang2022} find that tweets promoting repositories attract more stars, though the effect is weaker when the tweet author is affiliated with the promoted repository. A key difference between our setting and these prior studies is that software adoption involves a consequential technical decision---integrating a dependency into a codebase---where the cost of a poor choice (broken builds, security vulnerabilities, maintenance burden) provides strong motivation to evaluate quality directly rather than rely on peripheral cues.

% THEORETICAL INTERPRETATION
Software adoption may differ from these social settings in ways that attenuate the influence of peripheral metrics. The Elaboration Likelihood Model \citep{ELMppr, ELMbook} predicts that peripheral cues---star and download counts---carry the most weight when decision makers lack either the motivation or ability to evaluate the underlying quality. Developers choosing a new package for a project are motivated to evaluate quality and have at least some ability to do so through code documentation, commit frequency, contributor activity, and project responsiveness \citep{Dabbish2012, Tsay2014, BORGES2018112}, and thus may give less weight to peripheral metrics. The \cite{Muchnik2013} IT-null result is consistent with this account: IT-literate users, like developers, may process content via the central route rather than deferring to social cues. The finding by \cite{Fang2022} that signal source credibility moderates developer responses to endorsements further supports the view that developers are not passively influenced by peripheral metrics.

Signaling Theory \citep{Spence1973} offers a complementary lens. Signals lose credibility when they are easy to fake. GitHub stars are commercially available at scale for trivial cost, and the existence of these markets is widely known among developers, giving reason to discount star counts as reliable quality indicators \citep{Campbell1979, he2024}.

The ability to evaluate quality varies considerably across contexts, however. Assessing package safety, maintenance trajectory, and fitness for purpose remains difficult without downloading and inspecting the code, and sometimes even after doing so. In agentic coding workflows, where LLM-based tools select and install packages with limited human oversight, the scrutiny that might attenuate social proof effects is largely absent. Conversely, in corporate environments, organizations increasingly rely on software composition analysis tools and internal registries that gate which packages developers can install, substituting institutional review for individual judgment. The role of peripheral signals like stars is therefore neither uniformly decisive nor uniformly irrelevant, but depends on the decision-making environment. Our null results should not be read as evidence that social proof is irrelevant to software adoption, but rather that a single manipulated metric, at modest intensity, did not detectably shift behavior in our sample.

% IMPLICATIONS FOR PLATFORM DESIGN
\subsection{Implications for Platform Design}
These findings nonetheless carry practical implications. Even if stars did not shift adoption in our experiments, the threat of social proof manipulation persists as manipulation scales up and as decision-making contexts shift toward less scrutinized environments. Several concrete platform interventions could reduce the efficacy of such manipulation. First, GitHub could surface richer contextual information alongside star counts: star velocity over time (since sudden spikes are a common manipulation signature), the proportion of stargazers with substantive commit histories, and the account age distribution of stargazers. Second, platforms could more prominently integrate composite security health metrics. The OpenSSF Scorecard, for instance, aggregates automated checks across code review practices, dependency management, vulnerability disclosure, and CI/CD configuration into a single risk-weighted score, offering a harder-to-fake alternative to raw popularity counts. Third, the emerging infrastructure of build provenance and trusted publishing---now supported by PyPI, npm, RubyGems, and NuGet via OpenID Connect attestations---provides cryptographic verification of where and how a package was built, shifting trust from social signals to verifiable supply chain evidence. Fourth, beyond removing fake stars after the fact, GitHub could impose graduated consequences on repositories that engage in coordinated manipulation, including freezing or flagging repositories, particularly those distributing malicious payloads \citep{he2024}. Such enforcement reduces the expected payoff to astroturfing regardless of whether individual developers attend to these signals.

% ETHICS
\subsection{Ethical Considerations}
Like other field experiments \citep{Muchnik2013, Rijt2014}, this study implements \emph{in situ} manipulation of publicly visible metrics on live, real-world platforms without the prior consent of platform users---a design choice necessary for ecological validity. Our protocol was designed with reference to the \emph{ACM Publications Policy on Research Involving Human Participants and Subjects}, the \emph{ACM Code of Ethics and Professional Conduct}, and the \emph{Menlo Report} principles for ICT research \citep{acm-human-participants, acm-code-ethics, menlo-ieee-2012}. As \citet{Kohno2023} argue, IRB approval alone does not constitute sufficient ethical justification; we therefore draw on the consequentialist framework to assess whether the anticipated benefits of our design outweigh its potential harms.

The potential harms were narrow in scope and limited in duration. We added stars to newly published, low-visibility repositories and incremented download counts for a bounded period (see \cref{fig:median_downloads_pypi_experiment}). We did not collect personal or sensitive data, alter code, documentation, or dependencies, deploy malicious content, or expose users to security risks. The number of stars added was large enough to be visible for packages with minimal prior exposure, but modest enough to avoid triggering trending algorithms or drawing unusual scrutiny. Similarly, in the PyPI experiment, scripted downloads were sufficient to shift download tallies but unlikely to meaningfully mislead users or affect rankings at scale. Any reputational effect on treated repositories was small and temporary.

The potential benefits, by contrast, are substantial. Supply chain attacks through manipulated social proof represent a documented and growing threat \citep{Ohm2020, Ladisa2023, Ohm2023, he2024, Sonatype2024}. Establishing whether such manipulation actually shifts developer behavior is a prerequisite for designing effective countermeasures---and this question cannot be answered without \emph{in situ} experimentation on real platforms.

We nonetheless acknowledge the deontological tension inherent in this design: package maintainers were not consented as experimental units, a concern not dissolved by the null result. The conservative design choices described above represent our effort to minimize rights violations within the constraint that full prior consent would destroy ecological validity.

% LIMITATIONS
\subsection{Limitations}
The ethical constraints of a limited treatment scope, while necessary, have implications for statistical power. Although our analysis (\cref{sec:bounds}) provides evidence against large effects, we cannot exclude the possibility that a more intense or sustained manipulation---on the scale that commercial astroturfing campaigns would deploy---would produce a measurable behavioral change in developers. Real astroturfing campaigns maintain and escalate signals over time, whereas our treatment was a one-shot boost whose salience may decay in ways that sustained manipulation would not. Moreover, while we test for treatment effects that vary linearly over time, which likewise yield nulls, identifying dose-response non-linearities would require multiple treatment-intensity arms, and detecting threshold effects in time would require greater temporal granularity and statistical power than our experiment affords \citep{GelmanHillVehtari2020}.

We also tested for heterogeneous treatment effects by package complexity using repository size and documentation length as proxies, under the hypothesis that peripheral cues like stars carry more weight when direct evaluation is harder \citep{ELMppr, ELMbook}. We found no significant interactions; however, the confidence intervals on the interaction terms are wide enough that we cannot reject substantively meaningful differential effects (see \crefrange{tab:github_exp_medians_regtable_allhumaninstallers_high_size_mb}{tab:github_exp_medians_regtable_allhumaninstallers_high_processed_readme_len}). Our sample provides limited statistical power to detect interactions, which require roughly sixteen times the sample size needed for main effects under reasonable assumptions about relative effect sizes \citep{GelmanHillVehtari2020}. The absence of significant heterogeneity is therefore consistent with both a genuinely uniform null and with meaningful moderation that our design is underpowered to detect. These proxies are also crude measures of how difficult a package is to evaluate by inspection. Metrics that more directly capture evaluability, such as cyclomatic complexity, dependency depth, or test coverage, would provide sharper tests of the ELM prediction that peripheral cues matter most when direct assessment is difficult.

A further threat to validity concerns the outcome measure itself. We filter PyPI downloads to remove those by bot installers (\cref{sec:measures-pypi-downloads}), but it remains possible that our measures include downloads from automated CI/CD pipelines and container builds that install packages as part of testing and deployment workflows. This concern is mitigated by the fact that our sample consists of newly published, low-visibility packages that are unlikely to feature in many automated build pipelines. In any case, such automated downloads would not respond to star manipulation, so any residual contamination would attenuate treatment effects and bias toward the null rather than generate spurious effects.

Finally, the scope of our experiments limits generalizability. Both studies focus on newly published and low-visibility Python packages. Peripheral heuristics like star counts are theorized to be most influential precisely when richer evaluation signals are absent \citep{ELMppr, ELMbook}---which is the condition our sample reflects. Our null results are therefore unlikely to be reversed for established packages with longer track records and more credible social proof, though we note that stars may function differently in that context, serving as coordination signals or search-ranking inputs rather than as quality cues. The findings are also specific to the Python ecosystem and the 2023 period. Behavioral effects may differ across other language ecosystems and over time as platforms evolve.

% FUTURE DIRECTIONS
\subsection{Future Directions}
Several directions for future research follow. First, larger doses within ethical constraints could help address whether the null effects would hold under more intense manipulation. We note, however, that a recurring tension in this line of research is that ecologically valid field experiments necessitate live-platform manipulation without participant consent. One path toward better deontological alignment \citep{Kohno2023} might be opt-in experimental registries, where package maintainers and platform users consent in advance to participation. Another possible design is pairs of the same simulated but safe packages with varied manipulated metrics (and disguised names) to track differences in uptake without the ethical costs of live-platform intervention. Conjoint or discrete choice experiments offer yet another avenue: developers could be presented with hypothetical package profiles varying in stars, downloads, documentation quality, and maintenance recency, allowing researchers to estimate the marginal weight placed on each attribute without any platform manipulation at all. Such survey-experimental designs sacrifice ecological validity but permit far larger samples, sharper identification of interaction effects, and full control over the signal environment.

Second, the rise of agentic coding workflows introduces a qualitatively different decision-making process that merits dedicated investigation. When LLM-based tools select and install packages, the adoption decision is mediated by a model's training data and retrieval context rather than by a developer's direct evaluation of code quality. Studies that present coding agents with package selection tasks under experimentally varied star counts and download metrics could isolate whether these tools are more susceptible to peripheral cues than human developers. Such studies avoid the ethical costs of live manipulation entirely, since the decision maker is a model rather than an uninformed human participant.

Finally, future studies should examine other language ecosystems (e.g., npm), where supply chain attacks have been increasingly documented \citep{Ohm2020, Ladisa2023, Ohm2023, Sonatype2024}. Cross-ecosystem comparisons are particularly valuable because platforms differ in how prominently they surface social proof metrics and in the maturity of their institutional safeguards, such as software composition analysis tools and curated registries, offering natural variation in the conditions under which peripheral cues may influence adoption.

% CONCLUSION
\subsection{Conclusion}
We close with a note of caution, mindful of Carl Sagan's famous aphorism: ``The absence of evidence is not evidence of absence.'' The threat persists as long as social proof metrics remain gameable and corruptible, and people are insufficiently aware of their manipulability. As a recent study \citep{he2024} shows, the scale of fake stars on GitHub is in the millions across thousands of repositories between 2019 and 2024. Malware on PyPI is rising as Python packages are increasingly exploited through dependency-based attacks \citep{Ohm2023}. The growing adoption of agentic coding workflows, in which LLM-based tools select and install packages with limited human oversight, may further reduce the scrutiny that individual developers apply to adoption decisions, potentially increasing the efficacy of astroturfing at precisely the moment when its consequences are most severe. These trends highlight the threat that astroturfing poses to online security.

%TC:ignore

% \clearpage
\singlespacing
\small
\section*{Declarations}
\addcontentsline{toc}{section}{Declarations}

\subsection*{Funding}
Not applicable.

\subsection*{Ethical approval}
This study was reviewed by the Colorado State University Institutional Review Board (under CSU IRB Protocol 7411).

\subsection*{Data Availability Statement}
Code and data are published under an open-source license at \url{http://github.com/themains/social_proof_stars}.
% Code and data are published under an open-source license at \url{http://github.com/xxxxxxx/xxxxxx_xxxxx_xxxxx} [masked for anonymized submission]. 

\subsection*{Conflict of Interest}
The authors have no conflicts of interest to declare.

\clearpage
\bibliographystyle{apsr}
\bibliography{social_proof}
\clearpage

\appendix
\renewcommand{\thesection}{SI \arabic{section}}
\renewcommand\thetable{\thesection.\arabic{table}}  
\renewcommand\thefigure{\thesection.\arabic{figure}}
\counterwithin{figure}{section}
\counterwithin{table}{section}

\section{Supporting Information}\label{si}

\FloatBarrier

\subsection{GitHub Experiment}

\subsubsection{Balance tests: GitHub repositories characteristics}\label{sec:balance-tests-repository-package}

\begin{table}[ht]\small \setlength\tabcolsep{5 pt}
\caption{Balance tests: GitHub repositories characteristics}
\label{tab:baltest-repo-treated-01}
\begin{adjustbox}{max width=\textwidth}
%%% Table created in Stata by command iebaltab
%%% (https://github.com/worldbank/ietoolkit)
%%% (https://dimewiki.worldbank.org/iebaltab)
%%% The command was specified exactly like this: 
%%% iebaltab size_mb fork year_created subscribers_count has_issues forks open_issues n_topics python_lang , total groupvar(treated) star(.1 .05 .01) stats(pair(nrmd)) nonote grplabels( 0 Control @ 1 Treated @ ) order(0 1) control(0) grouplabels(0 "Control packages" @ 1 "Treated packages") rowlabels( size_mb "Repository size" @ year_created "Year created" @ fork "Forked = 1" @ subscribers_count "Subscribers" @ has_issues "Has issues = 1" @ forks "Number of forks" @ open_issues "Number of open issues" @ n_topics "Number of topics listed" @ python_lang "Python = 1" ) totallabel(Full sample) format(%9.2f) savetex(../output/baltest-repo-treated-01.tex) replace

\begin{tabular}{@{\extracolsep{5pt}}lcccccccc}
\\[-1.8ex]\hline \hline \\[-1.8ex]
 & \multicolumn{2}{c}{(1)}  & \multicolumn{2}{c}{(2)}  & \multicolumn{2}{c}{(3)}  & \multicolumn{2}{c}{(3)-(2)} \\
 & \multicolumn{2}{c}{Full sample}  & \multicolumn{2}{c}{Control packages}  & \multicolumn{2}{c}{Treated packages}  & \multicolumn{2}{c}{Pairwise t-test}  \\
Variable & N & Mean/(SE) & N & Mean/(SE) & N & Mean/(SE) & N & Normalized difference \\ \hline \\[-1.8ex] 
Repository size   & 582    & 7.85    & 485    & 8.44    & 97    & 4.89    & 582    & -0.14   \\
 &   & (1.22)  &   & (1.43)  &   & (1.63)  &   &  \\ [1ex]
Forked = 1   & 582    & 0.04    & 485    & 0.04    & 97    & 0.05    & 582    & 0.07   \\
 &   & (0.01)  &   & (0.01)  &   & (0.02)  &   &  \\ [1ex]
Year created   & 582    & 2022.31    & 485    & 2022.32    & 97    & 2022.28    & 582    & -0.02   \\
 &   & (0.08)  &   & (0.08)  &   & (0.20)  &   &  \\ [1ex]
Subscribers   & 582    & 4.33    & 485    & 3.84    & 97    & 6.75    & 582    & 0.18**   \\
 &   & (0.54)  &   & (0.50)  &   & (2.02)  &   &  \\ [1ex]
Has issues = 1   & 582    & 0.96    & 485    & 0.96    & 97    & 0.95    & 582    & -0.05   \\
 &   & (0.01)  &   & (0.01)  &   & (0.02)  &   &  \\ [1ex]
Number of forks   & 582    & 19.87    & 485    & 15.22    & 97    & 43.12    & 582    & 0.18**   \\
 &   & (4.78)  &   & (3.92)  &   & (20.89)  &   &  \\ [1ex]
Number of open issues   & 582    & 4.57    & 485    & 4.19    & 97    & 6.46    & 582    & 0.08   \\
 &   & (1.14)  &   & (1.26)  &   & (2.67)  &   &  \\ [1ex]
Number of topics listed   & 582    & 1.76    & 485    & 1.76    & 97    & 1.76    & 582    & 0.00   \\
 &   & (0.14)  &   & (0.15)  &   & (0.35)  &   &  \\ [1ex]
Python = 1   & 582    & 0.86    & 485    & 0.87    & 97    & 0.82    & 582    & -0.13   \\
 &   & (0.01)  &   & (0.02)  &   & (0.04)  &   &  \\ [1ex]
\hline \hline \\[-1.8ex]

\end{tabular}

\end{adjustbox}
\caption*{
\footnotesize 
\emph{Notes---}Table reports the balance in GitHub repository characteristics between the treated and control Python packages.
Column (1) reports the mean of repository characteristics. 
Column (2) reports the mean of the control packages.
Column (3) reports the mean of the treated packages.
Standard errors of the mean are in parentheses.
The last column reports the standardized mean difference (the difference in group means divided by the pooled standard deviation).  
The repository size is in megabytes. 
Forked indicates whether the repository was forked from another existing repository. 
Topics are optional labels for a GitHub repository (e.g., web application, encryption, Python). 
Python = 1 indicates Python is the primary detected language.
% e.g. TensorFlow's topics: machine-learning deep-neural-networks deep-learning neural-network tensorflow ml distributed
See \cref{tab:baltest-repo-treated-012} for the same table with low and high dosage treatments. 
See \cref{tab:baltest-readme-treated-01} for the same table reporting balance for package description and dependency balance. 
Significance levels: $^{***} p < .01; ^{**} p < .05; ^{*} p <.1$.}
\end{table}

\begin{table}[!ht]\small \setlength\tabcolsep{5 pt}
\caption{Balance tests: GitHub repositories characteristics (low-dosage and high-dosage treatment groups vs. control)}
\label{tab:baltest-repo-treated-012}
\begin{adjustbox}{max width=\textwidth}
%%% Table created in Stata by command iebaltab
%%% (https://github.com/worldbank/ietoolkit)
%%% (https://dimewiki.worldbank.org/iebaltab)
%%% The command was specified exactly like this: 
%%% iebaltab size_mb fork year_created subscribers_count has_issues forks open_issues n_topics python_lang , groupvar(treated2) star(.1 .05 .01) stats(pair(nrmd)) nonote grplabels( 0 Control @ 1 Treated (low) @ 2 Treated (high) @ ) order(0 1 2) control(0) grouplabels(0 "Control packages" @ 1 "Treated (low dose)" @ 2 "Treated (high dose)") rowlabels( year_created "Year created" @ fork "Forked = 1" @ size_mb "Repository size" @ has_issues "Has issues = 1" @ forks "Number of forks" @ open_issues "Number of open issues" @ subscribers_count "Subscribers" @ n_topics "Number of topics listed" @ python_lang "Python = 1" ) format(%9.2f) savetex(../output/baltest-repo-treated-012.tex) replace

\begin{tabular}{@{\extracolsep{5pt}}lcccccccccc}
\\[-1.8ex]\hline \hline \\[-1.8ex]
 & \multicolumn{2}{c}{(1)}  & \multicolumn{2}{c}{(2)}  & \multicolumn{2}{c}{(3)}  & \multicolumn{2}{c}{(2)-(1)} & \multicolumn{2}{c}{(3)-(1)} \\
 & \multicolumn{2}{c}{Control packages}  & \multicolumn{2}{c}{Treated (low dose)}  & \multicolumn{2}{c}{Treated (high dose)}  & \multicolumn{4}{c}{Pairwise t-test}  \\
Variable & N & Mean/(SE) & N & Mean/(SE) & N & Mean/(SE) & N & Normalized difference & N & Normalized difference \\ \hline \\[-1.8ex] 
Repository size   & 485    & 8.44    & 73    & 4.43    & 24    & 6.28    & 558    & -0.16    & 509    & -0.09   \\
 &   & (1.43)  &   & (1.98)  &   & (2.76)  &   &  &   &  \\ [1ex]
Forked = 1   & 485    & 0.04    & 73    & 0.04    & 24    & 0.08    & 558    & 0.02    & 509    & 0.19   \\
 &   & (0.01)  &   & (0.02)  &   & (0.06)  &   &  &   &  \\ [1ex]
Year created   & 485    & 2022.32    & 73    & 2022.49    & 24    & 2021.63    & 558    & 0.10    & 509    & -0.30*   \\
 &   & (0.08)  &   & (0.19)  &   & (0.55)  &   &  &   &  \\ [1ex]
Subscribers   & 485    & 3.84    & 73    & 4.45    & 24    & 13.75    & 558    & 0.05    & 509    & 0.42***   \\
 &   & (0.50)  &   & (1.63)  &   & (6.37)  &   &  &   &  \\ [1ex]
Has issues = 1   & 485    & 0.96    & 73    & 0.96    & 24    & 0.92    & 558    & 0.00    & 509    & -0.17   \\
 &   & (0.01)  &   & (0.02)  &   & (0.06)  &   &  &   &  \\ [1ex]
Number of forks   & 485    & 15.22    & 73    & 17.05    & 24    & 122.42    & 558    & 0.02    & 509    & 0.40***   \\
 &   & (3.92)  &   & (11.25)  &   & (76.12)  &   &  &   &  \\ [1ex]
Number of open issues   & 485    & 4.19    & 73    & 3.64    & 24    & 15.04    & 558    & -0.02    & 509    & 0.29*   \\
 &   & (1.26)  &   & (1.70)  &   & (9.39)  &   &  &   &  \\ [1ex]
Number of topics listed   & 485    & 1.76    & 73    & 1.84    & 24    & 1.54    & 558    & 0.02    & 509    & -0.08   \\
 &   & (0.15)  &   & (0.44)  &   & (0.49)  &   &  &   &  \\ [1ex]
Python = 1   & 485    & 0.87    & 73    & 0.82    & 24    & 0.83    & 558    & -0.13    & 509    & -0.10   \\
 &   & (0.02)  &   & (0.05)  &   & (0.08)  &   &  &   &  \\ [1ex]
\hline \hline \\[-1.8ex]

\end{tabular}

\end{adjustbox}
\caption*{
\footnotesize 
\emph{Notes---}
Same as \cref{tab:baltest-repo-treated-01}, except the treatment group is distinguished by low and high dosage treatment packages (see \cref{sec:gh-treatment-conditions}).
*** Significant at the 1 percent level. ** Significant at the 5 percent level. * Significant at the 10 percent level.
\cref{sec:balance-gharchive} reports balance tests for pre-treatment repository-level events from GitHub archive.
}
\end{table}

\clearpage
\FloatBarrier
% ~~~~~~~~~~~~~~~~~~~~~~~~~~~~~~~~~~~~~~~~~~~~~~~~~~~~~~~~~~~~~~
\subsubsection{Balance tests: PyPI package characteristics}\label{sec:balance-tests-pypi}

\begin{table}[!ht]\small \setlength\tabcolsep{5 pt}
\caption{Balance tests: PyPI package characteristics}
\label{tab:baltest-readme-treated-01}
\begin{adjustbox}{max width=\textwidth}
%%% Table created in Stata by command iebaltab
%%% (https://github.com/worldbank/ietoolkit)
%%% (https://dimewiki.worldbank.org/iebaltab)
%%% The command was specified exactly like this: 
%%% iebaltab raw_readme_len processed_readme_len n_requirements , total groupvar(treated) star(.1 .05 .01) stats(pair(nrmd)) nonote grplabels( 0 Control @ 1 Treated @ ) order(0 1) control(0) grouplabels(0 "Control packages" @ 1 "Treated packages") rowlabels( raw_readme_len "Package description length" @ processed_readme_len "Package description length (cleaned)" @ n_requirements "Number of dependencies" @ ) totallabel(Full sample) format(%9.2f) savetex(../output/baltest-readme-treated-01.tex) replace

\begin{tabular}{@{\extracolsep{5pt}}lcccccccc}
\\[-1.8ex]\hline \hline \\[-1.8ex]
 & \multicolumn{2}{c}{(1)}  & \multicolumn{2}{c}{(2)}  & \multicolumn{2}{c}{(3)}  & \multicolumn{2}{c}{(3)-(2)} \\
 & \multicolumn{2}{c}{Full sample}  & \multicolumn{2}{c}{Control packages}  & \multicolumn{2}{c}{Treated packages}  & \multicolumn{2}{c}{Pairwise t-test}  \\
Variable & N & Mean/(SE) & N & Mean/(SE) & N & Mean/(SE) & N & Normalized difference \\ \hline \\[-1.8ex] 
Package description length (raw)   & 622    & 2458.61    & 522    & 2516.53    & 100    & 2156.24    & 622    & -0.11   \\
 &   & (143.49)  &   & (161.85)  &   & (287.41)  &   &  \\ [1ex]
Package description length (cleaned)   & 622    & 2363.36    & 522    & 2416.14    & 100    & 2087.89    & 622    & -0.10   \\
 &   & (138.58)  &   & (156.14)  &   & (280.47)  &   &  \\ [1ex]
Number of dependencies   & 622    & 3.73    & 522    & 3.60    & 100    & 4.36    & 622    & 0.10   \\
 &   & (0.31)  &   & (0.34)  &   & (0.66)  &   &  \\ [1ex]
\hline \hline \\[-1.8ex]

\end{tabular}

\end{adjustbox}
\caption*{
\footnotesize 
\emph{Notes---}Table reports the balance in PyPI package characteristics between the treated and control Python packages.
Column (1) reports the mean of repository characteristics. 
Column (2) reports the overall mean.
Column (3) reports the mean of the treated packages.
Standard errors of the mean are in parentheses.
The last column reports the standardized mean differences.
Package description length is the length of the (optional) package description, which can include instructions on installation and usage, etc. 
These usually come from a \texttt{Readme} file.
The description may have been marked up to be rendered in HTML, so we do some basic cleaning of the raw text to convert HTML to text and report it in the table.
Number of dependencies is based on the number of requirements the package relies on.
See also \cref{tab:baltest-repo-treated-01} for balance tests in GitHub repository characteristics.
See \cref{tab:baltest-readme-treated-012} for the same table with low and high-dosage treatments.
Significance levels: $^{***} p < .01; ^{**} p < .05; ^{*} p <.1$. 
}
\end{table}

\begin{table}[!ht]\small \setlength\tabcolsep{5 pt}
\caption{Balance tests: PyPI package characteristics (low-dosage and high-dosage treatment groups vs. control)}
\label{tab:baltest-readme-treated-012}
\begin{adjustbox}{max width=\textwidth}
%%% Table created in Stata by command iebaltab
%%% (https://github.com/worldbank/ietoolkit)
%%% (https://dimewiki.worldbank.org/iebaltab)
%%% The command was specified exactly like this: 
%%% iebaltab raw_readme_len processed_readme_len n_requirements , groupvar(treated2) star(.1 .05 .01) stats(pair(nrmd)) nonote grplabels( 0 Control @ 1 Treated (low) @ 2 Treated (high) @ ) order(0 1 2) control(0) rowlabels( raw_readme_len "Package description length" @ processed_readme_len "Package description length (cleaned)" @ n_requirements "Number of dependencies" @ ) format(%9.2f) savetex(../output/baltest-readme-treated-012.tex) replace

\begin{tabular}{@{\extracolsep{5pt}}lcccccccccc}
\\[-1.8ex]\hline \hline \\[-1.8ex]
 & \multicolumn{2}{c}{(1)}  & \multicolumn{2}{c}{(2)}  & \multicolumn{2}{c}{(3)}  & \multicolumn{2}{c}{(2)-(1)} & \multicolumn{2}{c}{(3)-(1)} \\
 & \multicolumn{2}{c}{Control}  & \multicolumn{2}{c}{Treated (low)}  & \multicolumn{2}{c}{Treated (high)}  & \multicolumn{4}{c}{Pairwise t-test}  \\
Variable & N & Mean/(SE) & N & Mean/(SE) & N & Mean/(SE) & N & Normalized difference & N & Normalized difference \\ \hline \\[-1.8ex] 
Package description length   & 522    & 2516.53    & 75    & 2145.51    & 25    & 2188.44    & 597    & -0.11    & 547    & -0.11   \\
 &   & (161.85)  &   & (351.09)  &   & (471.91)  &   &  &   &  \\ [1ex]
Package description length (cleaned)   & 522    & 2416.14    & 75    & 2077.33    & 25    & 2119.56    & 597    & -0.10    & 547    & -0.10   \\
 &   & (156.14)  &   & (342.40)  &   & (462.04)  &   &  &   &  \\ [1ex]
Number of dependencies   & 522    & 3.60    & 75    & 4.55    & 25    & 3.80    & 597    & 0.13    & 547    & 0.03   \\
 &   & (0.34)  &   & (0.77)  &   & (1.35)  &   &  &   &  \\ [1ex]
\hline \hline \\[-1.8ex]

\end{tabular}

\end{adjustbox}
\caption*{
\footnotesize 
\emph{Notes---}
Same as \cref{tab:baltest-readme-treated-01}, except the treatment group is distinguished by low and high dosage treatment packages (see \cref{sec:gh-treatment-conditions}).
See also \cref{tab:baltest-repo-treated-01} and \cref{tab:baltest-repo-treated-012}.
Significance levels: $^{***} p < .01; ^{**} p < .05; ^{*} p <.1$. 
}
\end{table}

\clearpage
\FloatBarrier
% ~~~~~~~~~~~~~~~~~~~~~~~~~~~~~~~~~~~~~~~~~~~~~~~~~~~~~~~~~~~~~~
\subsubsection{Balance tests: GitHub user characteristics}\label{sec:balance-tests-user}

\begin{table}[!ht]\small \setlength\tabcolsep{5 pt}
\caption{Balance tests: GitHub user characteristics}
\label{tab:baltest-user-treated-01}
\begin{adjustbox}{max width=\textwidth}
%%% Table created in Stata by command iebaltab
%%% (https://github.com/worldbank/ietoolkit)
%%% (https://dimewiki.worldbank.org/iebaltab)
%%% The command was specified exactly like this: 
%%% iebaltab public_repos public_gists followers following year_created year_updated org list_co list_email list_blog list_bio bio_size , total groupvar(treated) star(.1 .05 .01) stats(pair(nrmd)) nonote grplabels( 0 Control @ 1 Treated @ ) order(0 1) control(0) grouplabels(0 "Control packages" @ 1 "Treated packages") rowlabels( public_repos "Number of repositories" @ public_gists "Number of gists" @ followers "Number of followers" @ following "Number of people followed" @ year_created "Year created" @ year_updated "Year updated" @ org "Organization = 1" @ list_co "List company = 1" @ list_email "List email = 1" @ list_blog "List personal site = 1" @ list_bio "List brief bio = 1" @ bio_size "Brief bio length" @ ) totallabel(Full sample) format(%9.2f) savetex(../output/baltest-user-treated-01.tex) replace

\begin{tabular}{@{\extracolsep{5pt}}lcccccccc}
\\[-1.8ex]\hline \hline \\[-1.8ex]
 & \multicolumn{2}{c}{(1)}  & \multicolumn{2}{c}{(2)}  & \multicolumn{2}{c}{(3)}  & \multicolumn{2}{c}{(3)-(2)} \\
 & \multicolumn{2}{c}{Full sample}  & \multicolumn{2}{c}{Control packages}  & \multicolumn{2}{c}{Treated packages}  & \multicolumn{2}{c}{Pairwise t-test}  \\
Variable & N & Mean/(SE) & N & Mean/(SE) & N & Mean/(SE) & N & Normalized difference \\ \hline \\[-1.8ex] 
Number of repositories   & 545    & 57.57    & 453    & 51.80    & 92    & 85.96    & 545    & 0.11   \\
 &   & (12.88)  &   & (14.06)  &   & (32.05)  &   &  \\ [1ex]
Number of gists   & 545    & 3.28    & 453    & 3.03    & 92    & 4.51    & 545    & 0.10   \\
 &   & (0.56)  &   & (0.57)  &   & (1.72)  &   &  \\ [1ex]
Number of followers   & 545    & 189.85    & 453    & 194.56    & 92    & 166.66    & 545    & -0.02   \\
 &   & (78.36)  &   & (93.34)  &   & (66.05)  &   &  \\ [1ex]
Number of people followed   & 545    & 11.40    & 453    & 10.81    & 92    & 14.32    & 545    & 0.11   \\
 &   & (1.44)  &   & (1.58)  &   & (3.45)  &   &  \\ [1ex]
Year created   & 545    & 2017.33    & 453    & 2017.43    & 92    & 2016.86    & 545    & -0.16   \\
 &   & (0.16)  &   & (0.18)  &   & (0.36)  &   &  \\ [1ex]
Year updated   & 545    & 2017.33    & 453    & 2017.43    & 92    & 2016.86    & 545    & -0.16   \\
 &   & (0.16)  &   & (0.18)  &   & (0.36)  &   &  \\ [1ex]
Organization = 1   & 545    & 0.22    & 453    & 0.22    & 92    & 0.25    & 545    & 0.07   \\
 &   & (0.02)  &   & (0.02)  &   & (0.05)  &   &  \\ [1ex]
List company = 1   & 545    & 0.27    & 453    & 0.26    & 92    & 0.34    & 545    & 0.17   \\
 &   & (0.02)  &   & (0.02)  &   & (0.05)  &   &  \\ [1ex]
List email = 1   & 545    & 0.32    & 453    & 0.32    & 92    & 0.34    & 545    & 0.04   \\
 &   & (0.02)  &   & (0.02)  &   & (0.05)  &   &  \\ [1ex]
List personal site = 1   & 545    & 0.46    & 453    & 0.45    & 92    & 0.51    & 545    & 0.12   \\
 &   & (0.02)  &   & (0.02)  &   & (0.05)  &   &  \\ [1ex]
List brief bio = 1   & 545    & 0.48    & 453    & 0.47    & 92    & 0.51    & 545    & 0.08   \\
 &   & (0.02)  &   & (0.02)  &   & (0.05)  &   &  \\ [1ex]
Brief bio length   & 545    & 28.44    & 453    & 28.94    & 92    & 25.98    & 545    & -0.08   \\
 &   & (1.75)  &   & (1.95)  &   & (3.84)  &   &  \\ [1ex]
\hline \hline \\[-1.8ex]

\end{tabular}

\end{adjustbox}
\caption*{
\footnotesize 
\emph{Notes---}Table reports the balance in GitHub user characteristics between the treated and control Python packages.
Column (1) reports the mean of user characteristics. 
Column (2) reports the mean of users of the control packages.
Column (3) reports the mean of users of the treated packages.
Standard errors of the mean are in parentheses.
The last column reports the standardized mean difference.
The year created and year updated indicate when the GitHub user account was created and updated, respectively.
The organization indicates whether the package is hosted in an organization account.
Brief bio length is the number of characters in the (optional) biography.
See \cref{tab:baltest-user-treated-012} for the same table with low and high-dosage treatments.
Significance levels: $^{***} p < .01; ^{**} p < .05; ^{*} p <.1$. 
}
\end{table}

\begin{table}[!ht]\small \setlength\tabcolsep{5 pt}
\caption{Balance tests: GitHub user characteristics (low-dosage and high-dosage treatment groups vs. control)}
\label{tab:baltest-user-treated-012}
\begin{adjustbox}{max width=\textwidth}
%%% Table created in Stata by command iebaltab
%%% (https://github.com/worldbank/ietoolkit)
%%% (https://dimewiki.worldbank.org/iebaltab)
%%% The command was specified exactly like this: 
%%% iebaltab public_repos public_gists followers following year_created year_updated org list_co list_email list_blog list_bio bio_size , groupvar(treated2) star(.1 .05 .01) stats(pair(nrmd)) nonote grplabels( 0 Control @ 1 Treated (low) @ 2 Treated (high) @ ) order(0 1 2) control(0) rowlabels( public_repos "Number of repositories" @ public_gists "Number of gists" @ followers "Number of followers" @ following "Number of people followed" @ year_created "Year created" @ year_updated "Year updated" @ org "Organization" @ list_co "List company" @ list_email "List email" @ list_blog "List personal site" @ list_bio "List brief bio" @ bio_size "Brief bio length" @ ) format(%9.2f) savetex(../output/baltest-user-treated-012.tex) replace

\begin{tabular}{@{\extracolsep{5pt}}lcccccccccc}
\\[-1.8ex]\hline \hline \\[-1.8ex]
 & \multicolumn{2}{c}{(1)}  & \multicolumn{2}{c}{(2)}  & \multicolumn{2}{c}{(3)}  & \multicolumn{2}{c}{(2)-(1)} & \multicolumn{2}{c}{(3)-(1)} \\
 & \multicolumn{2}{c}{Control}  & \multicolumn{2}{c}{Treated (low)}  & \multicolumn{2}{c}{Treated (high)}  & \multicolumn{4}{c}{Pairwise t-test}  \\
Variable & N & Mean/(SE) & N & Mean/(SE) & N & Mean/(SE) & N & Normalized difference & N & Normalized difference \\ \hline \\[-1.8ex] 
Number of repositories   & 453    & 51.80    & 70    & 93.61    & 22    & 61.59    & 523    & 0.13    & 475    & 0.04   \\
 &   & (14.06)  &   & (41.71)  &   & (19.73)  &   &  &   &  \\ [1ex]
Number of gists   & 453    & 3.03    & 70    & 3.00    & 22    & 9.32    & 523    & -0.00    & 475    & 0.28**   \\
 &   & (0.57)  &   & (1.09)  &   & (6.32)  &   &  &   &  \\ [1ex]
Number of followers   & 453    & 194.56    & 70    & 103.07    & 22    & 369.00    & 523    & -0.06    & 475    & 0.11   \\
 &   & (93.34)  &   & (40.03)  &   & (244.44)  &   &  &   &  \\ [1ex]
Number of people followed   & 453    & 10.81    & 70    & 16.80    & 22    & 6.41    & 523    & 0.17    & 475    & -0.18   \\
 &   & (1.58)  &   & (4.45)  &   & (2.21)  &   &  &   &  \\ [1ex]
Year created   & 453    & 2017.43    & 70    & 2017.06    & 22    & 2016.23    & 523    & -0.10    & 475    & -0.32   \\
 &   & (0.18)  &   & (0.40)  &   & (0.79)  &   &  &   &  \\ [1ex]
Year updated   & 453    & 2017.43    & 70    & 2017.06    & 22    & 2016.23    & 523    & -0.10    & 475    & -0.32   \\
 &   & (0.18)  &   & (0.40)  &   & (0.79)  &   &  &   &  \\ [1ex]
Organization   & 453    & 0.22    & 70    & 0.26    & 22    & 0.23    & 523    & 0.09    & 475    & 0.02   \\
 &   & (0.02)  &   & (0.05)  &   & (0.09)  &   &  &   &  \\ [1ex]
List company   & 453    & 0.26    & 70    & 0.29    & 22    & 0.50    & 523    & 0.06    & 475    & 0.51**   \\
 &   & (0.02)  &   & (0.05)  &   & (0.11)  &   &  &   &  \\ [1ex]
List email   & 453    & 0.32    & 70    & 0.34    & 22    & 0.32    & 523    & 0.05    & 475    & -0.00   \\
 &   & (0.02)  &   & (0.06)  &   & (0.10)  &   &  &   &  \\ [1ex]
List personal site   & 453    & 0.45    & 70    & 0.47    & 22    & 0.64    & 523    & 0.04    & 475    & 0.37*   \\
 &   & (0.02)  &   & (0.06)  &   & (0.10)  &   &  &   &  \\ [1ex]
List brief bio   & 453    & 0.47    & 70    & 0.51    & 22    & 0.50    & 523    & 0.08    & 475    & 0.05   \\
 &   & (0.02)  &   & (0.06)  &   & (0.11)  &   &  &   &  \\ [1ex]
Brief bio length   & 453    & 28.94    & 70    & 23.84    & 22    & 32.77    & 523    & -0.13    & 475    & 0.09   \\
 &   & (1.95)  &   & (4.25)  &   & (8.69)  &   &  &   &  \\ [1ex]
\hline \hline \\[-1.8ex]

\end{tabular}

\end{adjustbox}
\caption*{
\footnotesize
\emph{Notes---}Same as \cref{tab:baltest-user-treated-01}, except the treatment group is distinguished by low and high dosage treatment packages (see \cref{sec:gh-design}).
Significance levels: $^{***} p < .01; ^{**} p < .05; ^{*} p <.1$.
}
\end{table}

\clearpage
\FloatBarrier
% ~~~~~~~~~~~~~~~~~~~~~~~~~~~~~~~~~~~~~~~~~~~~~~~~~~~~~~~~~~~~~~
\subsubsection{Organic GitHub Starrers}\label{sec:organic-starrers}

\begin{table}[!ht]\centering \normalsize \setlength\tabcolsep{10 pt}
\caption{Summary statistics of organic starrers' GitHub characteristics}
\label{tab:organic_starrers_profile_summary}
\begin{adjustbox}{max width=\textwidth}
\begin{tabular}{@{\hspace{0\tabcolsep}}llrr@{\hspace{0\tabcolsep}}}
\toprule\toprule
&&\multicolumn{1}{c}{(1)} &\multicolumn{1}{c}{(2)}\\
&&\multicolumn{1}{c}{Mean (s.d.)}&\multicolumn{1}{c}{Median [IQR]}\\
% \cmidrule{3-4}
\midrule
Public repositories & {} &   35.0 (49.2) &  12.0 [6.0,61.0] \\
Public gists & {} &    9.4 (29.4) &    0.0 [0.0,4.0] \\
Followers & {} &  63.6 (160.4) &   9.0 [1.0,17.0] \\
Following & {} &   34.7 (90.5) &   7.0 [0.0,20.0] \\
Account age (years) & {} &     8.1 (3.5) &   8.5 [6.1,10.4] \\
Lists name & {} &     0.9 (0.4) & \multicolumn{1}{c}{---}                 \\
Lists company & {} &     0.4 (0.5) & \multicolumn{1}{c}{---}                 \\
Lists blog/website & {} &     0.6 (0.5) & \multicolumn{1}{c}{---}                \\
Lists geographical location & {} &     0.5 (0.5) & \multicolumn{1}{c}{---}                 \\
Lists brief biography & {} &     0.5 (0.5) & \multicolumn{1}{c}{---}                \\
Lists Twitter handle & {} &     0.3 (0.5) & \multicolumn{1}{c}{---}                 \\
\bottomrule\bottomrule
\end{tabular}
\end{adjustbox}
\caption*{
\footnotesize 
\emph{Notes---}Table shows summary statistics of the organic GitHub star givers listed in \cref{tab:organic_starrers_list} (corresponding to the high dosage treatment condition in the GitHub experiment, see \cref{sec:gh-treatment-conditions}). The last five rows are indicator variables.
}
\end{table}

\begin{landscape}% Landscape page
\begin{table}[!ht]\centering \normalsize \setlength\tabcolsep{10 pt}
\caption{List of GitHub organic starrers}
\label{tab:organic_starrers_list}
\begin{adjustbox}{max width=1.325\textwidth}
\begin{tabular}{@{\hspace{0\tabcolsep}}llrrrrrrrrrrrrrr@{\hspace{0\tabcolsep}}}
\toprule\toprule
&\multicolumn{1}{c}{(1)} &\multicolumn{1}{c}{(2)}&\multicolumn{1}{c}{(3)}&\multicolumn{1}{c}{(4)}&\multicolumn{1}{c}{(5)}&\multicolumn{1}{c}{(6)}&\multicolumn{1}{c}{(7)}\\
&\multicolumn{1}{r}{Created}&\multicolumn{1}{r}{Blog/website}&\multicolumn{1}{r}{Brief profile biography}&\multicolumn{1}{r}{Repos}&\multicolumn{1}{r}{Gists}&\multicolumn{1}{r}{Followers}&\multicolumn{1}{r}{Following}\\
\cmidrule{2-8}
     basharnaji & 2014-05-14 &                                          --- &                                                --- &   9 &   0 &   9 &  10 \\
      bmwhetter & 2016-09-29 &                                          --- & My name is Brian Whetter and I graduated with a... &   3 &   0 &   1 &   1 \\
   BonaventureR & 2017-09-20 &                            bonaventureraj.me &                                                --- &   6 &   0 &   3 &   3 \\
        c-s-ale & 2016-06-02 &                                      ox.work &       Lover of Machine Learning, D\&D, and Physics! &  28 &   0 &  13 &   2 \\
  chris-alexiuk & 2022-09-26 &                                          --- & I work on Ox and FourthBrain, and a bunch of co... &  19 &   0 &  10 &   0 \\
     danweitzel & 2013-01-15 &                        http://danweitzel.net &                                                --- &  11 &   5 &  13 &  25 \\
deepankermishra & 2014-10-17 &                                          --- &                                                --- &   5 &   0 &   5 &   9 \\
     dhingratul & 2013-06-21 &      https://www.linkedin.com/in/dhingratul/ & Machine Learning Engineer| Experienced research... &  82 &   0 &  95 &  64 \\
  dhruvarora-db & 2021-07-20 &                                          --- &                                                --- &   0 &   0 &   0 &   0 \\
        dwillis & 2008-03-04 &                         http://thescoop.org/ & I teach data journalism at the University of Ma... & 208 & 134 & 705 & 130 \\
    humanpranav & 2020-07-15 &                                          --- &                                             Hello! &   8 &   0 &   0 &   0 \\
  khurramnasser & 2012-07-09 &                                          --- &                                                --- &   0 &   0 &   0 &   0 \\
           LSYS & 2015-01-13 &                    https://www.lucasshen.com & Applied econs | Data science | Quantitative soc... &  13 &   4 &  17 & 407 \\
       Madhu009 & 2016-06-07 & https://www.linkedin.com/in/madhusanjeeviai/ &     AI, Blockchain, Mobile Dev, Web Dev,           &  61 &   0 & 186 &   7 \\
    NoahFinberg & 2014-08-27 &                                     domo.com & Data App Innovation @ Domo. Former Co-founder \&... &  12 &   4 &  10 &  20 \\
    pvbhanuteja & 2017-10-30 &                                   bhanu.cyou &                                                --- &  61 &  12 &   1 &   0 \\
     rajashekar & 2011-10-20 &                      https://rajashekar.dev/ &   Software Engineer, Learner, Developer, Prokopton &  67 &   9 &  23 &  33 \\
    sharmaamits & 2017-12-19 &                                          --- &                                                --- &   1 &   0 &   0 &   0 \\
   sjhangiani12 & 2015-05-25 &                          sharanjhangiani.com &                                                --- &  41 &   0 &   5 &   8 \\
        soodoku & 2011-04-11 &                         http://www.gsood.com &                                                --- &  93 &  29 & 239 &  10 \\
\bottomrule\bottomrule
\end{tabular}
\end{adjustbox}
\caption*{
\footnotesize 
\emph{Notes---}Table lists the 20 organic GitHub star givers (corresponding to the high dosage treatment condition in the GitHub experiment, see \cref{sec:gh-treatment-conditions}).
See \cref{tab:organic_starrers_profile_summary} for summary statistics of these starrers.
}
\end{table}
\end{landscape}

\FloatBarrier
% ~~~~~~~~~~~~~~~~~~~~~~~~~~~~~~~~~~~~~~~~~~~~~~~~~~~~~~~~~~~~~~
\subsubsection{Manipulation Check}
\label{sec:manipulation-check}

\begin{table}[!ht] \centering \normalsize \setlength\tabcolsep{10 pt} \setlength{\defaultaddspace}{0pt}
	\def\sym#1{\ifmmode^{#1}\else\(^{#1}\)\fi}
	\caption{GitHub experiment: Manipulation check.}
	\label{tab:github_exp_stars_regtable}
	\begin{adjustbox}{max width=\textwidth}
		\begin{tabular}{@{\hspace{0\tabcolsep}}l*{4}{D{.}{.}{-1}}@{\hspace{0\tabcolsep}}}
			\toprule\toprule
                &\multicolumn{1}{c}{(1)} &\multicolumn{1}{c}{(2)}&\multicolumn{1}{c}{(3)}&\multicolumn{1}{c}{(4)}\\
                \cmidrule(l){2-5}
                &\multicolumn{4}{c}{Outcome variable is: GitHub stars}\\                
                &\multicolumn{2}{c}{Difference in medians} &\multicolumn{2}{c}{Difference in means}\\
                \cmidrule(lr){2-3}\cmidrule(l){4-5}
                &\multicolumn{1}{c}{On May 12}&\multicolumn{1}{c}{On May 21}&\multicolumn{1}{c}{On May 12}&\multicolumn{1}{c}{On May 21}\\
			\midrule
Treatment (low dosage)&        0.00         &       20.00\sym{***}&       -0.27         &       18.43\sym{***}\\
            &      (0.27)         &      (0.27)         &      (2.76)         &      (2.55)         \\
            &\multicolumn{1}{c}{\text{[$-0.52\:\text{to}\:0.52$]}}         &\multicolumn{1}{c}{\text{[$19.48\:\text{to}\:20.52$]}}         &\multicolumn{1}{c}{\text{[$-5.69\:\text{to}\:5.15$]}}         &\multicolumn{1}{c}{\text{[$13.41\:\text{to}\:23.44$]}}         \\
            &\multicolumn{1}{c}{\text{$<p=1.000>$}}         &\multicolumn{1}{c}{\text{$<p=0.000>$}}         &\multicolumn{1}{c}{\text{$<p=0.923>$}}         &\multicolumn{1}{c}{\text{$<p=0.000>$}}         \\
Treatment (high dosage)&        0.00         &       69.00\sym{***}&        6.87         &       62.15\sym{***}\\
            &      (0.87)         &      (0.87)         &      (6.81)         &      (4.57)         \\
            &\multicolumn{1}{c}{\text{[$-1.70\:\text{to}\:1.70$]}}         &\multicolumn{1}{c}{\text{[$67.30\:\text{to}\:70.70$]}}         &\multicolumn{1}{c}{\text{[$-6.52\:\text{to}\:20.25$]}}         &\multicolumn{1}{c}{\text{[$53.16\:\text{to}\:71.13$]}}         \\
            &\multicolumn{1}{c}{\text{$<p=1.000>$}}         &\multicolumn{1}{c}{\text{$<p=0.000>$}}         &\multicolumn{1}{c}{\text{$<p=0.314>$}}         &\multicolumn{1}{c}{\text{$<p=0.000>$}}         \\
Constant    &        0.00         &        0.00         &        8.41\sym{***}&        8.69\sym{***}\\
            &      (0.10)         &      (0.10)         &      (1.07)         &      (1.08)         \\
            &\multicolumn{1}{c}{\text{[$-0.19\:\text{to}\:0.19$]}}         &\multicolumn{1}{c}{\text{[$-0.19\:\text{to}\:0.19$]}}         &\multicolumn{1}{c}{\text{[$6.31\:\text{to}\:10.52$]}}         &\multicolumn{1}{c}{\text{[$6.58\:\text{to}\:10.81$]}}         \\
            &\multicolumn{1}{c}{\text{$<p=1.000>$}}         &\multicolumn{1}{c}{\text{$<p=1.000>$}}         &\multicolumn{1}{c}{\text{$<p=0.000>$}}         &\multicolumn{1}{c}{\text{$<p=0.000>$}}         \\
\midrule
Median/Mean of outcome&         0.0         &         1.0         &         8.7         &        13.7         \\
Package observations&         585         &         585         &         585         &         585         \\
Day observations&           1         &           1         &           1         &           1         \\
Package-day observations&         585         &         585         &         585         &         585         \\

			\bottomrule
		\end{tabular}
	\end{adjustbox}
	\caption*{\footnotesize Note:
            The table presents manipulation checks for the treatment variable: GitHub stars by presenting differences in medians and means for the treatment and control groups.
            Column (1) reports differences in medians at the start of treatment (May 12).
            Column (2) reports differences in medians at the end of treatment (May 21).            
            Columns (1)--(2) correspond to \cref{fig:timeseries-stars-treated-012-medians}.
            Columns (3)--(4) report differences in means, corresponding to \cref{fig:timeseries-stars-treated-012}.
            % The sample period for GitHub stars is a 30-day period from 9 May 2023 to 7 June 2023.
            % The sample period for PyPI downloads is a 47-day period from 25 April 2023 to 10 June 2023.
            % The post period includes all dates after 12 May 2023, when intervention occurred (\cref{sec:gh-treatment-conditions}).
            \cref{sec:gh-treatment-conditions} describes the high- and low-dosage treatment conditions.
            Standard errors are clustered by packages.
    	Parentheses: standard errors.
    	Square brackets: 95\% confidence intervals.
    	Angle brackets: p-values.
    	Significance levels: $^+$ 0.1 $^*$ 0.05 $^{**}$ 0.01 $^{***}$ 0.001.
	}
\end{table}

\begin{figure}[ht]
\centering
    \includegraphics[width=\textwidth]{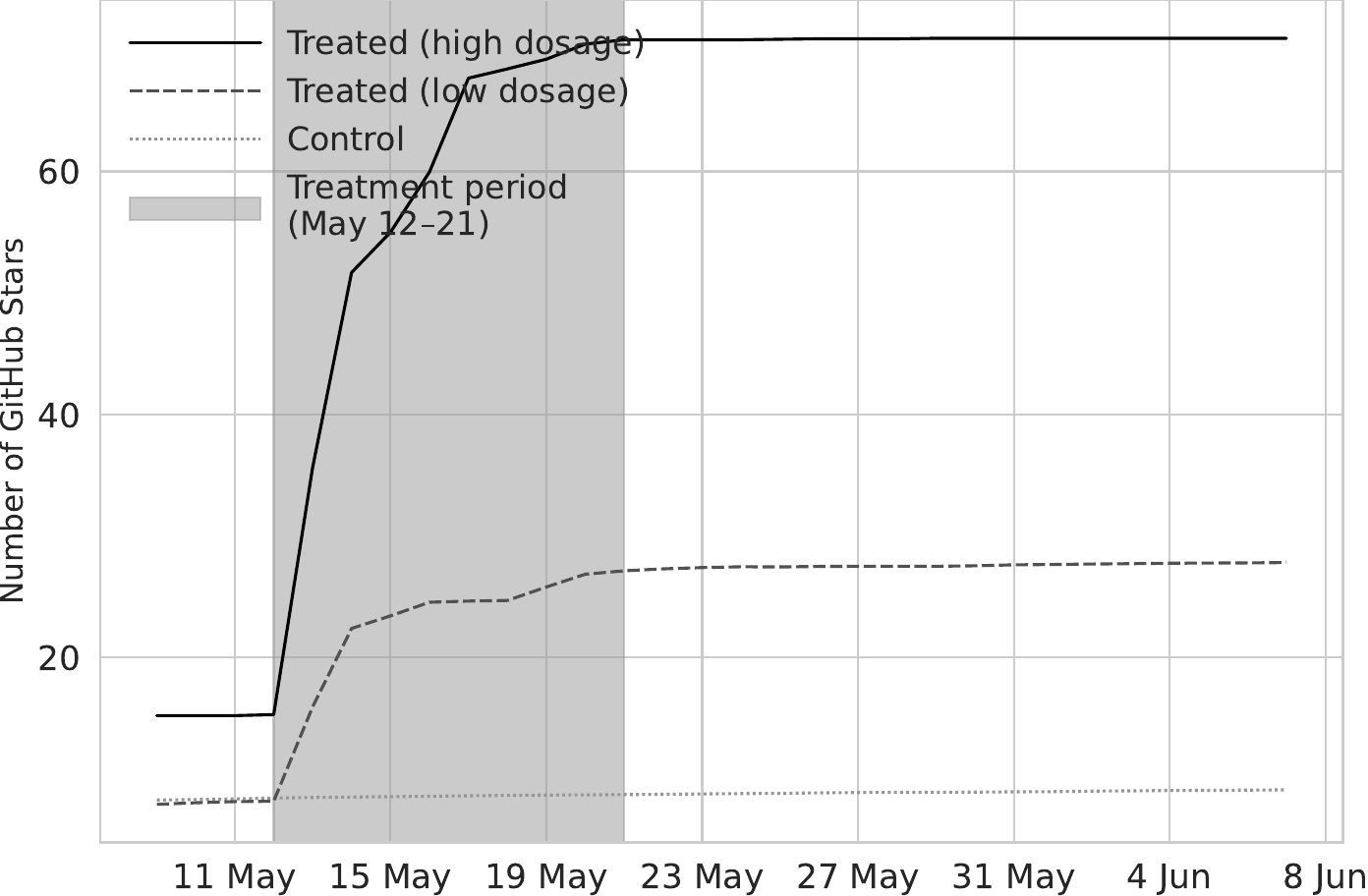}
\caption{
\textbf{Manipulation Check: GitHub Stars for Treated vs. Control}.
Same as \cref{fig:timeseries-stars-treated-012-medians}, except in means.
The figure reports the means version of \cref{fig:timeseries-stars-treated-012-medians}.
The figure plots the mean number of stars on a particular day for the three groups: high-dosage treatment (market and network stars), low-dosage treatment (market stars), and control (no stars). 
% for numbers, see https://github.com/soodoku/social_proof_stars/blob/main/github_exp/metrics-timeseries/src/plot_timeseries_stars.ipynb
In all, we plot data for 585 packages and 17,550 package days. The shaded vertical bar indicates the period during which the treatment was being applied. \cref{tab:github_exp_stars_regtable} presents estimates of the impact of treatment on downloads. 
}
\label{fig:timeseries-stars-treated-012}
\end{figure}

\clearpage
\FloatBarrier
% ~~~~~~~~~~~~~~~~~~~~~~~~~~~~~~~~~~~~~~~~~~~~~~~~~~~~~~~~~~~~~~
\subsubsection{Differences in Medians}
\label{sec:difference-medians}
\begin{table}[!ht] \centering \normalsize \setlength\tabcolsep{6 pt} \setlength{\defaultaddspace}{0pt}
	\def\sym#1{\ifmmode^{#1}\else\(^{#1}\)\fi}
	\caption{GitHub Experiment Results - ITT estimates for medians.}
	\label{tab:github_exp_medians_regtable}
	\begin{adjustbox}{max width=\textwidth}
		\begin{tabular}{@{\hspace{0\tabcolsep}}l*{11}{D{.}{.}{-1}}@{\hspace{0\tabcolsep}}}
			\toprule\toprule
                &\multicolumn{1}{c}{(1)} &\multicolumn{1}{c}{(2)} &\multicolumn{1}{c}{(3)}&\multicolumn{1}{c}{(4)}&\multicolumn{1}{c}{(5)}&\multicolumn{1}{c}{(6)}\\
                \cmidrule(l){2-7}
                &\multicolumn{6}{c}{Outcome variable is: PyPI downloads}\\
                &\multicolumn{1}{c}{On Jun 21}&\multicolumn{1}{c}{On Jul 21}&\multicolumn{1}{c}{On Aug 21}&\multicolumn{1}{c}{On Sep 21}&\multicolumn{1}{c}{On Oct 21}&\multicolumn{1}{c}{Full post period}\\			\midrule
Treatment (low dosage)&         8.0         &        25.0         &        28.0         &        20.0         &        20.0         &        10.3         \\
            &      (17.4)         &      (24.7)         &      (28.6)         &      (32.5)         &      (32.5)         &       (7.7)         \\
            &\multicolumn{1}{c}{\text{[$-26.2\:\text{to}\:42.2$]}}         &\multicolumn{1}{c}{\text{[$-23.4\:\text{to}\:73.4$]}}         &\multicolumn{1}{c}{\text{[$-28.1\:\text{to}\:84.1$]}}         &\multicolumn{1}{c}{\text{[$-43.9\:\text{to}\:83.9$]}}         &\multicolumn{1}{c}{\text{[$-43.9\:\text{to}\:83.9$]}}         &\multicolumn{1}{c}{\text{[$-4.8\:\text{to}\:25.3$]}}         \\
            &\multicolumn{1}{c}{\text{$<p=0.646>$}}         &\multicolumn{1}{c}{\text{$<p=0.311>$}}         &\multicolumn{1}{c}{\text{$<p=0.328>$}}         &\multicolumn{1}{c}{\text{$<p=0.539>$}}         &\multicolumn{1}{c}{\text{$<p=0.539>$}}         &\multicolumn{1}{c}{\text{$<p=0.183>$}}         \\
Treatment (high dosage)&        14.0         &        24.0         &         7.0         &        -7.0         &        -7.0         &        20.0         \\
            &      (50.7)         &      (50.0)         &      (50.7)         &      (49.3)         &      (49.3)         &      (16.3)         \\
            &\multicolumn{1}{c}{\text{[$-85.7\:\text{to}\:113.7$]}}         &\multicolumn{1}{c}{\text{[$-74.3\:\text{to}\:122.3$]}}         &\multicolumn{1}{c}{\text{[$-92.6\:\text{to}\:106.6$]}}         &\multicolumn{1}{c}{\text{[$-103.9\:\text{to}\:89.9$]}}         &\multicolumn{1}{c}{\text{[$-103.9\:\text{to}\:89.9$]}}         &\multicolumn{1}{c}{\text{[$-11.8\:\text{to}\:51.9$]}}         \\
            &\multicolumn{1}{c}{\text{$<p=0.783>$}}         &\multicolumn{1}{c}{\text{$<p=0.632>$}}         &\multicolumn{1}{c}{\text{$<p=0.890>$}}         &\multicolumn{1}{c}{\text{$<p=0.887>$}}         &\multicolumn{1}{c}{\text{$<p=0.887>$}}         &\multicolumn{1}{c}{\text{$<p=0.218>$}}         \\
Linear trend&                     &                     &                     &                     &                     &         0.7\sym{***}\\
            &                     &                     &                     &                     &                     &       (0.0)         \\
            &                     &                     &                     &                     &                     &\multicolumn{1}{c}{\text{[$0.6\:\text{to}\:0.8$]}}         \\
            &                     &                     &                     &                     &                     &\multicolumn{1}{c}{\text{$<p=0.000>$}}         \\
Treatment (low dosage)  $ \times$ Linear trend&                     &                     &                     &                     &                     &         0.2         \\
            &                     &                     &                     &                     &                     &       (0.4)         \\
            &                     &                     &                     &                     &                     &\multicolumn{1}{c}{\text{[$-0.6\:\text{to}\:0.9$]}}         \\
            &                     &                     &                     &                     &                     &\multicolumn{1}{c}{\text{$<p=0.695>$}}         \\
Treatment (high dosage) $ \times$ Linear trend&                     &                     &                     &                     &                     &        -0.1         \\
            &                     &                     &                     &                     &                     &       (0.7)         \\
            &                     &                     &                     &                     &                     &\multicolumn{1}{c}{\text{[$-1.5\:\text{to}\:1.3$]}}         \\
            &                     &                     &                     &                     &                     &\multicolumn{1}{c}{\text{$<p=0.910>$}}         \\
Constant    &        84.0\sym{***}&        98.0\sym{***}&       119.0\sym{***}&       140.0\sym{***}&       140.0\sym{***}&        56.8\sym{***}\\
            &       (4.8)         &       (6.3)         &       (8.2)         &      (10.0)         &      (10.0)         &       (4.1)         \\
            &\multicolumn{1}{c}{\text{[$74.6\:\text{to}\:93.4$]}}         &\multicolumn{1}{c}{\text{[$85.5\:\text{to}\:110.5$]}}         &\multicolumn{1}{c}{\text{[$103.0\:\text{to}\:135.0$]}}         &\multicolumn{1}{c}{\text{[$120.4\:\text{to}\:159.6$]}}         &\multicolumn{1}{c}{\text{[$120.4\:\text{to}\:159.6$]}}         &\multicolumn{1}{c}{\text{[$48.6\:\text{to}\:64.9$]}}         \\
            &\multicolumn{1}{c}{\text{$<p=0.000>$}}         &\multicolumn{1}{c}{\text{$<p=0.000>$}}         &\multicolumn{1}{c}{\text{$<p=0.000>$}}         &\multicolumn{1}{c}{\text{$<p=0.000>$}}         &\multicolumn{1}{c}{\text{$<p=0.000>$}}         &\multicolumn{1}{c}{\text{$<p=0.000>$}}         \\
\midrule
Median of outcome&        86.0         &       103.5         &       125.0         &       141.0         &       141.0         &       112.0         \\
Package observations&         622         &         622         &         622         &         622         &         622         &         622         \\
Day observations&           1         &           1         &           1         &           1         &           1         &         165         \\
Package-day observations&         622         &         622         &         622         &         622         &         622         &     102,630         \\

			\bottomrule
		\end{tabular}
	\end{adjustbox}
	\caption*{\footnotesize Note:
            The table presents Intention-to-Treat (ITT) estimates for median downloads in the GitHub experiment.
            Corresponds to \cref{fig:timeseries-downloads-treated-012-medians}.
            Columns (1)--(5) report snapshots of the difference in medians at various dates.
            Column (6) reports post-treatment differences in medians over the full post-treatment period of 165 days, allowing for heterogeneous treatment effects through a linear time trend.
    	Parentheses: standard errors.
    	Square brackets: 95\% confidence intervals.
    	Angle brackets: p-values.
            Significance levels: $^{***} p < .001; ^{**} p < .01; ^{*} p < .05; ^{+} p <.1$.
	}
\end{table}

\clearpage
\FloatBarrier
% ~~~~~~~~~~~~~~~~~~~~~~~~~~~~~~~~~~~~~~~~~~~~~~~~~~~~~~~~~~~~~~
\subsubsection{Differences in Means}
\label{sec:difference-means}

\begin{figure}[ht]
\centering
    \includegraphics[width=\textwidth]{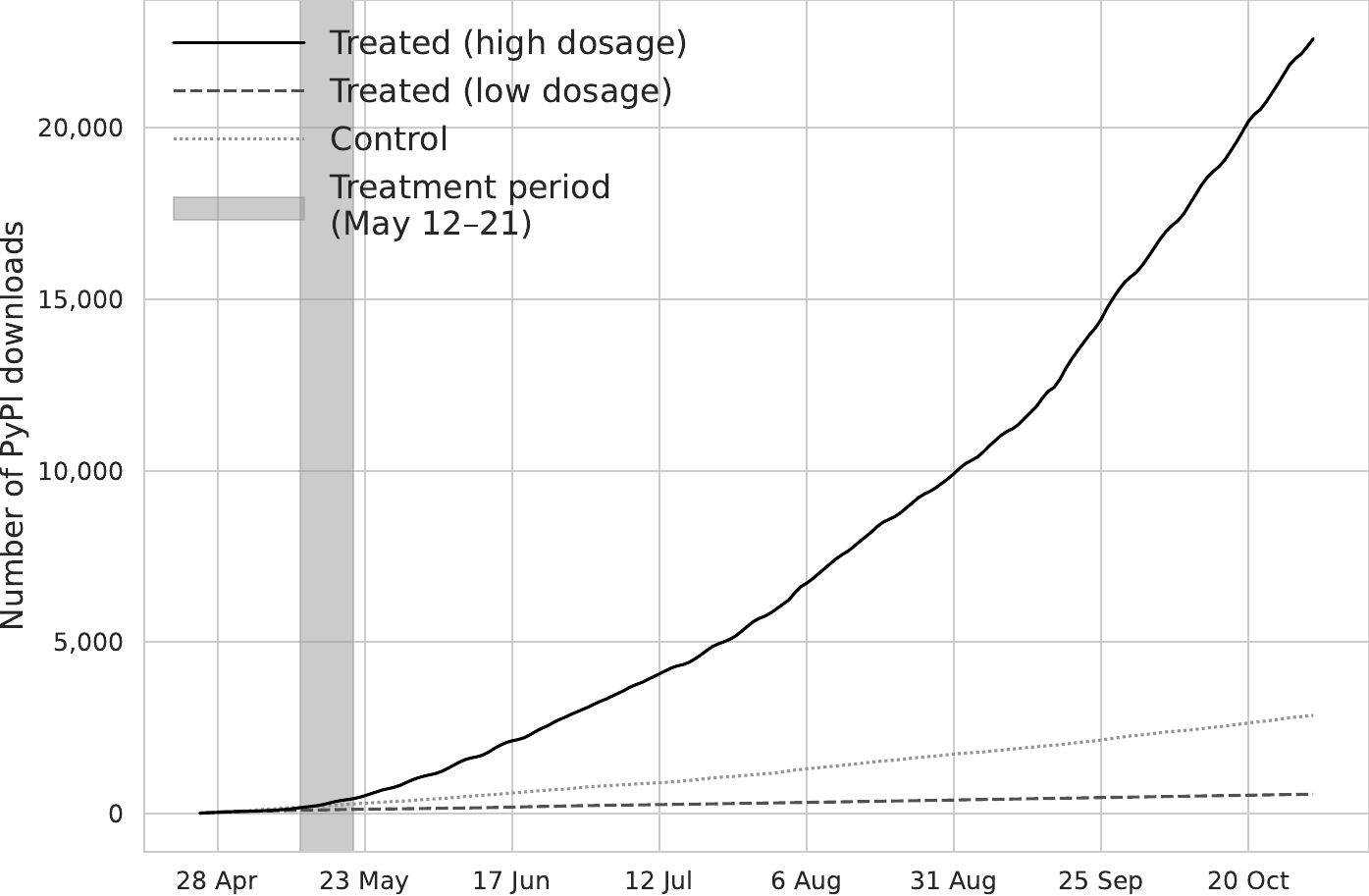}
\caption{
\textbf{Mean PyPI downloads for Treated vs. Control in GitHub Experiment.} 
    Same as \cref{fig:timeseries-downloads-treated-012-medians}, except for means.
    The figure plots the mean of cumulative downloads for each of the three groups. Each point is a day averaged within the group for 622 packages and 118,180 package days. 
    % for numbers, can look at https://github.com/soodoku/social_proof_stars/blob/main/github_exp/get_metrics/src/consolidate_pypi_downloads_allinstallers.ipynb
    Treatment is distinguished by low and high dosages (see \cref{sec:gh-design}). 
    The shaded vertical bar indicates the treatment period. 
    Downloads include only human downloads (\cref{tab:human_bot_download_classification}). 
    See also \crefrange{fig:timeseries-downloads-treated-012-individual}{fig:timeseries-downloads-treated-012-without-top2-extreme-outliers} for the time series of individual packages. \cref{tab:github_exp_regtable} reports the estimates of differences in means. 
}
\label{fig:timeseries-downloads-treated-012}
\end{figure}

\begin{figure}[ht]
\centering
    \includegraphics[width=\textwidth]{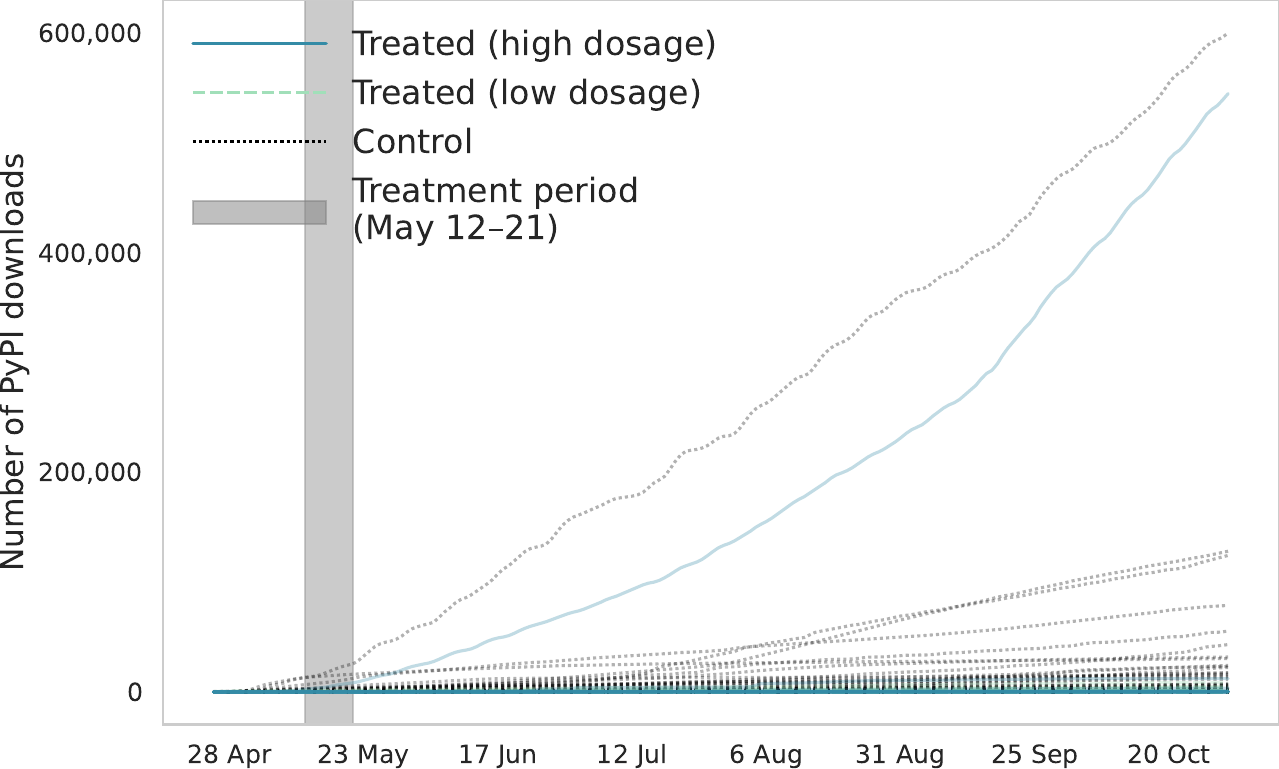}
\caption{
\textbf{Individual PyPI downloads for Treated vs Control.} Similar to \cref{fig:timeseries-downloads-treated-012}, but with the time series of the individual packages.
The four packages with the highest download trajectories ($>$ 100,000 total downloads) are identified and discussed in \cref{tab:outliers}.
See \cref{fig:timeseries-downloads-treated-012-without-top2-extreme-outliers} for the same figure with the two most extreme outliers removed.
}
\label{fig:timeseries-downloads-treated-012-individual}
\end{figure}

\begin{figure}[ht]
\centering
    \includegraphics[width=\textwidth]{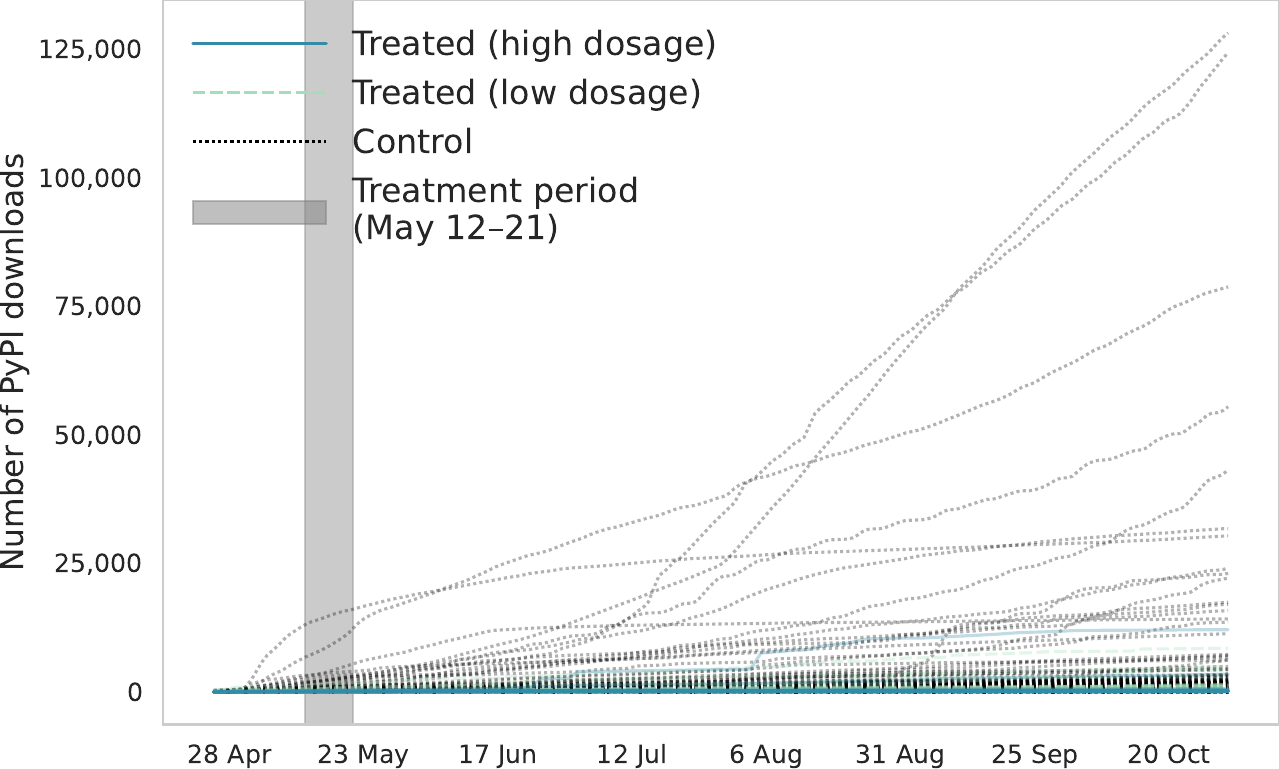}
\caption{
\textbf{Individual PyPI downloads for treated vs control.} Similar to \cref{fig:timeseries-downloads-treated-012-individual}, but without the top two extreme time series.
}
\label{fig:timeseries-downloads-treated-012-without-top2-extreme-outliers}
\end{figure}

While there appears to be a break in trend in \cref{fig:timeseries-downloads-treated-012}, we note three additional time series behaviors. First, the within-group variance is large, as will be evident in the estimates (\cref{tab:github_exp_regtable}). Second, this, at least in part, can be explained by a few extreme outliers (\crefrange{fig:timeseries-downloads-treated-012-individual}{fig:timeseries-downloads-treated-012-without-top2-extreme-outliers}). Third, and relatedly, we do not observe a similar trend break for the medians (\cref{fig:timeseries-downloads-treated-012-medians}).

\cref{tab:pypi_exp_regtable} reports the ITT estimates.
Approximately one month after intervention, on June 21, the low-dosage group's mean download tally is no higher than the control group ($p = .051$), while the high-dosage group's download tally is higher but statistically indistinguishable from the control group ($p = .495$, \cref{tab:github_exp_regtable}). 
%The estimates suggest that these differences in means increase over time (as seen in \cref{fig:timeseries-downloads-treated-012}) but remain statistically non-significant at conventional levels. 
Three months after intervention (on August 21), the low-dosage group's mean download tally is 1,180 lower than the control group ($p = .069$), while the high-dosage group's download tally is 7k higher than the control group ($p = .388$), with the standard errors of the estimates always in the same order of magnitude as the estimates (\cref{tab:github_exp_regtable}). We also estimate a model that allows the treatment effect to vary over time (column (6) of \cref{tab:github_exp_regtable}). Neither the low-dosage nor high-dosage group has a trend different from the control group. If anything, the low-dosage treatment group has a weaker trend than the control group ($p = .066$).

\cref{tab:outliers} lists four outlier packages, the highest trajectories in \cref{fig:timeseries-downloads-treated-012-individual}, together with their GitHub repository URLs and treatment assignment.
First, \textit{pyaesm} (\textit{ricmoo/pyaes}) is a lightweight implementation of the Advanced Encryption Standard (AES) encryption algorithm, requiring no external dependencies beyond Python’s built-in modules.
While the PyPI package was released on April 28, 2023 (\url{https://pypi.org/project/pyaesm/}), the underlying source code in the GitHub repository had existed for many years prior to the PyPI release.
Second, \textit{dghs-imgutils} (\textit{deepghs/imgutils}) implements ``anime-style image data processing'' using pretrained models, including those for clustering, object detection, LineArt generation, and character extraction.
The repository includes extensive documentation, visual examples, and demonstration code in its README, which may contribute to its visibility, engagement, and uptake.
Third, \textit{salt-analytics-framework} (\textit{github.com/saltstack/salt-analytics-framework}) is an extension to the Salt automation and orchestration system that adds support to ``collect, process and forward analytics/metrics data''.
Salt (\url{https://github.com/saltstack/salt}) is itself an established open-source infrastructure automation and configuration management system (with 15k+ stars and 82k/month downloads).
The observed growth in downloads of \textit{salt-analytics-framework}, therefore, very likely reflects ecosystem spillovers from Salt.
Finally, Firebase is a Google-managed platform that allows developers to build web and mobile applications.
\textit{firebase-functions} (\textit{firebase/firebase-functions-python}) is its official Python SDK for integrating with Google’s Firebase and Cloud Functions infrastructure.

Of the four, three are assigned to control (\textit{pyaesm}, \textit{dghs-imgutils}, \textit{salt-analytics-framework}), and so are unlikely to have been affected by our treatment.
Two (\textit{salt-analytics-framework}, \textit{firebase-functions}) have uptake in usage (downloads) that very likely reflect positive spillovers from large established ecosystems.
Only one (\textit{firebase-functions}) was assigned to our treatment group, but it is unlikely to have benefited from our manipulation.
Instead, the uptick in downloads coinciding with our treatment period is more likely to have arisen from Google's announcement of Python support for Firebase Cloud Functions at Google I/O on May 10 (2 days prior to our treatment implementation)---see, for example, \url{https://firebase.blog/posts/2023/05/whats-new-at-google-io/} and \url{https://techcrunch.com/2023/05/10/googles-firebase-gets-ai-extensions-opens-up-its-marketplace/}.

\begin{table}[!ht] \centering \normalsize \setlength\tabcolsep{2 pt} \setlength{\defaultaddspace}{0pt}
	\def\sym#1{\ifmmode^{#1}\else\(^{#1}\)\fi}
	\caption{GitHub Experiment Results - ITT estimates for means.}
	\label{tab:github_exp_regtable}
	\begin{adjustbox}{max width=\textwidth}
		\begin{tabular}{@{\hspace{0\tabcolsep}}l*{6}{D{.}{.}{-1}}@{\hspace{0\tabcolsep}}}
			\toprule\toprule
                &\multicolumn{1}{c}{(1)} &\multicolumn{1}{c}{(2)} &\multicolumn{1}{c}{(3)}&\multicolumn{1}{c}{(4)}&\multicolumn{1}{c}{(5)}&\multicolumn{1}{c}{(6)}\\
                \cmidrule(l){2-7}
                &\multicolumn{6}{c}{Outcome variable is: PyPI downloads}\\
                &\multicolumn{1}{c}{On Jun 21}&\multicolumn{1}{c}{On Jul 21}&\multicolumn{1}{c}{On Aug 21}&\multicolumn{1}{c}{On Sep 21}&\multicolumn{1}{c}{On Oct 21}&\multicolumn{1}{c}{Full post period}\\			\midrule
Treatment (low dosage)&      -260.1\sym{+}  &      -743.3\sym{+}  &    -1,180.3\sym{+}  &    -1,595.5\sym{+}  &    -2,101.2\sym{+}  &       -12.0         \\
            &     (132.8)         &     (424.7)         &     (648.3)         &     (855.0)         &   (1,135.2)         &      (76.4)         \\
            &\multicolumn{1}{c}{\text{[$-520.8\:\text{to}\:0.6$]}}         &\multicolumn{1}{c}{\text{[$-1,577.4\:\text{to}\:90.8$]}}         &\multicolumn{1}{c}{\text{[$-2,453.5\:\text{to}\:92.9$]}}         &\multicolumn{1}{c}{\text{[$-3,274.6\:\text{to}\:83.5$]}}         &\multicolumn{1}{c}{\text{[$-4,330.4\:\text{to}\:128.1$]}}         &\multicolumn{1}{c}{\text{[$-162.0\:\text{to}\:138.0$]}}         \\
            &\multicolumn{1}{c}{\text{$<p=0.051>$}}         &\multicolumn{1}{c}{\text{$<p=0.081>$}}         &\multicolumn{1}{c}{\text{$<p=0.069>$}}         &\multicolumn{1}{c}{\text{$<p=0.062>$}}         &\multicolumn{1}{c}{\text{$<p=0.065>$}}         &\multicolumn{1}{c}{\text{$<p=0.875>$}}         \\
Treatment (high dosage)&       679.6         &     3,727.3         &     7,035.0         &    11,213.5         &    17,550.9         &    -2,303.5         \\
            &     (995.8)         &   (4,496.9)         &   (8,150.4)         &  (12,796.6)         &  (19,813.3)         &   (2,202.8)         \\
            &\multicolumn{1}{c}{\text{[$-1,276.0\:\text{to}\:2,635.2$]}}         &\multicolumn{1}{c}{\text{[$-5,103.8\:\text{to}\:12,558.3$]}}         &\multicolumn{1}{c}{\text{[$-8,970.8\:\text{to}\:23,040.8$]}}         &\multicolumn{1}{c}{\text{[$-13,916.4\:\text{to}\:36,343.4$]}}         &\multicolumn{1}{c}{\text{[$-21,358.5\:\text{to}\:56,460.3$]}}         &\multicolumn{1}{c}{\text{[$-6,629.4\:\text{to}\:2,022.4$]}}         \\
            &\multicolumn{1}{c}{\text{$<p=0.495>$}}         &\multicolumn{1}{c}{\text{$<p=0.408>$}}         &\multicolumn{1}{c}{\text{$<p=0.388>$}}         &\multicolumn{1}{c}{\text{$<p=0.381>$}}         &\multicolumn{1}{c}{\text{$<p=0.376>$}}         &\multicolumn{1}{c}{\text{$<p=0.296>$}}         \\
Linear trend&                     &                     &                     &                     &                     &        15.9\sym{*}  \\
            &                     &                     &                     &                     &                     &       (7.1)         \\
            &                     &                     &                     &                     &                     &\multicolumn{1}{c}{\text{[$2.0\:\text{to}\:29.9$]}}         \\
            &                     &                     &                     &                     &                     &\multicolumn{1}{c}{\text{$<p=0.025>$}}         \\
Treatment (low dosage)  $ \times$ Linear trend&                     &                     &                     &                     &                     &       -13.2\sym{+}  \\
            &                     &                     &                     &                     &                     &       (7.1)         \\
            &                     &                     &                     &                     &                     &\multicolumn{1}{c}{\text{[$-27.2\:\text{to}\:0.9$]}}         \\
            &                     &                     &                     &                     &                     &\multicolumn{1}{c}{\text{$<p=0.066>$}}         \\
Treatment (high dosage) $ \times$ Linear trend&                     &                     &                     &                     &                     &       116.0         \\
            &                     &                     &                     &                     &                     &     (124.3)         \\
            &                     &                     &                     &                     &                     &\multicolumn{1}{c}{\text{[$-128.2\:\text{to}\:360.1$]}}         \\
            &                     &                     &                     &                     &                     &\multicolumn{1}{c}{\text{$<p=0.351>$}}         \\
Constant    &       406.7\sym{**} &     1,022.0\sym{*}  &     1,543.0\sym{*}  &     2,043.8\sym{*}  &     2,643.2\sym{*}  &       123.6\sym{+}  \\
            &     (127.3)         &     (419.1)         &     (641.5)         &     (846.6)         &   (1,126.5)         &      (70.6)         \\
            &\multicolumn{1}{c}{\text{[$156.8\:\text{to}\:656.6$]}}         &\multicolumn{1}{c}{\text{[$198.9\:\text{to}\:1,845.1$]}}         &\multicolumn{1}{c}{\text{[$283.1\:\text{to}\:2,802.9$]}}         &\multicolumn{1}{c}{\text{[$381.2\:\text{to}\:3,706.3$]}}         &\multicolumn{1}{c}{\text{[$431.0\:\text{to}\:4,855.4$]}}         &\multicolumn{1}{c}{\text{[$-15.0\:\text{to}\:262.2$]}}         \\
            &\multicolumn{1}{c}{\text{$<p=0.001>$}}         &\multicolumn{1}{c}{\text{$<p=0.015>$}}         &\multicolumn{1}{c}{\text{$<p=0.016>$}}         &\multicolumn{1}{c}{\text{$<p=0.016>$}}         &\multicolumn{1}{c}{\text{$<p=0.019>$}}         &\multicolumn{1}{c}{\text{$<p=0.080>$}}         \\
\midrule
R$^2$       &     0.00333         &     0.00648         &     0.00880         &      0.0113         &      0.0137         &      0.0140         \\
Mean of outcome&       402.7         &     1,082.2         &     1,683.4         &     2,302.1         &     3,095.3         &     1,586.5         \\
Package observations&         622         &         622         &         622         &         622         &         622         &         622         \\
Day observations&           1         &           1         &           1         &           1         &           1         &         165         \\
Package-day observations&         622         &         622         &         622         &         622         &         622         &     102,630         \\

			\bottomrule
		\end{tabular}
	\end{adjustbox}
	\caption*{\footnotesize Note:
            Similar to \cref{tab:github_exp_medians_regtable}, but for differences in means.
            The table presents Intention-to-Treat (ITT) estimates for mean downloads in the GitHub experiment.
            Corresponds to \cref{fig:timeseries-downloads-treated-012}.
            Columns (1)--(5) report snapshots of the difference in mean at various dates.
            Column (6) reports post-treatment differences in means over the full post-treatment period of 165 days, allowing for heterogeneous treatment effects through a linear time trend.
    	Parentheses: standard errors.
    	Square brackets: 95\% confidence intervals.
    	Angle brackets: p-values.
            Significance levels: $^{***} p < .001; ^{**} p < .01; ^{*} p < .05; ^{+} p <.1$.
	}
\end{table}

\FloatBarrier
\clearpage

% https://github.com/themains/social_proof_stars/blob/main/github_exp/metrics-timeseries/src/plot_timeseries_pypi_downloads.ipynb
% see df_dl.query("tt_downloads > 100_000")["pkg"].unique()
% array(['dghs-imgutils', 'firebase-functions', 'pyaesm', 'salt-analytics-framework'], dtype=object)
\begin{table}
\centering 
\normalsize
\setlength\tabcolsep{8 pt}
\caption{Outlier Packages with Anomalously High Download Counts}
\label{tab:outliers}
\begin{adjustbox}{max width=\textwidth}
\begin{tabular}{@{\hspace{0\tabcolsep}}llllr@{\hspace{0\tabcolsep}}}
\toprule\toprule
% \multicolumn{1}{c}{}&
% \multicolumn{1}{c}{(1)}&
% \multicolumn{1}{c}{(2)}&
% \multicolumn{1}{c}{(3)}&
% \multicolumn{1}{c}{(4)}\\
\multicolumn{1}{l}{}&
\multicolumn{1}{l}{Package}&
\multicolumn{1}{l}{GitHub Repository URL}&
\multicolumn{1}{l}{Assignment}&
Total Downloads\\
\midrule
1 & pyaesm                   & \url{github.com/ricmoo/pyaes}                        & Control               & 123{,}198 \\
2 & dghs-imgutils            & \url{github.com/deepghs/imgutils}                    & Control               & 118{,}853 \\
3 & salt-analytics-framework & \url{github.com/saltstack/salt-analytics-framework}  & Control               & 580{,}519 \\
4 & firebase-functions       & \url{github.com/firebase/firebase-functions-python}  & Treated (High-dosage) & 525{,}229 \\
\bottomrule
\end{tabular}
\end{adjustbox}
\caption*{
\footnotesize
\emph{Notes---}Outliers with more than 100,000 total downloads (see \cref{fig:timeseries-downloads-treated-012-individual}). Total downloads measured as of the end of the sample period on Oct 31, 2023.
}
\end{table}
% \begin{table}
% \centering 
% \normalsize
% \setlength\tabcolsep{8 pt}
% \caption{Outlier packages with anomalously high download counts}
% \label{tab:outliers}
% \begin{adjustbox}{max width=\textwidth}
% \begin{tabular}{@{\hspace{0\tabcolsep}}llll@{\hspace{0\tabcolsep}}}
% \toprule\toprule
% \multicolumn{1}{c}{}&
% \multicolumn{1}{c}{(1)}&
% \multicolumn{1}{c}{(2)}&
% \multicolumn{1}{c}{(3)}\\
% \multicolumn{1}{c}{}&
% \multicolumn{1}{c}{Package}&
% \multicolumn{1}{c}{GitHub Repository URL}&
% \multicolumn{1}{c}{Assignment}\\
% \midrule
% 1 & pyaesm                   & \url{github.com/ricmoo/pyaes}                       & Control \\
% 2 & dghs-imgutils            & \url{github.com/deepghs/imgutils}                   & Control \\
% 3 & salt-analytics-framework & \url{github.com/saltstack/salt-analytics-framework}  & Control \\
% 4 & firebase-functions       & \url{github.com/firebase/firebase-functions-python}  & Treated (High-dosage) \\
% \bottomrule
% \end{tabular}
% \end{adjustbox}
% \caption*{
% \footnotesize
% \emph{Notes---}Outliers with more than 100,000 total downloads (see \cref{fig:timeseries-downloads-treated-012-individual}).
% }
% \end{table}

\FloatBarrier

\begin{table}[!ht] \centering \normalsize \setlength\tabcolsep{3 pt} \setlength{\defaultaddspace}{0pt}
\def\sym#1{\ifmmode^{#1}\else\(^{#1}\)\fi}
\caption{GitHub Experiment Results - LATE estimates for means.}
\label{tab:github_exp_regtable_allhumaninstallers_late}
\begin{adjustbox}{max width=\textwidth}
    \begin{tabular}{@{\hspace{0\tabcolsep}}l*{6}{D{.}{.}{-1}}@{\hspace{0\tabcolsep}}}
    \toprule\toprule
        &\multicolumn{1}{c}{(1)} &\multicolumn{1}{c}{(2)} &\multicolumn{1}{c}{(3)}&\multicolumn{1}{c}{(4)}&\multicolumn{1}{c}{(5)}&\multicolumn{1}{c}{(6)}\\
        \cmidrule(l){2-7}
        &\multicolumn{6}{c}{Outcome variable is: PyPI downloads}\\
        &\multicolumn{1}{c}{On Jun 21}&\multicolumn{1}{c}{On Jul 21}&\multicolumn{1}{c}{On Aug 21}&\multicolumn{1}{c}{On Sep 21}&\multicolumn{1}{c}{On Oct 21}&\multicolumn{1}{c}{Full post period}\\
    \midrule
Received treatment&       213.9         &       859.3         &     1,963.1         &     3,660.5         &     6,317.4         &    -1,299.7         \\
            &   (1,347.3)         &   (2,662.2)         &   (4,660.9)         &   (7,275.5)         &  (11,067.7)         &   (1,245.9)         \\
            &\multicolumn{1}{c}{\text{[$-2,426.8\:\text{to}\:2,854.7$]}}         &\multicolumn{1}{c}{\text{[$-4,358.5\:\text{to}\:6,077.1$]}}         &\multicolumn{1}{c}{\text{[$-7,172.1\:\text{to}\:11,098.3$]}}         &\multicolumn{1}{c}{\text{[$-10,599.3\:\text{to}\:17,920.2$]}}         &\multicolumn{1}{c}{\text{[$-15,375.0\:\text{to}\:28,009.7$]}}         &\multicolumn{1}{c}{\text{[$-3,741.6\:\text{to}\:1,142.2$]}}         \\
            &\multicolumn{1}{c}{\text{$<p=0.874>$}}         &\multicolumn{1}{c}{\text{$<p=0.747>$}}         &\multicolumn{1}{c}{\text{$<p=0.674>$}}         &\multicolumn{1}{c}{\text{$<p=0.615>$}}         &\multicolumn{1}{c}{\text{$<p=0.568>$}}         &\multicolumn{1}{c}{\text{$<p=0.297>$}}         \\
Linear trend&                     &                     &                     &                     &                     &        15.9\sym{*}  \\
            &                     &                     &                     &                     &                     &       (7.1)         \\
            &                     &                     &                     &                     &                     &\multicolumn{1}{c}{\text{[$2.0\:\text{to}\:29.8$]}}         \\
            &                     &                     &                     &                     &                     &\multicolumn{1}{c}{\text{$<p=0.025>$}}         \\
Received treatment  $ \times$ Linear trend&                     &                     &                     &                     &                     &        42.5         \\
            &                     &                     &                     &                     &                     &      (71.4)         \\
            &                     &                     &                     &                     &                     &\multicolumn{1}{c}{\text{[$-97.5\:\text{to}\:182.5$]}}         \\
            &                     &                     &                     &                     &                     &\multicolumn{1}{c}{\text{$<p=0.552>$}}         \\
Constant    &       655.7\sym{**} &     1,041.0\sym{*}  &     1,556.9\sym{*}  &     2,066.6\sym{*}  &     2,664.7\sym{*}  &       123.6\sym{+}  \\
            &     (249.4)         &     (429.5)         &     (645.6)         &     (856.4)         &   (1,135.8)         &      (70.5)         \\
            &\multicolumn{1}{c}{\text{[$166.9\:\text{to}\:1,144.4$]}}         &\multicolumn{1}{c}{\text{[$199.2\:\text{to}\:1,882.8$]}}         &\multicolumn{1}{c}{\text{[$291.5\:\text{to}\:2,822.2$]}}         &\multicolumn{1}{c}{\text{[$388.0\:\text{to}\:3,745.2$]}}         &\multicolumn{1}{c}{\text{[$438.6\:\text{to}\:4,890.9$]}}         &\multicolumn{1}{c}{\text{[$-14.6\:\text{to}\:261.8$]}}         \\
            &\multicolumn{1}{c}{\text{$<p=0.009>$}}         &\multicolumn{1}{c}{\text{$<p=0.015>$}}         &\multicolumn{1}{c}{\text{$<p=0.016>$}}         &\multicolumn{1}{c}{\text{$<p=0.016>$}}         &\multicolumn{1}{c}{\text{$<p=0.019>$}}         &\multicolumn{1}{c}{\text{$<p=0.080>$}}         \\
\midrule
Mean of outcome&       671.2         &     1,103.2         &     1,698.9         &     2,331.4         &     3,121.8         &     1,586.5         \\
Package observations&         622         &         622         &         622         &         622         &         622         &         622         \\
Day observations&           1         &           1         &           1         &           1         &           1         &         165         \\
Package-day observations&         622         &         622         &         622         &         622         &         622         &     102,630         \\

    \bottomrule
    \end{tabular}
\end{adjustbox}
\caption*{\footnotesize Note:
        The table presents Local Average Treatment Effect (LATE) estimates for mean downloads in the GitHub experiment.
        Compliers (those who ``Received treatment'') are defined as those receiving at least 20 stars at the end of the treatment window.
        LATE estimates are from instrumental variable regressions, instrumenting compliers with the random treatment assignment.
        Corresponds to \cref{fig:timeseries-downloads-treated-012}.
        Columns (1)--(5) report snapshots of the difference in mean at various dates.
        Column (6) reports post-treatment differences in means over the full post-treatment period of 165 days, allowing for heterogeneous treatment effects through a linear time trend.
	Parentheses: standard errors.
	Square brackets: 95\% confidence intervals.
	Angle brackets: p-values.
        Significance levels: $^{***} p < .001; ^{**} p < .01; ^{*} p < .05; ^{+} p <.1$.
}
\end{table}

\FloatBarrier
% ~~~~~~~~~~~~~~~~~~~~~~~~~~~~~~~~~~~~~~~~~~~~~~~~~~~~~~~~~~~~~~
\subsubsection{Heterogeneity by Package Complexity}
\label{sec:het-analyses}

\begin{table}[!ht]
\centering \normalsize \setlength\tabcolsep{6 pt} \setlength{\defaultaddspace}{0pt}
	\def\sym#1{\ifmmode^{#1}\else\(^{#1}\)\fi}
	\caption{GitHub Experiment Results - ITT estimates by codebase size.}
	\label{tab:github_exp_medians_regtable_allhumaninstallers_high_size_mb}
	\begin{adjustbox}{max width=\textwidth}
		\begin{tabular}{@{\hspace{0\tabcolsep}}l*{5}{D{.}{.}{-1}}@{\hspace{0\tabcolsep}}}
			\toprule\toprule
                &\multicolumn{1}{c}{(1)} &\multicolumn{1}{c}{(2)} &\multicolumn{1}{c}{(3)}&\multicolumn{1}{c}{(4)}&\multicolumn{1}{c}{(5)}\\
                \cmidrule{2-6}
                &\multicolumn{5}{c}{Outcome variable is: PyPI downloads}\\
                &\multicolumn{1}{c}{On Jun 21}&\multicolumn{1}{c}{On Jul 21}&\multicolumn{1}{c}{On Aug 21}&\multicolumn{1}{c}{On Sep 21}&\multicolumn{1}{c}{On Oct 21}\\			
                \midrule
Treatment (low dosage)&        -4.0         &         0.0         &       -12.0         &       -26.0         &       -23.0         \\
            &      (14.6)         &      (20.2)         &      (22.7)         &      (27.7)         &      (31.2)         \\
            &\multicolumn{1}{c}{\text{[$-32.6\:\text{to}\:24.6$]}}         &\multicolumn{1}{c}{\text{[$-39.7\:\text{to}\:39.7$]}}         &\multicolumn{1}{c}{\text{[$-56.6\:\text{to}\:32.6$]}}         &\multicolumn{1}{c}{\text{[$-80.5\:\text{to}\:28.5$]}}         &\multicolumn{1}{c}{\text{[$-84.2\:\text{to}\:38.2$]}}         \\
            &\multicolumn{1}{c}{\text{$<p=0.784>$}}         &\multicolumn{1}{c}{\text{$<p=1.000>$}}         &\multicolumn{1}{c}{\text{$<p=0.598>$}}         &\multicolumn{1}{c}{\text{$<p=0.349>$}}         &\multicolumn{1}{c}{\text{$<p=0.461>$}}         \\
Treatment (high dosage)&        44.0         &        32.0         &        14.0         &        -8.0         &       -15.0         \\
            &      (58.7)         &      (66.7)         &      (94.0)         &      (92.9)         &      (93.4)         \\
            &\multicolumn{1}{c}{\text{[$-71.3\:\text{to}\:159.3$]}}         &\multicolumn{1}{c}{\text{[$-99.0\:\text{to}\:163.0$]}}         &\multicolumn{1}{c}{\text{[$-170.6\:\text{to}\:198.6$]}}         &\multicolumn{1}{c}{\text{[$-190.4\:\text{to}\:174.4$]}}         &\multicolumn{1}{c}{\text{[$-198.4\:\text{to}\:168.4$]}}         \\
            &\multicolumn{1}{c}{\text{$<p=0.454>$}}         &\multicolumn{1}{c}{\text{$<p=0.631>$}}         &\multicolumn{1}{c}{\text{$<p=0.882>$}}         &\multicolumn{1}{c}{\text{$<p=0.931>$}}         &\multicolumn{1}{c}{\text{$<p=0.872>$}}         \\
Large repository size&        14.0         &        16.0         &        22.0         &        10.0         &        21.0         \\
            &       (8.9)         &      (12.7)         &      (16.5)         &      (20.5)         &      (25.7)         \\
            &\multicolumn{1}{c}{\text{[$-3.5\:\text{to}\:31.5$]}}         &\multicolumn{1}{c}{\text{[$-9.0\:\text{to}\:41.0$]}}         &\multicolumn{1}{c}{\text{[$-10.3\:\text{to}\:54.3$]}}         &\multicolumn{1}{c}{\text{[$-30.2\:\text{to}\:50.2$]}}         &\multicolumn{1}{c}{\text{[$-29.4\:\text{to}\:71.4$]}}         \\
            &\multicolumn{1}{c}{\text{$<p=0.117>$}}         &\multicolumn{1}{c}{\text{$<p=0.210>$}}         &\multicolumn{1}{c}{\text{$<p=0.182>$}}         &\multicolumn{1}{c}{\text{$<p=0.625>$}}         &\multicolumn{1}{c}{\text{$<p=0.414>$}}         \\
Treatment (low) $\times$ Large repository size&        65.0         &        82.0         &        94.0         &       124.0\sym{+}  &       115.0\sym{+}  \\
            &      (45.4)         &      (56.5)         &      (57.3)         &      (65.1)         &      (69.7)         \\
            &\multicolumn{1}{c}{\text{[$-24.2\:\text{to}\:154.2$]}}         &\multicolumn{1}{c}{\text{[$-28.9\:\text{to}\:192.9$]}}         &\multicolumn{1}{c}{\text{[$-18.6\:\text{to}\:206.6$]}}         &\multicolumn{1}{c}{\text{[$-3.8\:\text{to}\:251.8$]}}         &\multicolumn{1}{c}{\text{[$-21.8\:\text{to}\:251.8$]}}         \\
            &\multicolumn{1}{c}{\text{$<p=0.153>$}}         &\multicolumn{1}{c}{\text{$<p=0.147>$}}         &\multicolumn{1}{c}{\text{$<p=0.101>$}}         &\multicolumn{1}{c}{\text{$<p=0.057>$}}         &\multicolumn{1}{c}{\text{$<p=0.099>$}}         \\
Treatment (high) $\times$ Large repository size&       -52.0         &       -28.0         &         5.0         &        58.0         &       107.0         \\
            &      (92.2)         &      (97.3)         &     (128.7)         &     (125.0)         &     (131.8)         \\
            &\multicolumn{1}{c}{\text{[$-233.1\:\text{to}\:129.1$]}}         &\multicolumn{1}{c}{\text{[$-219.1\:\text{to}\:163.1$]}}         &\multicolumn{1}{c}{\text{[$-247.8\:\text{to}\:257.8$]}}         &\multicolumn{1}{c}{\text{[$-187.6\:\text{to}\:303.6$]}}         &\multicolumn{1}{c}{\text{[$-151.9\:\text{to}\:365.9$]}}         \\
            &\multicolumn{1}{c}{\text{$<p=0.573>$}}         &\multicolumn{1}{c}{\text{$<p=0.774>$}}         &\multicolumn{1}{c}{\text{$<p=0.969>$}}         &\multicolumn{1}{c}{\text{$<p=0.643>$}}         &\multicolumn{1}{c}{\text{$<p=0.417>$}}         \\
Constant    &        74.0\sym{***}&        90.0\sym{***}&       108.0\sym{***}&       134.0\sym{***}&       149.0\sym{***}\\
            &       (6.1)         &       (8.1)         &      (10.3)         &      (11.7)         &      (15.2)         \\
            &\multicolumn{1}{c}{\text{[$62.1\:\text{to}\:85.9$]}}         &\multicolumn{1}{c}{\text{[$74.0\:\text{to}\:106.0$]}}         &\multicolumn{1}{c}{\text{[$87.7\:\text{to}\:128.3$]}}         &\multicolumn{1}{c}{\text{[$111.0\:\text{to}\:157.0$]}}         &\multicolumn{1}{c}{\text{[$119.2\:\text{to}\:178.8$]}}         \\
            &\multicolumn{1}{c}{\text{$<p=0.000>$}}         &\multicolumn{1}{c}{\text{$<p=0.000>$}}         &\multicolumn{1}{c}{\text{$<p=0.000>$}}         &\multicolumn{1}{c}{\text{$<p=0.000>$}}         &\multicolumn{1}{c}{\text{$<p=0.000>$}}         \\
\midrule
Median of outcome&          84         &         100         &         119         &         139         &         161         \\
Package observations&         585         &         585         &         585         &         585         &         585         \\
Day observations&           1         &           1         &           1         &           1         &           1         \\
Package-day observations&         585         &         585         &         585         &         585         &         585         \\

			\bottomrule
		\end{tabular}
	\end{adjustbox}
	\caption*{\footnotesize Note:
            Similar to \cref{tab:github_exp_medians_regtable}, but modeled to allow for differential effects by repository complexity.
            The `Large repository size` moderator indicates packages with an above-median size (total size of all files in the GitHub repository, including the codebase, in MB).
            Columns (1)-–(5) report differences in medians at monthly snapshots. 
    	    Parentheses: standard errors.
        	Square brackets: 95\% confidence intervals.
        	Angle brackets: p-values.
            Significance levels: $^{***} p < .001; ^{**} p < .01; ^{*} p < .05; ^{+} p <.1$.
	}
\end{table}

\begin{table}[!ht]
\centering \normalsize \setlength\tabcolsep{6 pt} \setlength{\defaultaddspace}{0pt}
	\def\sym#1{\ifmmode^{#1}\else\(^{#1}\)\fi}
	\caption{GitHub Experiment Results - ITT estimates by \textit{readme} documentation length.}
	\label{tab:github_exp_medians_regtable_allhumaninstallers_high_processed_readme_len}
	\begin{adjustbox}{max width=\textwidth}
		\begin{tabular}{@{\hspace{0\tabcolsep}}l*{5}{D{.}{.}{-1}}@{\hspace{0\tabcolsep}}}
			\toprule\toprule
                &\multicolumn{1}{c}{(1)} &\multicolumn{1}{c}{(2)} &\multicolumn{1}{c}{(3)}&\multicolumn{1}{c}{(4)}&\multicolumn{1}{c}{(5)}\\
                \cmidrule{2-6}                &\multicolumn{5}{c}{Outcome variable is: PyPI downloads}\\
                &\multicolumn{1}{c}{On Jun 21}&\multicolumn{1}{c}{On Jul 21}&\multicolumn{1}{c}{On Aug 21}&\multicolumn{1}{c}{On Sep 21}&\multicolumn{1}{c}{On Oct 21}\\			
                \midrule
Treatment (low dosage)&         4.0         &         4.0         &        10.0         &        -8.0         &       -23.0         \\
            &      (20.0)         &      (32.1)         &      (35.0)         &      (43.2)         &      (50.1)         \\
            &\multicolumn{1}{c}{\text{[$-35.3\:\text{to}\:43.3$]}}         &\multicolumn{1}{c}{\text{[$-59.1\:\text{to}\:67.1$]}}         &\multicolumn{1}{c}{\text{[$-58.8\:\text{to}\:78.8$]}}         &\multicolumn{1}{c}{\text{[$-92.9\:\text{to}\:76.9$]}}         &\multicolumn{1}{c}{\text{[$-121.3\:\text{to}\:75.3$]}}         \\
            &\multicolumn{1}{c}{\text{$<p=0.842>$}}         &\multicolumn{1}{c}{\text{$<p=0.901>$}}         &\multicolumn{1}{c}{\text{$<p=0.775>$}}         &\multicolumn{1}{c}{\text{$<p=0.853>$}}         &\multicolumn{1}{c}{\text{$<p=0.646>$}}         \\
Treatment (high dosage)&        62.0         &        80.0         &       116.0\sym{+}  &        96.0         &        81.0\sym{+}  \\
            &      (61.5)         &      (57.9)         &      (64.6)         &      (68.8)         &      (44.8)         \\
            &\multicolumn{1}{c}{\text{[$-58.8\:\text{to}\:182.8$]}}         &\multicolumn{1}{c}{\text{[$-33.6\:\text{to}\:193.6$]}}         &\multicolumn{1}{c}{\text{[$-10.9\:\text{to}\:242.9$]}}         &\multicolumn{1}{c}{\text{[$-39.1\:\text{to}\:231.1$]}}         &\multicolumn{1}{c}{\text{[$-6.9\:\text{to}\:168.9$]}}         \\
            &\multicolumn{1}{c}{\text{$<p=0.314>$}}         &\multicolumn{1}{c}{\text{$<p=0.167>$}}         &\multicolumn{1}{c}{\text{$<p=0.073>$}}         &\multicolumn{1}{c}{\text{$<p=0.163>$}}         &\multicolumn{1}{c}{\text{$<p=0.071>$}}         \\
Long readme documentation&         2.0         &        11.0         &        28.0\sym{+}  &        28.0         &        29.0         \\
            &      (10.2)         &      (12.0)         &      (15.1)         &      (19.8)         &      (25.1)         \\
            &\multicolumn{1}{c}{\text{[$-17.9\:\text{to}\:21.9$]}}         &\multicolumn{1}{c}{\text{[$-12.6\:\text{to}\:34.6$]}}         &\multicolumn{1}{c}{\text{[$-1.7\:\text{to}\:57.7$]}}         &\multicolumn{1}{c}{\text{[$-10.8\:\text{to}\:66.8$]}}         &\multicolumn{1}{c}{\text{[$-20.3\:\text{to}\:78.3$]}}         \\
            &\multicolumn{1}{c}{\text{$<p=0.844>$}}         &\multicolumn{1}{c}{\text{$<p=0.361>$}}         &\multicolumn{1}{c}{\text{$<p=0.065>$}}         &\multicolumn{1}{c}{\text{$<p=0.157>$}}         &\multicolumn{1}{c}{\text{$<p=0.248>$}}         \\
Treatment (low) $\times$ Long readme documentation&        10.0         &        19.0         &        42.0         &        74.0         &       114.0         \\
            &      (27.3)         &      (44.8)         &      (52.5)         &      (62.1)         &      (90.2)         \\
            &\multicolumn{1}{c}{\text{[$-43.7\:\text{to}\:63.7$]}}         &\multicolumn{1}{c}{\text{[$-69.0\:\text{to}\:107.0$]}}         &\multicolumn{1}{c}{\text{[$-61.1\:\text{to}\:145.1$]}}         &\multicolumn{1}{c}{\text{[$-47.9\:\text{to}\:195.9$]}}         &\multicolumn{1}{c}{\text{[$-63.2\:\text{to}\:291.2$]}}         \\
            &\multicolumn{1}{c}{\text{$<p=0.715>$}}         &\multicolumn{1}{c}{\text{$<p=0.672>$}}         &\multicolumn{1}{c}{\text{$<p=0.424>$}}         &\multicolumn{1}{c}{\text{$<p=0.234>$}}         &\multicolumn{1}{c}{\text{$<p=0.207>$}}         \\
Treatment (high) $\times$ Long readme documentation&       -68.0         &       -75.0         &      -128.0\sym{+}  &      -117.0         &       -98.0         \\
            &      (68.9)         &      (59.8)         &      (69.6)         &      (83.3)         &      (93.2)         \\
            &\multicolumn{1}{c}{\text{[$-203.3\:\text{to}\:67.3$]}}         &\multicolumn{1}{c}{\text{[$-192.4\:\text{to}\:42.4$]}}         &\multicolumn{1}{c}{\text{[$-264.8\:\text{to}\:8.8$]}}         &\multicolumn{1}{c}{\text{[$-280.6\:\text{to}\:46.6$]}}         &\multicolumn{1}{c}{\text{[$-281.1\:\text{to}\:85.1$]}}         \\
            &\multicolumn{1}{c}{\text{$<p=0.324>$}}         &\multicolumn{1}{c}{\text{$<p=0.210>$}}         &\multicolumn{1}{c}{\text{$<p=0.067>$}}         &\multicolumn{1}{c}{\text{$<p=0.161>$}}         &\multicolumn{1}{c}{\text{$<p=0.294>$}}         \\
Constant    &        84.0\sym{***}&        94.0\sym{***}&       106.0\sym{***}&       126.0\sym{***}&       149.0\sym{***}\\
            &       (8.4)         &      (10.3)         &      (11.6)         &      (13.1)         &      (17.8)         \\
            &\multicolumn{1}{c}{\text{[$67.6\:\text{to}\:100.4$]}}         &\multicolumn{1}{c}{\text{[$73.9\:\text{to}\:114.1$]}}         &\multicolumn{1}{c}{\text{[$83.2\:\text{to}\:128.8$]}}         &\multicolumn{1}{c}{\text{[$100.3\:\text{to}\:151.7$]}}         &\multicolumn{1}{c}{\text{[$114.0\:\text{to}\:184.0$]}}         \\
            &\multicolumn{1}{c}{\text{$<p=0.000>$}}         &\multicolumn{1}{c}{\text{$<p=0.000>$}}         &\multicolumn{1}{c}{\text{$<p=0.000>$}}         &\multicolumn{1}{c}{\text{$<p=0.000>$}}         &\multicolumn{1}{c}{\text{$<p=0.000>$}}         \\
\midrule
Median of outcome&          86         &         104         &         126         &       141.5         &       163.5         \\
Package observations&         622         &         622         &         622         &         622         &         622         \\
Day observations&           1         &           1         &           1         &           1         &           1         \\
Package-day observations&         622         &         622         &         622         &         622         &         622         \\

			\bottomrule
		\end{tabular}
	\end{adjustbox}
	\caption*{\footnotesize Note:
            Similar to \cref{tab:github_exp_medians_regtable}, but modeled to allow for differential effects by repository complexity.
            The `Long description` moderator indicates packages with an above-median \textit{readme} documentation length (cleaned for HTML markup; raw text is similarly null).
            Columns (1)-–(5) report differences in medians at monthly snapshots. 
    	    Parentheses: standard errors.
        	Square brackets: 95\% confidence intervals.
        	Angle brackets: p-values.
            Significance levels: $^{***} p < .001; ^{**} p < .01; ^{*} p < .05; ^{+} p <.1$.
	}
\end{table}

\clearpage
\FloatBarrier
% ~~~~~~~~~~~~~~~~~~~~~~~~~~~~~~~~~~~~~~~~~~~~~~~~~~~~~~~~~~~~~~~~~~~~~~~~~~~~~~~
% ~~~~~~~~~~~~~~~~~~~~~~~~~~~~~~~~~~~~~~~~~~~~~~~~~~~~~~~~~~~~~~~~~~~~~~~~~~~~~~~
\subsection{Balance Tests of GitHub Archive Events}
\label{sec:balance-gharchive}
% ~~~~~~~~~~~~~~~~~~~~~~~~~~~~~~~~~~~~~~~~~~~~~~~~~~~~~~~~~~~~~~~~~~~~~~~~~~~~~~~
% ~~~~~~~~~~~~~~~~~~~~~~~~~~~~~~~~~~~~~~~~~~~~~~~~~~~~~~~~~~~~~~~~~~~~~~~~~~~~~~~
In this section, we construct pre-treatment balance tables using historical GitHub event data from the GitHub Archive Project.
Specifically, we aggregated cumulative event counts for each repository in the pre-treatment period across seven GitHub activity metrics: stars (WatchEvent), forks, pushes, pull requests, issues opened, issues closed, and releases.

For the GitHub experiment, we observe no pre-treatment difference between the treated and control groups (\cref{tab:baltest-gharchive-pre-treated-01}).
When we split the treatment group by dosage (high vs. low, \cref{sec:gh-treatment-conditions}), we observe a pre-treatment difference between the high-dosage group and the control group on issues and forks (\cref{tab:baltest-gharchive-pre-treated-012}).%
\footnote{Including these seven pre-treatment metrics as baseline covariates does not change the null findings in \cref{tab:github_exp_medians_regtable}.}
However, we attribute these differences to the outliers in the small high-dosage sample for two reasons: (i) median pre-treatment differences are null, and (ii) the mean differences in the high-dosage group are driven by a single repository (\textit{deepmind/mujoco})---removing it removes all statistically significant difference.

For the PyPI experiment, we likewise observe no pre-treatment difference across any of the seven GitHub activities (\cref{tab:baltest-gharchive-pre-treated}).

\begin{table}[ht]
\footnotesize \setlength\tabcolsep{6 pt}
\caption{GitHub experiment: Balance tests of repository events}
\label{tab:baltest-gharchive-pre-treated-01}
\begin{adjustbox}{max width=\textwidth}
%%% Table created in Stata by command iebaltab
%%% (https://github.com/worldbank/ietoolkit)
%%% (https://dimewiki.worldbank.org/iebaltab)
%%% The command was specified exactly like this: 
%%% iebaltab watchevent_pre pushevent_pre pullrequestevent_pre issues_opened_pre issues_closed_pre forkevent_pre releaseevent_pre , total groupvar(treated) star(.1 .05 .01) stats(pair(nrmd)) nonote grplabels( 0 Control @ 1 Treated @ ) order(0 1) control(0) grouplabels(0 "Control repositories" @ 1 "Treated repositories") rowlabels( forkevent_pre "Fork events" @ pullrequestevent_pre "Pull request events" @ pushevent_pre "Push events" @ releaseevent_pre "Release events" @ watchevent_pre "Stars" @ issues_opened_pre "Issues opened" @ issues_closed_pre "Issues closed" ) totallabel(Full sample) format(%9.2f) savetex(../output/baltest-gharchive-pre-treated-01.tex) replace

\begin{tabular}{@{\extracolsep{5pt}}lcccccccc}
\\[-1.8ex]\hline \hline \\[-1.8ex]
 & \multicolumn{2}{c}{(1)}  & \multicolumn{2}{c}{(2)}  & \multicolumn{2}{c}{(3)}  & \multicolumn{2}{c}{(3)-(2)} \\
 & \multicolumn{2}{c}{Full sample}  & \multicolumn{2}{c}{Control repositories}  & \multicolumn{2}{c}{Treated repositories}  & \multicolumn{2}{c}{Pairwise t-test}  \\
Variable & N & Mean/(SE) & N & Mean/(SE) & N & Mean/(SE) & N & Normalized difference \\ \hline \\[-1.8ex] 
Stars   & 582    & 11.72    & 485    & 11.85    & 97    & 11.11    & 582    & -0.01   \\
 &   & (3.43)  &   & (3.95)  &   & (5.63)  &   &  \\ [1ex]
Push events   & 582    & 30.61    & 485    & 29.45    & 97    & 36.42    & 582    & 0.06   \\
 &   & (5.04)  &   & (5.77)  &   & (9.16)  &   &  \\ [1ex]
Pull request events   & 582    & 11.62    & 485    & 10.51    & 97    & 17.16    & 582    & 0.12   \\
 &   & (2.47)  &   & (2.79)  &   & (4.88)  &   &  \\ [1ex]
Issues opened   & 582    & 2.10    & 485    & 1.86    & 97    & 3.31    & 582    & 0.12   \\
 &   & (0.44)  &   & (0.43)  &   & (1.50)  &   &  \\ [1ex]
Issues closed   & 582    & 1.87    & 485    & 1.62    & 97    & 3.13    & 582    & 0.12   \\
 &   & (0.48)  &   & (0.48)  &   & (1.55)  &   &  \\ [1ex]
Fork events   & 582    & 1.88    & 485    & 1.70    & 97    & 2.80    & 582    & 0.11   \\
 &   & (0.37)  &   & (0.38)  &   & (1.10)  &   &  \\ [1ex]
Release events   & 582    & 1.70    & 485    & 1.72    & 97    & 1.63    & 582    & -0.02   \\
 &   & (0.19)  &   & (0.22)  &   & (0.40)  &   &  \\ [1ex]
\hline \hline \\[-1.8ex]

\end{tabular}

\end{adjustbox}
\caption*{
\footnotesize 
\emph{Notes---}%
This table reports pre-treatment balance tests for GitHub repository event activity constructed from the 2023 GitHub Archive data.
2023 event counts are aggregated at the repository level for the period prior to the treatment window (May 12--20, 2023).
% Table reports the balance in GitHub repository characteristics between the treated and control Python packages.
% Column (1) reports the mean of repository characteristics. 
% Column (2) reports the mean of the control packages.
% Column (3) reports the mean of the treated packages.
% Standard errors of the mean are in parentheses.
% The last column reports the standardized mean difference. 
% The repository size is in megabytes. 
% Forked indicates whether the repository was forked from another existing repository. 
% Topics are optional labels for a GitHub repository (e.g., web application, encryption, Python). 
% Python = 1 indicates Python is the primary detected language.
% % e.g. TensorFlow's topics: machine-learning deep-neural-networks deep-learning neural-network tensorflow ml distributed
% See \cref{tab:baltest-repo-treated-012} for the same table with low and high dosage treatments. 
% See \cref{tab:baltest-readme-treated-01} for the same table reporting balance for package description and dependency balance. 
% Significance levels: $^{***} p < .01; ^{**} p < .05; ^{*} p <.1$.
}
\end{table}

\begin{table}[!ht]
\footnotesize \setlength\tabcolsep{5 pt}
\caption{GitHub experiment: Balance tests of repository events (low-dosage and high-dosage treatment groups vs. control)}
\label{tab:baltest-gharchive-pre-treated-012}
\begin{adjustbox}{max width=\textwidth}
%%% Table created in Stata by command iebaltab
%%% (https://github.com/worldbank/ietoolkit)
%%% (https://dimewiki.worldbank.org/iebaltab)
%%% The command was specified exactly like this: 
%%% iebaltab watchevent_pre pushevent_pre pullrequestevent_pre issues_opened_pre issues_closed_pre forkevent_pre releaseevent_pre , groupvar(treated2) star(.1 .05 .01) stats(pair(nrmd)) nonote grplabels( 0 Control @ 1 Treated (low) @ 2 Treated (high) @ ) order(0 1 2) control(0) grouplabels( 0 "Control repositories" @ 1 "Treated (low dose)" @ 2 "Treated (high dose)" ) rowlabels( forkevent_pre "Fork events" @ pullrequestevent_pre "Pull request events" @ pushevent_pre "Push events" @ releaseevent_pre "Release events" @ watchevent_pre "Stars" @ issues_opened_pre "Issues opened" @ issues_closed_pre "Issues closed" ) format(%9.2f) savetex(../output/baltest-gharchive-pre-treated-012.tex) replace

\begin{tabular}{@{\extracolsep{5pt}}lcccccccccc}
\\[-1.8ex]\hline \hline \\[-1.8ex]
 & \multicolumn{2}{c}{(1)}  & \multicolumn{2}{c}{(2)}  & \multicolumn{2}{c}{(3)}  & \multicolumn{2}{c}{(2)-(1)} & \multicolumn{2}{c}{(3)-(1)} \\
 & \multicolumn{2}{c}{Control repositories}  & \multicolumn{2}{c}{Treated (low dose)}  & \multicolumn{2}{c}{Treated (high dose)}  & \multicolumn{4}{c}{Pairwise t-test}  \\
Variable & N & Mean/(SE) & N & Mean/(SE) & N & Mean/(SE) & N & Normalized difference & N & Normalized difference \\ \hline \\[-1.8ex] 
Stars   & 485    & 11.85    & 73    & 4.62    & 24    & 30.88    & 558    & -0.12    & 509    & 0.19   \\
 &   & (3.95)  &   & (1.68)  &   & (22.05)  &   &  &   &  \\ [1ex]
Push events   & 485    & 29.45    & 73    & 29.97    & 24    & 56.04    & 558    & 0.00    & 509    & 0.25   \\
 &   & (5.77)  &   & (10.71)  &   & (17.35)  &   &  &   &  \\ [1ex]
Pull request events   & 485    & 10.51    & 73    & 11.77    & 24    & 33.58    & 558    & 0.02    & 509    & 0.35*   \\
 &   & (2.79)  &   & (4.43)  &   & (14.14)  &   &  &   &  \\ [1ex]
Issues opened   & 485    & 1.86    & 73    & 1.78    & 24    & 7.96    & 558    & -0.01    & 509    & 0.31***   \\
 &   & (0.43)  &   & (0.94)  &   & (5.31)  &   &  &   &  \\ [1ex]
Issues closed   & 485    & 1.62    & 73    & 1.45    & 24    & 8.25    & 558    & -0.02    & 509    & 0.32***   \\
 &   & (0.48)  &   & (0.90)  &   & (5.61)  &   &  &   &  \\ [1ex]
Fork events   & 485    & 1.70    & 73    & 1.22    & 24    & 7.63    & 558    & -0.07    & 509    & 0.39***   \\
 &   & (0.38)  &   & (0.56)  &   & (4.01)  &   &  &   &  \\ [1ex]
Release events   & 485    & 1.72    & 73    & 1.88    & 24    & 0.88    & 558    & 0.03    & 509    & -0.24   \\
 &   & (0.22)  &   & (0.52)  &   & (0.31)  &   &  &   &  \\ [1ex]
\hline \hline \\[-1.8ex]

\end{tabular}

\end{adjustbox}
\caption*{
\footnotesize 
\emph{Notes---}
Same as \cref{tab:baltest-gharchive-pre-treated-01}, except the treatment group is distinguished by low and high dosage treatment packages (see \cref{sec:gh-treatment-conditions}).
*** Significant at the 1 percent level. ** Significant at the 5 percent level. * Significant at the 10 percent level.
}
\end{table}

\begin{table}[ht]
\footnotesize \setlength\tabcolsep{6 pt}
\caption{PyPI experiment: Balance tests of repository events}
\label{tab:baltest-gharchive-pre-treated}
\begin{adjustbox}{max width=\textwidth}
%%% Table created in Stata by command iebaltab
%%% (https://github.com/worldbank/ietoolkit)
%%% (https://dimewiki.worldbank.org/iebaltab)
%%% The command was specified exactly like this: 
%%% iebaltab watchevent_pre pushevent_pre pullrequestevent_pre issues_opened_pre issues_closed_pre forkevent_pre releaseevent_pre , browse total groupvar(treatment) star(.1 .05 .01) stats(pair(nrmd)) nonote grplabels( 0 Control @ 1 Treated @ ) order(0 1) control(0) grouplabels(0 "Control repositories" @ 1 "Treated repositories") rowlabels( forkevent_pre "Fork events" @ pullrequestevent_pre "Pull request events" @ pushevent_pre "Push events" @ releaseevent_pre "Release events" @ watchevent_pre "Stars" @ issues_opened_pre "Issues opened" @ issues_closed_pre "Issues closed" ) totallabel(Full sample) format(%9.2f) savetex(../tabs/baltest-gharchive-pre-treated.tex) replace

\begin{tabular}{@{\extracolsep{5pt}}lcccccccc}
\\[-1.8ex]\hline \hline \\[-1.8ex]
 & \multicolumn{2}{c}{(1)}  & \multicolumn{2}{c}{(2)}  & \multicolumn{2}{c}{(3)}  & \multicolumn{2}{c}{(3)-(2)} \\
 & \multicolumn{2}{c}{Full sample}  & \multicolumn{2}{c}{Control repositories}  & \multicolumn{2}{c}{Treated repositories}  & \multicolumn{2}{c}{Pairwise t-test}  \\
Variable & N & Mean/(SE) & N & Mean/(SE) & N & Mean/(SE) & N & Normalized difference \\ \hline \\[-1.8ex] 
Stars   & 17872    & 24.41    & 14241    & 24.51    & 3631    & 24.04    & 17872    & -0.00   \\
 &   & (4.45)  &   & (5.48)  &   & (4.23)  &   &  \\ [1ex]
Push events   & 17872    & 44.30    & 14241    & 43.03    & 3631    & 49.30    & 17872    & 0.02   \\
 &   & (2.50)  &   & (2.74)  &   & (6.05)  &   &  \\ [1ex]
Pull request events   & 17872    & 31.26    & 14241    & 29.67    & 3631    & 37.48    & 17872    & 0.03*   \\
 &   & (1.74)  &   & (1.87)  &   & (4.42)  &   &  \\ [1ex]
Issues opened   & 17872    & 6.49    & 14241    & 6.27    & 3631    & 7.34    & 17872    & 0.01   \\
 &   & (0.61)  &   & (0.72)  &   & (0.98)  &   &  \\ [1ex]
Issues closed   & 17872    & 5.35    & 14241    & 5.16    & 3631    & 6.11    & 17872    & 0.02   \\
 &   & (0.45)  &   & (0.53)  &   & (0.81)  &   &  \\ [1ex]
Fork events   & 17872    & 7.20    & 14241    & 7.11    & 3631    & 7.53    & 17872    & 0.00   \\
 &   & (1.17)  &   & (1.44)  &   & (1.05)  &   &  \\ [1ex]
Release events   & 17872    & 1.29    & 14241    & 1.25    & 3631    & 1.47    & 17872    & 0.02   \\
 &   & (0.10)  &   & (0.11)  &   & (0.23)  &   &  \\ [1ex]
\hline \hline \\[-1.8ex]

\end{tabular}

\end{adjustbox}
\caption*{
\footnotesize 
\emph{Notes---}%
This table reports pre-treatment balance tests for the PyPI experiment GitHub repository event activity constructed from the 2023 GitHub Archive data.
2023 event counts are aggregated at the repository level for the period prior to the treatment window (Jun 3--8, 2023).
Significance levels: $^{***} p < .01; ^{**} p < .05; ^{*} p <.1$.
}
\end{table}

% ~~~~~~~~~~~~~~~~~~~~~~~~~~~~~~~~~~~~~~~~~~~~~~~~~~~~~~~~~~~~~~~~~~~~~~~~~~~~~~~
% ~~~~~~~~~~~~~~~~~~~~~~~~~~~~~~~~~~~~~~~~~~~~~~~~~~~~~~~~~~~~~~~~~~~~~~~~~~~~~~~
\clearpage
\FloatBarrier
\subsection{PyPI Experiment}
% ~~~~~~~~~~~~~~~~~~~~~~~~~~~~~~~~~~~~~~~~~~~~~~~~~~~~~~~~~~~~~~~~~~~~~~~~~~~~~~~
% ~~~~~~~~~~~~~~~~~~~~~~~~~~~~~~~~~~~~~~~~~~~~~~~~~~~~~~~~~~~~~~~~~~~~~~~~~~~~~~~
\begin{table}[ht] \centering \normalsize \setlength\tabcolsep{6 pt} \setlength{\defaultaddspace}{0pt}
	\def\sym#1{\ifmmode^{#1}\else\(^{#1}\)\fi}
\caption{Classification of human vs bot downloads by package installer.}
	\label{tab:human_bot_download_classification}
	\begin{adjustbox}{max width=\textwidth}
\begin{tabular}{@{\hspace{0\tabcolsep}}ll@{\hspace{0\tabcolsep}}}
\toprule
Installer Name & Type \\
\midrule
pip & Human \\
Browser & Bot \\
Bandersnatch & Bot \\
setuptools & Human \\
Nexus & Human \\
requests & Bot \\
devpi & Bot \\
pdm & Human \\
Homebrew & Human \\
Artifactory & Human \\
OS & Human \\
Bazel & Human \\
pex & Human \\
conda & Human \\
chaquopy & Human \\
\bottomrule
\end{tabular}
\end{adjustbox}
\end{table}

\begin{figure}[ht]
\centering
    \includegraphics[width=\textwidth]{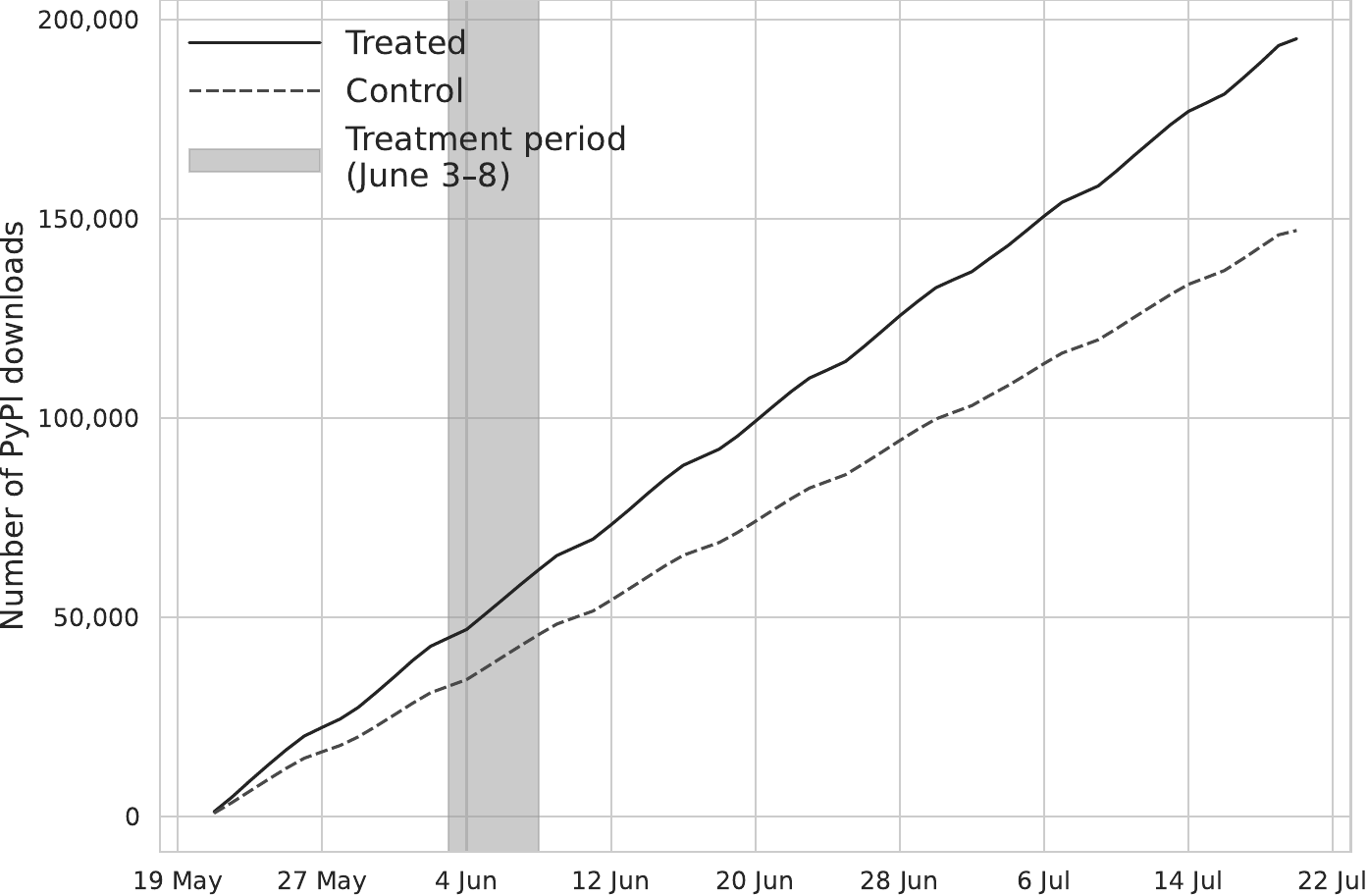}
\caption{
\textbf{Mean PyPI downloads for Treated vs. Control in PyPI Experiment.}
Same as \cref{fig:median_downloads_pypi_experiment}, except in means. The figure shows trends in mean daily downloads for the treated packages ($n = 4,814$) and control group packages ($n = 19,102$) for 1,458,876 package-day observations. 
% for numbers, see https://github.com/soodoku/social_proof_stars/blob/main/pydownloads/scripts/consolidate_pypi_downloads.ipynb
The shaded vertical bar indicates the treatment period. 
Downloads include only human downloads (\cref{tab:human_bot_download_classification}).
\cref{tab:pypi_exp_regtable} reports the estimates in the differences in means.
}
\label{fig:mean_downloads_pypi_experiment}
\end{figure}

\begin{table}[!ht] \centering \normalsize \setlength\tabcolsep{6 pt} \setlength{\defaultaddspace}{0pt}
	\def\sym#1{\ifmmode^{#1}\else\(^{#1}\)\fi}
	\caption{PyPI Experiment Results - ITT estimates for medians and means.}
	\label{tab:pypi_exp_regtable}
	\begin{adjustbox}{max width=\textwidth}
		\begin{tabular}{@{\hspace{0\tabcolsep}}l*{4}{D{.}{.}{-1}}@{\hspace{0\tabcolsep}}}
			\toprule\toprule
                &\multicolumn{1}{c}{(1)} &\multicolumn{1}{c}{(2)}&\multicolumn{1}{c}{(3)} &\multicolumn{1}{c}{(4)}\\
                \cmidrule(l){2-5}
                &\multicolumn{4}{c}{Outcome variable is: PyPI downloads}\\
                &\multicolumn{2}{c}{Diff. in medians}&\multicolumn{2}{c}{Diff. means}\\
                \cmidrule(lr){2-3}\cmidrule(l){4-5}
                &\multicolumn{1}{c}{Jun 22}&\multicolumn{1}{c}{Full post period}&\multicolumn{1}{c}{Jun 22}&\multicolumn{1}{c}{Full post period}\\
			\midrule
Treatment group&        83.0\sym{***}&        86.7\sym{***}&    26,888.8         &    16,127.6         \\
            &       (0.9)         &       (0.4)         &  (52,957.6)         &  (30,751.4)         \\
            &\multicolumn{1}{c}{\text{[$81.3\:\text{to}\:84.7$]}}         &\multicolumn{1}{c}{\text{[$85.8\:\text{to}\:87.5$]}}         &\multicolumn{1}{c}{\text{[$-76,911.6\:\text{to}\:130,689.1$]}}         &\multicolumn{1}{c}{\text{[$-44,147.1\:\text{to}\:76,402.3$]}}         \\
            &\multicolumn{1}{c}{\text{$<p=0.000>$}}         &\multicolumn{1}{c}{\text{$<p=0.000>$}}         &\multicolumn{1}{c}{\text{$<p=0.612>$}}         &\multicolumn{1}{c}{\text{$<p=0.600>$}}         \\
Linear trend&                     &         0.9\sym{***}&                     &     2,436.4\sym{***}\\
            &                     &       (0.0)         &                     &     (551.9)         \\
            &                     &\multicolumn{1}{c}{\text{[$0.9\:\text{to}\:1.0$]}}         &                     &\multicolumn{1}{c}{\text{[$1,354.7\:\text{to}\:3,518.0$]}}         \\
            &                     &\multicolumn{1}{c}{\text{$<p=0.000>$}}         &                     &\multicolumn{1}{c}{\text{$<p=0.000>$}}         \\
Treatment group  $ \times$ Linear trend&                     &        -0.2\sym{***}&                     &       748.0         \\
            &                     &       (0.0)         &                     &   (1,558.3)         \\
            &                     &\multicolumn{1}{c}{\text{[$-0.3\:\text{to}\:-0.2$]}}         &                     &\multicolumn{1}{c}{\text{[$-2,306.3\:\text{to}\:3,802.3$]}}         \\
            &                     &\multicolumn{1}{c}{\text{$<p=0.000>$}}         &                     &\multicolumn{1}{c}{\text{$<p=0.631>$}}         \\
Constant    &        33.0\sym{***}&        18.1\sym{***}&    79,904.2\sym{***}&    45,253.6\sym{***}\\
            &       (0.5)         &       (0.3)         &  (18,079.3)         &  (10,235.3)         \\
            &\multicolumn{1}{c}{\text{[$32.1\:\text{to}\:33.9$]}}         &\multicolumn{1}{c}{\text{[$17.5\:\text{to}\:18.7$]}}         &\multicolumn{1}{c}{\text{[$44,467.6\:\text{to}\:115,340.8$]}}         &\multicolumn{1}{c}{\text{[$25,191.7\:\text{to}\:65,315.5$]}}         \\
            &\multicolumn{1}{c}{\text{$<p=0.000>$}}         &\multicolumn{1}{c}{\text{$<p=0.000>$}}         &\multicolumn{1}{c}{\text{$<p=0.000>$}}         &\multicolumn{1}{c}{\text{$<p=0.000>$}}         \\
\midrule
Median/Mean of outcome&          43         &          49         &      85,317         &     104,119         \\
Package observations&      23,916         &      23,916         &      23,916         &      23,916         \\
Day observations&           1         &          42         &           1         &          42         \\
Package-day observations&      23,916         &   1,004,472         &      23,916         &   1,004,472         \\

			\bottomrule
		\end{tabular}
	\end{adjustbox}
	\caption*{\footnotesize Note:
            The table presents pre-to-post treatment changes in PyPI downloads.
            Column (1)-(2) reports post-treatment differences in medians 14 days after intervention occurred (22 Jun 2023).
            Column (3)--(4) reports post-treatment differences in means.
            Columns (2) and (4) allow for heterogeneous treatment effects through a linear time trend over the 42 days.
            Columns (1)--(2) correspond to \cref{fig:median_downloads_pypi_experiment}.
            Columns (3)--(4) do the same for differences in means and correspond to \cref{fig:mean_downloads_pypi_experiment}.
            % The sample period for GitHub stars is a 30-day period from 9 May 2023 to 7 June 2023.
            % The sample period for PyPI downloads is a 47-day period from 25 April 2023 to 10 June 2023.
            % The post period includes all dates after 12 May 2023, when intervention occurred (\cref{sec:gh-treatment-conditions}).
            Standard errors are clustered by packages.
        	Parentheses: standard errors.
        	Square brackets: 95\% confidence intervals.
        	Angle brackets: p-values.
        	Significance levels: $^{***} p < .001; ^{**} p < .01; ^{*} p < .05; ^{+} p <.1$.
	}
\end{table}

\begin{table}[!ht] \centering \small \setlength\tabcolsep{14 pt} \setlength{\defaultaddspace}{0pt}
\def\sym#1{\ifmmode^{#1}\else\(^{#1}\)\fi}
\caption{PyPI Experiment Results - LATE estimates for means.}
\label{tab:pypi_exp_regtable_late}
\begin{adjustbox}{max width=\textwidth}
	\begin{tabular}{@{\hspace{0\tabcolsep}}l*{2}{D{.}{.}{-1}}@{\hspace{0\tabcolsep}}}
		\toprule\toprule
            &\multicolumn{1}{c}{(1)} &\multicolumn{1}{c}{(2)}\\
            \cmidrule(l){2-3}
            &\multicolumn{2}{c}{Outcome variable is: PyPI downloads}\\
            &\multicolumn{1}{c}{Jun 22}&\multicolumn{1}{c}{Full post period}\\
		\midrule
Received treatment&    52,427.1         &    31,445.3         \\
            & (103,217.8)         &  (59,946.3)         \\
            &\multicolumn{1}{c}{\text{[$-149876.1\:\text{to}\:254,730.3$]}}         &\multicolumn{1}{c}{\text{[$-86,047.3\:\text{to}\:148,937.9$]}}         \\
            &\multicolumn{1}{c}{\text{$<p=0.612>$}}         &\multicolumn{1}{c}{\text{$<p=0.600>$}}         \\
Linear trend&                     &     2,436.4\sym{***}\\
            &                     &     (551.8)         \\
            &                     &\multicolumn{1}{c}{\text{[$1,354.8\:\text{to}\:3,518.0$]}}         \\
            &                     &\multicolumn{1}{c}{\text{$<p=0.000>$}}         \\
Received treatment  $ \times$ Linear trend&                     &     1,458.4         \\
            &                     &   (3,037.7)         \\
            &                     &\multicolumn{1}{c}{\text{[$-4,495.4\:\text{to}\:7,412.2$]}}         \\
            &                     &\multicolumn{1}{c}{\text{$<p=0.631>$}}         \\
Constant    &    79,904.2\sym{***}&    45,253.6\sym{***}\\
            &  (18,078.4)         &  (10,235.1)         \\
            &\multicolumn{1}{c}{\text{[$44,471.3\:\text{to}\:115,337.1$]}}         &\multicolumn{1}{c}{\text{[$25,193.2\:\text{to}\:65,314.0$]}}         \\
            &\multicolumn{1}{c}{\text{$<p=0.000>$}}         &\multicolumn{1}{c}{\text{$<p=0.000>$}}         \\
\midrule
Mean of outcome&      85,317         &     104,119         \\
Package observations&      23,916         &      23,916         \\
Day observations&           1         &          42         \\
Package-day observations&      23,916         &   1,004,472         \\

		\bottomrule
	\end{tabular}
\end{adjustbox}
\caption*{\small Note:
        The table presents LATE estimates for mean downloads in the PyPI experiment.
        Compliers (those who ``Received treatment'') are defined as those receiving at least 50 downloads at the end of the treatment window.
        LATE estimates are from instrumental variable regressions, instrumenting compliers with the random treatment assignment.
        Column (1) reports differences in means 14 days after intervention occurred (22 Jun 2023).
        Column (2) reports post-treatment differences in means over the 42 days, allowing for heterogeneous treatment effects through a linear time trend.
        Columns (1)--(2) correspond to \cref{fig:mean_downloads_pypi_experiment}.
        Standard errors are clustered by packages.
    	Parentheses: standard errors.
    	Square brackets: 95\% confidence intervals.
    	Angle brackets: p-values.
    	Significance levels: $^{***} p < .001; ^{**} p < .01; ^{*} p < .05; ^{+} p <.1$.
}
\end{table}

% ~~~~~~~~~~~~~~~~~~~~~~~~~~~~~~~~~~~~~~~~~~~~~~~~~~~~~~~~~~~~~~~~~~~~~~~~~~~~~~~
% ~~~~~~~~~~~~~~~~~~~~~~~~~~~~~~~~~~~~~~~~~~~~~~~~~~~~~~~~~~~~~~~~~~~~~~~~~~~~~~~]
\clearpage
\FloatBarrier
\begin{landscape}
\subsection{GitHub Events as Outcomes}
\label{sec:gharchive-outcomes}

In this section, we use the seven GitHub activity metrics (stars, forks, pushes, pull requests, opened issues, closed issues, and releases) as outcomes.
Specifically, we estimate whether treated groups have a difference in those metrics at a post-treatment snapshot (July 1) and whether there is a linear trend across the entire post-treatment period (to the end of 2023).

For the GitHub experiment, we find no post-treatment difference in any metric, except that treated groups have more stars, which is another confirmation that manipulation was in place (\cref{tab:gh_exp_regtable_gharchive_means}).
We find similarly null results when we look at the post-treatment snapshot on December 31.
Unlike \cref{tab:github_exp_medians_regtable}, we do not estimate median differences since the median for most GitHub activity metrics is zero in both groups throughout the observation window (causing the estimator to degenerate).

For the PyPI experiment, we find similarly null post-treatment effects across all seven metrics (\cref{tab:pypi_exp_regtable_gharchive_means}).

\begin{table}[!ht] \centering \small \setlength\tabcolsep{2pt} \setlength{\defaultaddspace}{0pt}
	\def\sym#1{\ifmmode^{#1}\else\(^{#1}\)\fi}
	\caption{GitHub Experiment: Post-treatment differences in GitHub repository event activity (ITT).}
	\label{tab:gh_exp_regtable_gharchive_means}
	\begin{adjustbox}{max width=1.325\textwidth}
		\begin{tabular}{@{\hspace{0\tabcolsep}}l*{14}{D{.}{.}{-1}}@{\hspace{0\tabcolsep}}}
			\toprule\toprule
			&\multicolumn{1}{c}{(1)}&\multicolumn{1}{c}{(2)}&\multicolumn{1}{c}{(3)}&\multicolumn{1}{c}{(4)}&\multicolumn{1}{c}{(5)}&\multicolumn{1}{c}{(6)}&\multicolumn{1}{c}{(7)}&\multicolumn{1}{c}{(8)}&\multicolumn{1}{c}{(9)}&\multicolumn{1}{c}{(10)}&\multicolumn{1}{c}{(11)}&\multicolumn{1}{c}{(12)}&\multicolumn{1}{c}{(13)}&\multicolumn{1}{c}{(14)}\\
			\cmidrule(l){2-15}
			&\multicolumn{2}{c}{Stars}&\multicolumn{2}{c}{Push events}&\multicolumn{2}{c}{Pull requests}&\multicolumn{2}{c}{Issues opened}&\multicolumn{2}{c}{Issues closed}&\multicolumn{2}{c}{Fork events}&\multicolumn{2}{c}{Release events}\\
			\cmidrule(lr){2-3}\cmidrule(lr){4-5}\cmidrule(lr){6-7}\cmidrule(lr){8-9}\cmidrule(lr){10-11}\cmidrule(lr){12-13}\cmidrule(l){14-15}
			&\multicolumn{1}{c}{Jul 1}&\multicolumn{1}{c}{Full}&\multicolumn{1}{c}{Jul 1}&\multicolumn{1}{c}{Full}&\multicolumn{1}{c}{Jul 1}&\multicolumn{1}{c}{Full}&\multicolumn{1}{c}{Jul 1}&\multicolumn{1}{c}{Full}&\multicolumn{1}{c}{Jul 1}&\multicolumn{1}{c}{Full}&\multicolumn{1}{c}{Jul 1}&\multicolumn{1}{c}{Full}&\multicolumn{1}{c}{Jul 1}&\multicolumn{1}{c}{Full}\\
			\midrule
Treatment group&        29.1\sym{**} &        29.8\sym{**} &         8.7         &         6.2         &         7.8         &         5.9         &         1.9         &         1.4         &         1.9         &         1.6         &         1.4         &         1.3         &        -0.4         &        -0.2         \\
            &      (10.2)         &       (9.2)         &      (13.7)         &      (11.2)         &       (7.6)         &       (6.1)         &       (2.3)         &       (2.0)         &       (2.2)         &       (1.9)         &       (1.6)         &       (1.4)         &       (0.5)         &       (0.5)         \\
            &\multicolumn{1}{c}{\text{[$9.1\:\text{to}\:49.1$]}}         &\multicolumn{1}{c}{\text{[$11.8\:\text{to}\:47.8$]}}         &\multicolumn{1}{c}{\text{[$-18.2\:\text{to}\:35.5$]}}         &\multicolumn{1}{c}{\text{[$-15.7\:\text{to}\:28.2$]}}         &\multicolumn{1}{c}{\text{[$-7.1\:\text{to}\:22.7$]}}         &\multicolumn{1}{c}{\text{[$-6.0\:\text{to}\:17.9$]}}         &\multicolumn{1}{c}{\text{[$-2.6\:\text{to}\:6.4$]}}         &\multicolumn{1}{c}{\text{[$-2.5\:\text{to}\:5.2$]}}         &\multicolumn{1}{c}{\text{[$-2.5\:\text{to}\:6.2$]}}         &\multicolumn{1}{c}{\text{[$-2.2\:\text{to}\:5.4$]}}         &\multicolumn{1}{c}{\text{[$-1.8\:\text{to}\:4.6$]}}         &\multicolumn{1}{c}{\text{[$-1.5\:\text{to}\:4.1$]}}         &\multicolumn{1}{c}{\text{[$-1.4\:\text{to}\:0.6$]}}         &\multicolumn{1}{c}{\text{[$-1.2\:\text{to}\:0.7$]}}         \\
            &\multicolumn{1}{c}{\text{$<p=0.004>$}}         &\multicolumn{1}{c}{\text{$<p=0.001>$}}         &\multicolumn{1}{c}{\text{$<p=0.527>$}}         &\multicolumn{1}{c}{\text{$<p=0.577>$}}         &\multicolumn{1}{c}{\text{$<p=0.302>$}}         &\multicolumn{1}{c}{\text{$<p=0.329>$}}         &\multicolumn{1}{c}{\text{$<p=0.410>$}}         &\multicolumn{1}{c}{\text{$<p=0.478>$}}         &\multicolumn{1}{c}{\text{$<p=0.404>$}}         &\multicolumn{1}{c}{\text{$<p=0.416>$}}         &\multicolumn{1}{c}{\text{$<p=0.380>$}}         &\multicolumn{1}{c}{\text{$<p=0.351>$}}         &\multicolumn{1}{c}{\text{$<p=0.447>$}}         &\multicolumn{1}{c}{\text{$<p=0.621>$}}         \\
Linear trend&                     &         0.1\sym{***}&                     &         0.1\sym{***}&                     &         0.1\sym{***}&                     &         0.0\sym{***}&                     &         0.0\sym{***}&                     &         0.0\sym{***}&                     &         0.0\sym{***}\\
            &                     &       (0.0)         &                     &       (0.0)         &                     &       (0.0)         &                     &       (0.0)         &                     &       (0.0)         &                     &       (0.0)         &                     &       (0.0)         \\
            &                     &\multicolumn{1}{c}{\text{[$0.0\:\text{to}\:0.1$]}}         &                     &\multicolumn{1}{c}{\text{[$0.1\:\text{to}\:0.2$]}}         &                     &\multicolumn{1}{c}{\text{[$0.0\:\text{to}\:0.1$]}}         &                     &\multicolumn{1}{c}{\text{[$0.0\:\text{to}\:0.0$]}}         &                     &\multicolumn{1}{c}{\text{[$0.0\:\text{to}\:0.0$]}}         &                     &\multicolumn{1}{c}{\text{[$0.0\:\text{to}\:0.0$]}}         &                     &\multicolumn{1}{c}{\text{[$0.0\:\text{to}\:0.0$]}}         \\
            &                     &\multicolumn{1}{c}{\text{$<p=0.000>$}}         &                     &\multicolumn{1}{c}{\text{$<p=0.000>$}}         &                     &\multicolumn{1}{c}{\text{$<p=0.000>$}}         &                     &\multicolumn{1}{c}{\text{$<p=0.000>$}}         &                     &\multicolumn{1}{c}{\text{$<p=0.000>$}}         &                     &\multicolumn{1}{c}{\text{$<p=0.000>$}}         &                     &\multicolumn{1}{c}{\text{$<p=0.000>$}}         \\
Treatment group  $ \times$ Linear trend&                     &        -0.0         &                     &         0.1         &                     &         0.0         &                     &         0.0         &                     &         0.0         &                     &         0.0         &                     &        -0.0         \\
            &                     &       (0.0)         &                     &       (0.1)         &                     &       (0.0)         &                     &       (0.0)         &                     &       (0.0)         &                     &       (0.0)         &                     &       (0.0)         \\
            &                     &\multicolumn{1}{c}{\text{[$-0.1\:\text{to}\:0.1$]}}         &                     &\multicolumn{1}{c}{\text{[$-0.1\:\text{to}\:0.2$]}}         &                     &\multicolumn{1}{c}{\text{[$-0.0\:\text{to}\:0.1$]}}         &                     &\multicolumn{1}{c}{\text{[$-0.0\:\text{to}\:0.0$]}}         &                     &\multicolumn{1}{c}{\text{[$-0.0\:\text{to}\:0.0$]}}         &                     &\multicolumn{1}{c}{\text{[$-0.0\:\text{to}\:0.0$]}}         &                     &\multicolumn{1}{c}{\text{[$-0.0\:\text{to}\:0.0$]}}         \\
            &                     &\multicolumn{1}{c}{\text{$<p=0.850>$}}         &                     &\multicolumn{1}{c}{\text{$<p=0.381>$}}         &                     &\multicolumn{1}{c}{\text{$<p=0.246>$}}         &                     &\multicolumn{1}{c}{\text{$<p=0.319>$}}         &                     &\multicolumn{1}{c}{\text{$<p=0.427>$}}         &                     &\multicolumn{1}{c}{\text{$<p=0.373>$}}         &                     &\multicolumn{1}{c}{\text{$<p=0.124>$}}         \\
Constant    &        18.1\sym{**} &        14.8\sym{**} &        39.8\sym{***}&        33.9\sym{***}&        15.4\sym{***}&        12.5\sym{***}&         2.8\sym{***}&         2.2\sym{***}&         2.3\sym{***}&         1.9\sym{***}&         2.6\sym{***}&         2.1\sym{***}&         2.3\sym{***}&         2.0\sym{***}\\
            &       (5.6)         &       (4.8)         &       (7.2)         &       (6.3)         &       (3.6)         &       (3.1)         &       (0.6)         &       (0.5)         &       (0.6)         &       (0.5)         &       (0.6)         &       (0.5)         &       (0.3)         &       (0.2)         \\
            &\multicolumn{1}{c}{\text{[$7.2\:\text{to}\:29.0$]}}         &\multicolumn{1}{c}{\text{[$5.4\:\text{to}\:24.2$]}}         &\multicolumn{1}{c}{\text{[$25.6\:\text{to}\:53.9$]}}         &\multicolumn{1}{c}{\text{[$21.5\:\text{to}\:46.3$]}}         &\multicolumn{1}{c}{\text{[$8.3\:\text{to}\:22.4$]}}         &\multicolumn{1}{c}{\text{[$6.4\:\text{to}\:18.6$]}}         &\multicolumn{1}{c}{\text{[$1.7\:\text{to}\:4.0$]}}         &\multicolumn{1}{c}{\text{[$1.3\:\text{to}\:3.2$]}}         &\multicolumn{1}{c}{\text{[$1.1\:\text{to}\:3.6$]}}         &\multicolumn{1}{c}{\text{[$0.8\:\text{to}\:2.9$]}}         &\multicolumn{1}{c}{\text{[$1.5\:\text{to}\:3.8$]}}         &\multicolumn{1}{c}{\text{[$1.2\:\text{to}\:3.0$]}}         &\multicolumn{1}{c}{\text{[$1.7\:\text{to}\:2.9$]}}         &\multicolumn{1}{c}{\text{[$1.5\:\text{to}\:2.5$]}}         \\
            &\multicolumn{1}{c}{\text{$<p=0.001>$}}         &\multicolumn{1}{c}{\text{$<p=0.002>$}}         &\multicolumn{1}{c}{\text{$<p=0.000>$}}         &\multicolumn{1}{c}{\text{$<p=0.000>$}}         &\multicolumn{1}{c}{\text{$<p=0.000>$}}         &\multicolumn{1}{c}{\text{$<p=0.000>$}}         &\multicolumn{1}{c}{\text{$<p=0.000>$}}         &\multicolumn{1}{c}{\text{$<p=0.000>$}}         &\multicolumn{1}{c}{\text{$<p=0.000>$}}         &\multicolumn{1}{c}{\text{$<p=0.000>$}}         &\multicolumn{1}{c}{\text{$<p=0.000>$}}         &\multicolumn{1}{c}{\text{$<p=0.000>$}}         &\multicolumn{1}{c}{\text{$<p=0.000>$}}         &\multicolumn{1}{c}{\text{$<p=0.000>$}}         \\
\midrule
Median/Mean of outcome&           1         &           1         &          12         &          13         &           0         &           0         &           0         &           0         &           0         &           0         &           0         &           0         &           0         &           0         \\
Repository observations&         582         &         582         &         582         &         582         &         582         &         582         &         582         &         582         &         582         &         582         &         582         &         582         &         582         &         582         \\
Day observations&           1         &          42         &           1         &          42         &           1         &          42         &           1         &          42         &           1         &          42         &           1         &          42         &           1         &          42         \\
Repository-day observations&         582         &     130,950         &         582         &     130,950         &         582         &     130,950         &         582         &     130,950         &         582         &     130,950         &         582         &     130,950         &         582         &     130,950         \\

            \bottomrule
		\end{tabular}
	\end{adjustbox}
	\caption*{
        \scriptsize Note:
		The table presents post-treatment differences in means for cumulative GitHub repository event counts constructed from the 2023 GitHub Archive data.
		Odd-numbered columns report differences in means at a snapshot approximately six weeks after intervention ended (1 Jul 2023).
		Even-numbered columns report post-treatment differences in means over the full post-treatment period, allowing for heterogeneous treatment effects through a linear time trend.
		Parentheses: standard errors.
		Square brackets: 95\% confidence intervals.
		Angle brackets: p-values.
		Significance levels: $^{***} p < .001; ^{**} p < .01; ^{*} p < .05; ^{+} p <.1$.
	}
\end{table}

\begin{table}[!ht] \centering \small \setlength\tabcolsep{2pt} \setlength{\defaultaddspace}{0pt}
	\def\sym#1{\ifmmode^{#1}\else\(^{#1}\)\fi}
	\caption{GitHub Experiment: Post-treatment differences in GitHub repository event activity (ITT)  (low-dosage and high-dosage treatment groups vs. control).}
	\label{tab:gh_exp_regtable_gharchive_means_dosage}
	\begin{adjustbox}{max width=1.325\textwidth}
		\begin{tabular}{@{\hspace{0\tabcolsep}}l*{14}{D{.}{.}{-1}}@{\hspace{0\tabcolsep}}}
			\toprule\toprule
			&\multicolumn{1}{c}{(1)}&\multicolumn{1}{c}{(2)}&\multicolumn{1}{c}{(3)}&\multicolumn{1}{c}{(4)}&\multicolumn{1}{c}{(5)}&\multicolumn{1}{c}{(6)}&\multicolumn{1}{c}{(7)}&\multicolumn{1}{c}{(8)}&\multicolumn{1}{c}{(9)}&\multicolumn{1}{c}{(10)}&\multicolumn{1}{c}{(11)}&\multicolumn{1}{c}{(12)}&\multicolumn{1}{c}{(13)}&\multicolumn{1}{c}{(14)}\\
			\cmidrule(l){2-15}
			&\multicolumn{2}{c}{Stars}&\multicolumn{2}{c}{Push events}&\multicolumn{2}{c}{Pull requests}&\multicolumn{2}{c}{Issues opened}&\multicolumn{2}{c}{Issues closed}&\multicolumn{2}{c}{Fork events}&\multicolumn{2}{c}{Release events}\\
			\cmidrule(lr){2-3}\cmidrule(lr){4-5}\cmidrule(lr){6-7}\cmidrule(lr){8-9}\cmidrule(lr){10-11}\cmidrule(lr){12-13}\cmidrule(l){14-15}
			&\multicolumn{1}{c}{Jul 1}&\multicolumn{1}{c}{Full}&\multicolumn{1}{c}{Jul 1}&\multicolumn{1}{c}{Full}&\multicolumn{1}{c}{Jul 1}&\multicolumn{1}{c}{Full}&\multicolumn{1}{c}{Jul 1}&\multicolumn{1}{c}{Full}&\multicolumn{1}{c}{Jul 1}&\multicolumn{1}{c}{Full}&\multicolumn{1}{c}{Jul 1}&\multicolumn{1}{c}{Full}&\multicolumn{1}{c}{Jul 1}&\multicolumn{1}{c}{Full}\\
			\midrule
Low dosage  &         9.5         &        10.7\sym{*}  &         0.1         &        -1.9         &         1.0         &         0.0         &        -0.5         &        -0.8         &        -0.4         &        -0.6         &        -0.7         &        -0.6         &        -0.2         &        -0.0         \\
            &       (6.2)         &       (5.2)         &      (15.2)         &      (12.1)         &       (7.2)         &       (5.7)         &       (1.4)         &       (0.9)         &       (1.4)         &       (0.9)         &       (1.0)         &       (0.8)         &       (0.6)         &       (0.6)         \\
            &\multicolumn{1}{c}{\text{[$-2.6\:\text{to}\:21.6$]}}         &\multicolumn{1}{c}{\text{[$0.5\:\text{to}\:21.0$]}}         &\multicolumn{1}{c}{\text{[$-29.7\:\text{to}\:29.9$]}}         &\multicolumn{1}{c}{\text{[$-25.8\:\text{to}\:21.9$]}}         &\multicolumn{1}{c}{\text{[$-13.1\:\text{to}\:15.2$]}}         &\multicolumn{1}{c}{\text{[$-11.1\:\text{to}\:11.1$]}}         &\multicolumn{1}{c}{\text{[$-3.2\:\text{to}\:2.3$]}}         &\multicolumn{1}{c}{\text{[$-2.7\:\text{to}\:1.0$]}}         &\multicolumn{1}{c}{\text{[$-3.1\:\text{to}\:2.3$]}}         &\multicolumn{1}{c}{\text{[$-2.4\:\text{to}\:1.2$]}}         &\multicolumn{1}{c}{\text{[$-2.6\:\text{to}\:1.3$]}}         &\multicolumn{1}{c}{\text{[$-2.1\:\text{to}\:0.9$]}}         &\multicolumn{1}{c}{\text{[$-1.4\:\text{to}\:1.1$]}}         &\multicolumn{1}{c}{\text{[$-1.2\:\text{to}\:1.1$]}}         \\
            &\multicolumn{1}{c}{\text{$<p=0.124>$}}         &\multicolumn{1}{c}{\text{$<p=0.041>$}}         &\multicolumn{1}{c}{\text{$<p=0.996>$}}         &\multicolumn{1}{c}{\text{$<p=0.874>$}}         &\multicolumn{1}{c}{\text{$<p=0.884>$}}         &\multicolumn{1}{c}{\text{$<p=0.999>$}}         &\multicolumn{1}{c}{\text{$<p=0.748>$}}         &\multicolumn{1}{c}{\text{$<p=0.392>$}}         &\multicolumn{1}{c}{\text{$<p=0.758>$}}         &\multicolumn{1}{c}{\text{$<p=0.509>$}}         &\multicolumn{1}{c}{\text{$<p=0.490>$}}         &\multicolumn{1}{c}{\text{$<p=0.416>$}}         &\multicolumn{1}{c}{\text{$<p=0.783>$}}         &\multicolumn{1}{c}{\text{$<p=0.949>$}}         \\
High dosage &        88.7\sym{**} &        87.8\sym{**} &        34.8         &        31.0         &        28.5         &        24.0         &         9.0         &         8.1         &         8.8         &         8.3         &         7.8         &         7.2         &        -1.1\sym{*}  &        -0.9\sym{+}  \\
            &      (31.8)         &      (28.3)         &      (25.0)         &      (20.0)         &      (19.6)         &      (15.3)         &       (8.2)         &       (7.1)         &       (7.9)         &       (7.1)         &       (5.6)         &       (4.9)         &       (0.5)         &       (0.4)         \\
            &\multicolumn{1}{c}{\text{[$26.2\:\text{to}\:151.1$]}}         &\multicolumn{1}{c}{\text{[$32.3\:\text{to}\:143.3$]}}         &\multicolumn{1}{c}{\text{[$-14.3\:\text{to}\:83.8$]}}         &\multicolumn{1}{c}{\text{[$-8.2\:\text{to}\:70.3$]}}         &\multicolumn{1}{c}{\text{[$-10.1\:\text{to}\:67.0$]}}         &\multicolumn{1}{c}{\text{[$-6.0\:\text{to}\:54.0$]}}         &\multicolumn{1}{c}{\text{[$-7.1\:\text{to}\:25.2$]}}         &\multicolumn{1}{c}{\text{[$-5.9\:\text{to}\:22.1$]}}         &\multicolumn{1}{c}{\text{[$-6.7\:\text{to}\:24.3$]}}         &\multicolumn{1}{c}{\text{[$-5.6\:\text{to}\:22.1$]}}         &\multicolumn{1}{c}{\text{[$-3.2\:\text{to}\:18.9$]}}         &\multicolumn{1}{c}{\text{[$-2.4\:\text{to}\:16.8$]}}         &\multicolumn{1}{c}{\text{[$-2.1\:\text{to}\:-0.1$]}}         &\multicolumn{1}{c}{\text{[$-1.7\:\text{to}\:0.0$]}}         \\
            &\multicolumn{1}{c}{\text{$<p=0.005>$}}         &\multicolumn{1}{c}{\text{$<p=0.002>$}}         &\multicolumn{1}{c}{\text{$<p=0.164>$}}         &\multicolumn{1}{c}{\text{$<p=0.121>$}}         &\multicolumn{1}{c}{\text{$<p=0.147>$}}         &\multicolumn{1}{c}{\text{$<p=0.116>$}}         &\multicolumn{1}{c}{\text{$<p=0.272>$}}         &\multicolumn{1}{c}{\text{$<p=0.257>$}}         &\multicolumn{1}{c}{\text{$<p=0.264>$}}         &\multicolumn{1}{c}{\text{$<p=0.242>$}}         &\multicolumn{1}{c}{\text{$<p=0.164>$}}         &\multicolumn{1}{c}{\text{$<p=0.140>$}}         &\multicolumn{1}{c}{\text{$<p=0.033>$}}         &\multicolumn{1}{c}{\text{$<p=0.055>$}}         \\
Linear trend&                     &         0.1\sym{***}&                     &         0.1\sym{***}&                     &         0.1\sym{***}&                     &         0.0\sym{***}&                     &         0.0\sym{***}&                     &         0.0\sym{***}&                     &         0.0\sym{***}\\
            &                     &       (0.0)         &                     &       (0.0)         &                     &       (0.0)         &                     &       (0.0)         &                     &       (0.0)         &                     &       (0.0)         &                     &       (0.0)         \\
            &                     &\multicolumn{1}{c}{\text{[$0.0\:\text{to}\:0.1$]}}         &                     &\multicolumn{1}{c}{\text{[$0.1\:\text{to}\:0.2$]}}         &                     &\multicolumn{1}{c}{\text{[$0.0\:\text{to}\:0.1$]}}         &                     &\multicolumn{1}{c}{\text{[$0.0\:\text{to}\:0.0$]}}         &                     &\multicolumn{1}{c}{\text{[$0.0\:\text{to}\:0.0$]}}         &                     &\multicolumn{1}{c}{\text{[$0.0\:\text{to}\:0.0$]}}         &                     &\multicolumn{1}{c}{\text{[$0.0\:\text{to}\:0.0$]}}         \\
            &                     &\multicolumn{1}{c}{\text{$<p=0.000>$}}         &                     &\multicolumn{1}{c}{\text{$<p=0.000>$}}         &                     &\multicolumn{1}{c}{\text{$<p=0.000>$}}         &                     &\multicolumn{1}{c}{\text{$<p=0.000>$}}         &                     &\multicolumn{1}{c}{\text{$<p=0.000>$}}         &                     &\multicolumn{1}{c}{\text{$<p=0.000>$}}         &                     &\multicolumn{1}{c}{\text{$<p=0.000>$}}         \\
Low dosage $ \times$ Linear trend&                     &        -0.0         &                     &         0.0         &                     &         0.0         &                     &         0.0         &                     &         0.0         &                     &         0.0         &                     &        -0.0         \\
            &                     &       (0.0)         &                     &       (0.1)         &                     &       (0.0)         &                     &       (0.0)         &                     &       (0.0)         &                     &       (0.0)         &                     &       (0.0)         \\
            &                     &\multicolumn{1}{c}{\text{[$-0.1\:\text{to}\:0.0$]}}         &                     &\multicolumn{1}{c}{\text{[$-0.1\:\text{to}\:0.2$]}}         &                     &\multicolumn{1}{c}{\text{[$-0.0\:\text{to}\:0.1$]}}         &                     &\multicolumn{1}{c}{\text{[$-0.0\:\text{to}\:0.0$]}}         &                     &\multicolumn{1}{c}{\text{[$-0.0\:\text{to}\:0.0$]}}         &                     &\multicolumn{1}{c}{\text{[$-0.0\:\text{to}\:0.0$]}}         &                     &\multicolumn{1}{c}{\text{[$-0.0\:\text{to}\:0.0$]}}         \\
            &                     &\multicolumn{1}{c}{\text{$<p=0.354>$}}         &                     &\multicolumn{1}{c}{\text{$<p=0.644>$}}         &                     &\multicolumn{1}{c}{\text{$<p=0.536>$}}         &                     &\multicolumn{1}{c}{\text{$<p=0.545>$}}         &                     &\multicolumn{1}{c}{\text{$<p=0.719>$}}         &                     &\multicolumn{1}{c}{\text{$<p=0.979>$}}         &                     &\multicolumn{1}{c}{\text{$<p=0.192>$}}         \\
High dosage $ \times$ Linear trend&                     &         0.0         &                     &         0.1         &                     &         0.1         &                     &         0.0         &                     &         0.0         &                     &         0.0         &                     &        -0.0         \\
            &                     &       (0.1)         &                     &       (0.1)         &                     &       (0.1)         &                     &       (0.0)         &                     &       (0.0)         &                     &       (0.0)         &                     &       (0.0)         \\
            &                     &\multicolumn{1}{c}{\text{[$-0.1\:\text{to}\:0.2$]}}         &                     &\multicolumn{1}{c}{\text{[$-0.1\:\text{to}\:0.4$]}}         &                     &\multicolumn{1}{c}{\text{[$-0.1\:\text{to}\:0.3$]}}         &                     &\multicolumn{1}{c}{\text{[$-0.0\:\text{to}\:0.1$]}}         &                     &\multicolumn{1}{c}{\text{[$-0.0\:\text{to}\:0.1$]}}         &                     &\multicolumn{1}{c}{\text{[$-0.0\:\text{to}\:0.1$]}}         &                     &\multicolumn{1}{c}{\text{[$-0.0\:\text{to}\:0.0$]}}         \\
            &                     &\multicolumn{1}{c}{\text{$<p=0.494>$}}         &                     &\multicolumn{1}{c}{\text{$<p=0.298>$}}         &                     &\multicolumn{1}{c}{\text{$<p=0.241>$}}         &                     &\multicolumn{1}{c}{\text{$<p=0.284>$}}         &                     &\multicolumn{1}{c}{\text{$<p=0.294>$}}         &                     &\multicolumn{1}{c}{\text{$<p=0.180>$}}         &                     &\multicolumn{1}{c}{\text{$<p=0.107>$}}         \\
Constant    &        18.1\sym{**} &        14.8\sym{**} &        39.8\sym{***}&        33.9\sym{***}&        15.4\sym{***}&        12.5\sym{***}&         2.8\sym{***}&         2.2\sym{***}&         2.3\sym{***}&         1.9\sym{***}&         2.6\sym{***}&         2.1\sym{***}&         2.3\sym{***}&         2.0\sym{***}\\
            &       (5.6)         &       (4.8)         &       (7.2)         &       (6.3)         &       (3.6)         &       (3.1)         &       (0.6)         &       (0.5)         &       (0.6)         &       (0.5)         &       (0.6)         &       (0.5)         &       (0.3)         &       (0.2)         \\
            &\multicolumn{1}{c}{\text{[$7.2\:\text{to}\:29.0$]}}         &\multicolumn{1}{c}{\text{[$5.4\:\text{to}\:24.2$]}}         &\multicolumn{1}{c}{\text{[$25.6\:\text{to}\:53.9$]}}         &\multicolumn{1}{c}{\text{[$21.5\:\text{to}\:46.3$]}}         &\multicolumn{1}{c}{\text{[$8.3\:\text{to}\:22.4$]}}         &\multicolumn{1}{c}{\text{[$6.4\:\text{to}\:18.6$]}}         &\multicolumn{1}{c}{\text{[$1.7\:\text{to}\:4.0$]}}         &\multicolumn{1}{c}{\text{[$1.3\:\text{to}\:3.2$]}}         &\multicolumn{1}{c}{\text{[$1.1\:\text{to}\:3.6$]}}         &\multicolumn{1}{c}{\text{[$0.8\:\text{to}\:2.9$]}}         &\multicolumn{1}{c}{\text{[$1.5\:\text{to}\:3.8$]}}         &\multicolumn{1}{c}{\text{[$1.2\:\text{to}\:3.0$]}}         &\multicolumn{1}{c}{\text{[$1.7\:\text{to}\:2.9$]}}         &\multicolumn{1}{c}{\text{[$1.5\:\text{to}\:2.5$]}}         \\
            &\multicolumn{1}{c}{\text{$<p=0.001>$}}         &\multicolumn{1}{c}{\text{$<p=0.002>$}}         &\multicolumn{1}{c}{\text{$<p=0.000>$}}         &\multicolumn{1}{c}{\text{$<p=0.000>$}}         &\multicolumn{1}{c}{\text{$<p=0.000>$}}         &\multicolumn{1}{c}{\text{$<p=0.000>$}}         &\multicolumn{1}{c}{\text{$<p=0.000>$}}         &\multicolumn{1}{c}{\text{$<p=0.000>$}}         &\multicolumn{1}{c}{\text{$<p=0.000>$}}         &\multicolumn{1}{c}{\text{$<p=0.000>$}}         &\multicolumn{1}{c}{\text{$<p=0.000>$}}         &\multicolumn{1}{c}{\text{$<p=0.000>$}}         &\multicolumn{1}{c}{\text{$<p=0.000>$}}         &\multicolumn{1}{c}{\text{$<p=0.000>$}}         \\
\midrule
Median/Mean of outcome&           1         &           1         &          12         &          13         &           0         &           0         &           0         &           0         &           0         &           0         &           0         &           0         &           0         &           0         \\
Repository observations&         582         &         582         &         582         &         582         &         582         &         582         &         582         &         582         &         582         &         582         &         582         &         582         &         582         &         582         \\
Day observations&           1         &          42         &           1         &          42         &           1         &          42         &           1         &          42         &           1         &          42         &           1         &          42         &           1         &          42         \\
Repository-day observations&         582         &     130,950         &         582         &     130,950         &         582         &     130,950         &         582         &     130,950         &         582         &     130,950         &         582         &     130,950         &         582         &     130,950         \\

            \bottomrule
		\end{tabular}
	\end{adjustbox}
	\caption*{
        \scriptsize Note:
		The table presents post-treatment differences in means for cumulative GitHub repository event counts constructed from the 2023 GitHub Archive data.
		Odd-numbered columns report differences in means at a snapshot approximately six weeks after intervention ended (1 Jul 2023).
		Even-numbered columns report post-treatment differences in means over the full post-treatment period, allowing for heterogeneous treatment effects through a linear time trend.
		Parentheses: standard errors.
		Square brackets: 95\% confidence intervals.
		Angle brackets: p-values.
		Significance levels: $^{***} p < .001; ^{**} p < .01; ^{*} p < .05; ^{+} p <.1$.
	}
\end{table}

\begin{table}[!ht] \centering \small \setlength\tabcolsep{2pt} \setlength{\defaultaddspace}{0pt}
	\def\sym#1{\ifmmode^{#1}\else\(^{#1}\)\fi}
	\caption{PyPI Experiment: Post-treatment differences in GitHub repository event activity (ITT).}
	\label{tab:pypi_exp_regtable_gharchive_means}
	\begin{adjustbox}{max width=1.325\textwidth}
		\begin{tabular}{@{\hspace{0\tabcolsep}}l*{14}{D{.}{.}{-1}}@{\hspace{0\tabcolsep}}}
			\toprule\toprule
			&\multicolumn{1}{c}{(1)}&\multicolumn{1}{c}{(2)}&\multicolumn{1}{c}{(3)}&\multicolumn{1}{c}{(4)}&\multicolumn{1}{c}{(5)}&\multicolumn{1}{c}{(6)}&\multicolumn{1}{c}{(7)}&\multicolumn{1}{c}{(8)}&\multicolumn{1}{c}{(9)}&\multicolumn{1}{c}{(10)}&\multicolumn{1}{c}{(11)}&\multicolumn{1}{c}{(12)}&\multicolumn{1}{c}{(13)}&\multicolumn{1}{c}{(14)}\\
			\cmidrule(l){2-15}
			&\multicolumn{2}{c}{Stars}&\multicolumn{2}{c}{Push events}&\multicolumn{2}{c}{Pull requests}&\multicolumn{2}{c}{Issues opened}&\multicolumn{2}{c}{Issues closed}&\multicolumn{2}{c}{Fork events}&\multicolumn{2}{c}{Release events}\\
			\cmidrule(lr){2-3}\cmidrule(lr){4-5}\cmidrule(lr){6-7}\cmidrule(lr){8-9}\cmidrule(lr){10-11}\cmidrule(lr){12-13}\cmidrule(l){14-15}
			&\multicolumn{1}{c}{Jul 1}&\multicolumn{1}{c}{Full}&\multicolumn{1}{c}{Jul 1}&\multicolumn{1}{c}{Full}&\multicolumn{1}{c}{Jul 1}&\multicolumn{1}{c}{Full}&\multicolumn{1}{c}{Jul 1}&\multicolumn{1}{c}{Full}&\multicolumn{1}{c}{Jul 1}&\multicolumn{1}{c}{Full}&\multicolumn{1}{c}{Jul 1}&\multicolumn{1}{c}{Full}&\multicolumn{1}{c}{Jul 1}&\multicolumn{1}{c}{Full}\\
			\midrule
Treatment group&         0.0         &        -1.0         &         7.7         &         6.7         &         9.2         &         8.0         &         1.3         &         1.0         &         1.2         &         1.0         &         0.6         &         0.4         &         0.3         &         0.2         \\
            &       (7.7)         &       (7.4)         &       (7.5)         &       (6.7)         &       (5.6)         &       (4.9)         &       (1.4)         &       (1.2)         &       (1.1)         &       (1.0)         &       (2.0)         &       (1.9)         &       (0.3)         &       (0.3)         \\
            &\multicolumn{1}{c}{\text{[$-15.1\:\text{to}\:15.1$]}}         &\multicolumn{1}{c}{\text{[$-15.4\:\text{to}\:13.5$]}}         &\multicolumn{1}{c}{\text{[$-6.9\:\text{to}\:22.4$]}}         &\multicolumn{1}{c}{\text{[$-6.3\:\text{to}\:19.8$]}}         &\multicolumn{1}{c}{\text{[$-1.8\:\text{to}\:20.2$]}}         &\multicolumn{1}{c}{\text{[$-1.6\:\text{to}\:17.5$]}}         &\multicolumn{1}{c}{\text{[$-1.4\:\text{to}\:4.0$]}}         &\multicolumn{1}{c}{\text{[$-1.4\:\text{to}\:3.4$]}}         &\multicolumn{1}{c}{\text{[$-1.0\:\text{to}\:3.4$]}}         &\multicolumn{1}{c}{\text{[$-0.9\:\text{to}\:3.0$]}}         &\multicolumn{1}{c}{\text{[$-3.3\:\text{to}\:4.5$]}}         &\multicolumn{1}{c}{\text{[$-3.4\:\text{to}\:4.1$]}}         &\multicolumn{1}{c}{\text{[$-0.3\:\text{to}\:0.8$]}}         &\multicolumn{1}{c}{\text{[$-0.3\:\text{to}\:0.7$]}}         \\
            &\multicolumn{1}{c}{\text{$<p=0.999>$}}         &\multicolumn{1}{c}{\text{$<p=0.897>$}}         &\multicolumn{1}{c}{\text{$<p=0.300>$}}         &\multicolumn{1}{c}{\text{$<p=0.314>$}}         &\multicolumn{1}{c}{\text{$<p=0.100>$}}         &\multicolumn{1}{c}{\text{$<p=0.103>$}}         &\multicolumn{1}{c}{\text{$<p=0.347>$}}         &\multicolumn{1}{c}{\text{$<p=0.423>$}}         &\multicolumn{1}{c}{\text{$<p=0.283>$}}         &\multicolumn{1}{c}{\text{$<p=0.301>$}}         &\multicolumn{1}{c}{\text{$<p=0.766>$}}         &\multicolumn{1}{c}{\text{$<p=0.853>$}}         &\multicolumn{1}{c}{\text{$<p=0.330>$}}         &\multicolumn{1}{c}{\text{$<p=0.377>$}}         \\
Linear trend&                     &         0.1\sym{***}&                     &         0.3\sym{***}&                     &         0.2\sym{***}&                     &         0.0\sym{***}&                     &         0.0\sym{***}&                     &         0.0\sym{***}&                     &         0.0\sym{***}\\
            &                     &       (0.0)         &                     &       (0.0)         &                     &       (0.0)         &                     &       (0.0)         &                     &       (0.0)         &                     &       (0.0)         &                     &       (0.0)         \\
            &                     &\multicolumn{1}{c}{\text{[$0.1\:\text{to}\:0.1$]}}         &                     &\multicolumn{1}{c}{\text{[$0.2\:\text{to}\:0.3$]}}         &                     &\multicolumn{1}{c}{\text{[$0.2\:\text{to}\:0.2$]}}         &                     &\multicolumn{1}{c}{\text{[$0.0\:\text{to}\:0.0$]}}         &                     &\multicolumn{1}{c}{\text{[$0.0\:\text{to}\:0.0$]}}         &                     &\multicolumn{1}{c}{\text{[$0.0\:\text{to}\:0.0$]}}         &                     &\multicolumn{1}{c}{\text{[$0.0\:\text{to}\:0.0$]}}         \\
            &                     &\multicolumn{1}{c}{\text{$<p=0.000>$}}         &                     &\multicolumn{1}{c}{\text{$<p=0.000>$}}         &                     &\multicolumn{1}{c}{\text{$<p=0.000>$}}         &                     &\multicolumn{1}{c}{\text{$<p=0.000>$}}         &                     &\multicolumn{1}{c}{\text{$<p=0.000>$}}         &                     &\multicolumn{1}{c}{\text{$<p=0.000>$}}         &                     &\multicolumn{1}{c}{\text{$<p=0.000>$}}         \\
Treatment group  $ \times$ Linear trend&                     &         0.0         &                     &         0.1         &                     &         0.1         &                     &         0.0         &                     &         0.0         &                     &         0.0         &                     &         0.0         \\
            &                     &       (0.0)         &                     &       (0.0)         &                     &       (0.0)         &                     &       (0.0)         &                     &       (0.0)         &                     &       (0.0)         &                     &       (0.0)         \\
            &                     &\multicolumn{1}{c}{\text{[$-0.0\:\text{to}\:0.1$]}}         &                     &\multicolumn{1}{c}{\text{[$-0.0\:\text{to}\:0.1$]}}         &                     &\multicolumn{1}{c}{\text{[$-0.0\:\text{to}\:0.1$]}}         &                     &\multicolumn{1}{c}{\text{[$-0.0\:\text{to}\:0.0$]}}         &                     &\multicolumn{1}{c}{\text{[$-0.0\:\text{to}\:0.0$]}}         &                     &\multicolumn{1}{c}{\text{[$-0.0\:\text{to}\:0.0$]}}         &                     &\multicolumn{1}{c}{\text{[$-0.0\:\text{to}\:0.0$]}}         \\
            &                     &\multicolumn{1}{c}{\text{$<p=0.251>$}}         &                     &\multicolumn{1}{c}{\text{$<p=0.177>$}}         &                     &\multicolumn{1}{c}{\text{$<p=0.106>$}}         &                     &\multicolumn{1}{c}{\text{$<p=0.128>$}}         &                     &\multicolumn{1}{c}{\text{$<p=0.189>$}}         &                     &\multicolumn{1}{c}{\text{$<p=0.138>$}}         &                     &\multicolumn{1}{c}{\text{$<p=0.251>$}}         \\
Constant    &        28.0\sym{***}&        25.7\sym{***}&        51.2\sym{***}&        44.9\sym{***}&        35.3\sym{***}&        30.8\sym{***}&         7.2\sym{***}&         6.4\sym{***}&         6.1\sym{***}&         5.4\sym{***}&         8.1\sym{***}&         7.5\sym{***}&         1.4\sym{***}&         1.3\sym{***}\\
            &       (5.9)         &       (5.9)         &       (3.1)         &       (2.8)         &       (2.2)         &       (1.9)         &       (0.8)         &       (0.7)         &       (0.6)         &       (0.5)         &       (1.6)         &       (1.6)         &       (0.1)         &       (0.1)         \\
            &\multicolumn{1}{c}{\text{[$16.3\:\text{to}\:39.6$]}}         &\multicolumn{1}{c}{\text{[$14.2\:\text{to}\:37.3$]}}         &\multicolumn{1}{c}{\text{[$45.1\:\text{to}\:57.4$]}}         &\multicolumn{1}{c}{\text{[$39.5\:\text{to}\:50.4$]}}         &\multicolumn{1}{c}{\text{[$31.0\:\text{to}\:39.6$]}}         &\multicolumn{1}{c}{\text{[$27.0\:\text{to}\:34.6$]}}         &\multicolumn{1}{c}{\text{[$5.7\:\text{to}\:8.7$]}}         &\multicolumn{1}{c}{\text{[$5.0\:\text{to}\:7.8$]}}         &\multicolumn{1}{c}{\text{[$4.9\:\text{to}\:7.2$]}}         &\multicolumn{1}{c}{\text{[$4.3\:\text{to}\:6.5$]}}         &\multicolumn{1}{c}{\text{[$5.0\:\text{to}\:11.2$]}}         &\multicolumn{1}{c}{\text{[$4.4\:\text{to}\:10.5$]}}         &\multicolumn{1}{c}{\text{[$1.2\:\text{to}\:1.7$]}}         &\multicolumn{1}{c}{\text{[$1.1\:\text{to}\:1.5$]}}         \\
            &\multicolumn{1}{c}{\text{$<p=0.000>$}}         &\multicolumn{1}{c}{\text{$<p=0.000>$}}         &\multicolumn{1}{c}{\text{$<p=0.000>$}}         &\multicolumn{1}{c}{\text{$<p=0.000>$}}         &\multicolumn{1}{c}{\text{$<p=0.000>$}}         &\multicolumn{1}{c}{\text{$<p=0.000>$}}         &\multicolumn{1}{c}{\text{$<p=0.000>$}}         &\multicolumn{1}{c}{\text{$<p=0.000>$}}         &\multicolumn{1}{c}{\text{$<p=0.000>$}}         &\multicolumn{1}{c}{\text{$<p=0.000>$}}         &\multicolumn{1}{c}{\text{$<p=0.000>$}}         &\multicolumn{1}{c}{\text{$<p=0.000>$}}         &\multicolumn{1}{c}{\text{$<p=0.000>$}}         &\multicolumn{1}{c}{\text{$<p=0.000>$}}         \\
\midrule
Median/Mean of outcome&           0         &           0         &           0         &           0         &           0         &           0         &           0         &           0         &           0         &           0         &           0         &           0         &           0         &           0         \\
Package observations&      23,916         &      23,916         &      23,916         &      23,916         &      23,916         &      23,916         &      23,916         &      23,916         &      23,916         &      23,916         &      23,916         &      23,916         &      23,916         &      23,916         \\
Day observations&           1         &          23         &           1         &          23         &           1         &          23         &           1         &          23         &           1         &          23         &           1         &          23         &           1         &          23         \\
Package-day observations&      17,872         &   3,681,632         &      17,872         &   3,681,632         &      17,872         &   3,681,632         &      17,872         &   3,681,632         &      17,872         &   3,681,632         &      17,872         &   3,681,632         &      17,872         &   3,681,632         \\

			\bottomrule
		\end{tabular}
	\end{adjustbox}
	\caption*{
		\scriptsize Note:
		The table presents post-treatment differences in means for cumulative GitHub repository event counts constructed from the 2023 GitHub Archive data.
		Odd-numbered columns report differences in means at a snapshot approximately three weeks after intervention ended (1 Jul 2023).
		Even-numbered columns report post-treatment differences in means over the full post-treatment period, allowing for heterogeneous treatment effects through a linear time trend.
		Standard errors are clustered by package.
		Parentheses: standard errors.
		Square brackets: 95\% confidence intervals.
		Angle brackets: p-values.
		Significance levels: $^{***} p < .001; ^{**} p < .01; ^{*} p < .05; ^{+} p <.1$.
	}
\end{table}
\end{landscape}

% ~~~~~~~~~~~~~~~~~~~~~~~~~~~~~~~~~~~~~~~~~~~~~~~~~~~~~~~~~~~~~~~~~~~~~~~~~~~~~~~
% ~~~~~~~~~~~~~~~~~~~~~~~~~~~~~~~~~~~~~~~~~~~~~~~~~~~~~~~~~~~~~~~~~~~~~~~~~~~~~~~]
\clearpage
\FloatBarrier
\subsection{PyPI Observational Analysis}
\label{sec:var}
In this section, we present observational evidence using historical download data to understand the impact of human and bot downloads on human downloads. Using vector auto-regressive (VAR) models, we find that past human downloads predict future human downloads.

\subsubsection{Sample and Data}
We started by retrieving the full enumeration of all Python packages available from the PyPI repository. The index is extensive, with 458,274 packages at our retrieval time. We randomly sampled fifty thousand packages to keep querying costs and data processing tractable. We then queried daily downloads over a 3-month period (going back 91 days from our query date) from BigQuery and successfully retrieved n= 
 40,565.\footnote{\url{https://warehouse.pypa.io/api-reference/bigquery-datasets.html}.} We have the daily downloads split by package installer for each package. We classify installers into bot versus human downloads (\cref{sec:measures-pypi-downloads}). 
 %for numbers, see https://github.com/soodoku/social_proof_stars/blob/main/pydownloads/scripts/02b_var_one_by_one_50kpypi-90days.ipynb
 We filter our dataset to packages with at least 50 human downloads over the three months, which gives us 13,481 packages and 1,226,771 package-day observations. So, our final sample is of Python packages that have been downloaded somewhat consistently over the last three months.

\subsubsection{Research Design}
To estimate how past downloads predict future downloads, we estimate a VAR model for each package to understand the non-experimental correlational evidence. Specifically, we want to understand the extent to which past downloads can explain future downloads.

The VAR model of each package has two equations: one where human downloads at time $t$ is the outcome, and the other where bot downloads at time $t$ is the outcome. Both outcomes are modeled as a function of their past values (e.g., human downloads) and the past values of the other (e.g., bot downloads) up to a select number of temporal lag ($t-p$). The maximum number of lags is three weeks (21 days) to allow downloads up to three weeks old to affect present downloads. The Akaike Information Criterion then determines the number of lags ($p$) for each package. We implement this using \emph{statsmodels} \citep{seabold2010statsmodels}. Some packages do not have a fitted model because of numerical instability or unit roots. For packages that successfully converge with an optimal lag of at least a day (n = 6630), we further do a Granger causality test where the null is that past values of a time series do not collectively predict future values. From this, we get n = 6620 p-values from the Granger test (10 models fail to compute from numerical issues, \cite{seabold2010statsmodels}). Having a p-value that rejects the null at conventional levels does not mean that past downloads cause future downloads.

\begin{figure}[ht]
    \centering
    
    % First row
    \begin{subfigure}[b]{0.495\textwidth}
        \centering
        \includegraphics[width=\textwidth]{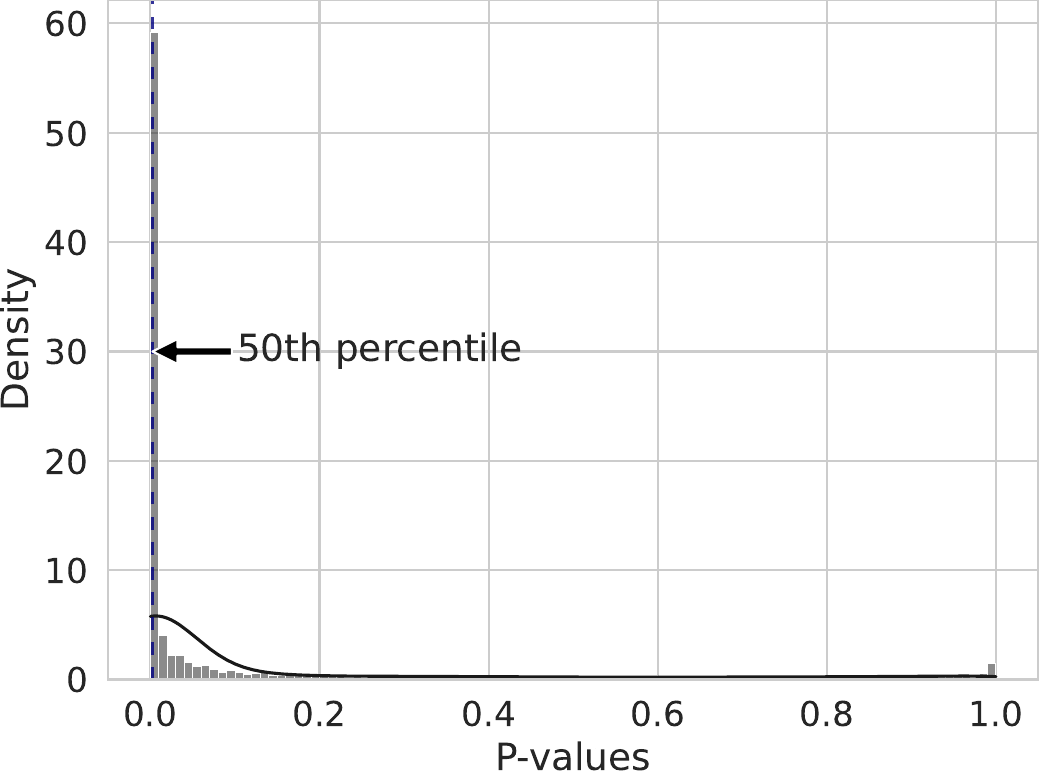}
        \caption{Human $\rightarrow$ human}
        \label{fig:50kpypi_90days_human2human}
    \end{subfigure}
    \hfill % add some space between images
    \begin{subfigure}[b]{0.495\textwidth}
        \centering
        \includegraphics[width=\textwidth]{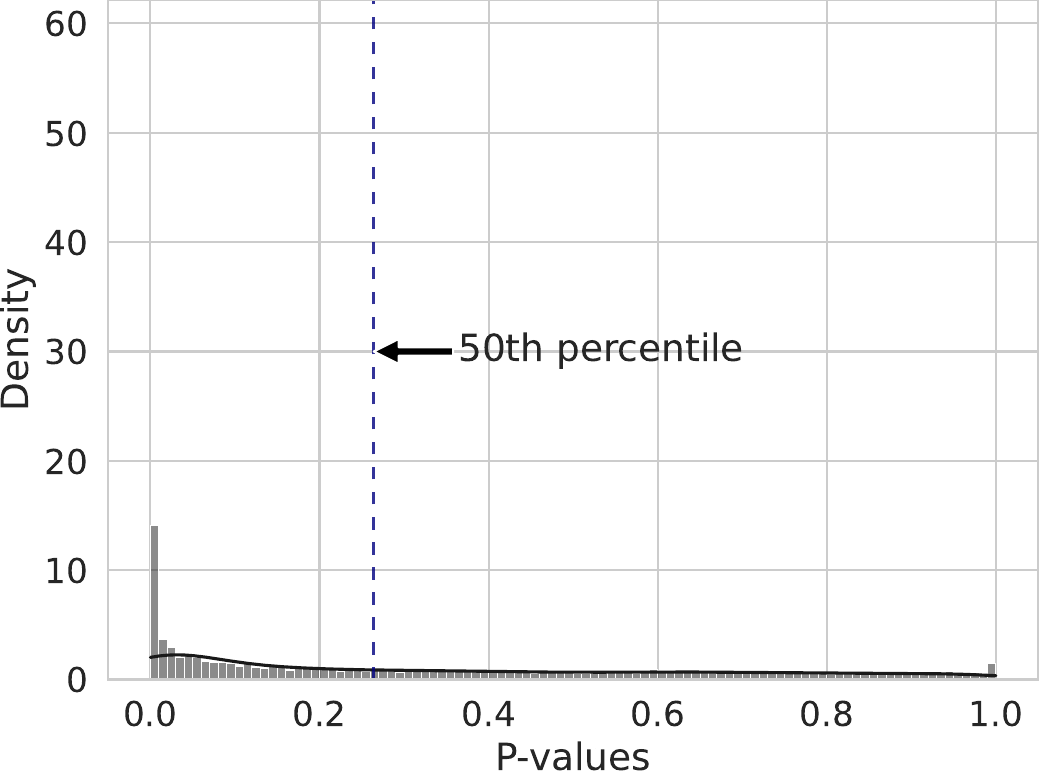}
        \caption{Human $\rightarrow$ bot}
        \label{fig:50kpypi_90days_human2bot}
    \end{subfigure}

    % Second row
    \begin{subfigure}[b]{0.495\textwidth}
        \centering
        \includegraphics[width=\textwidth]{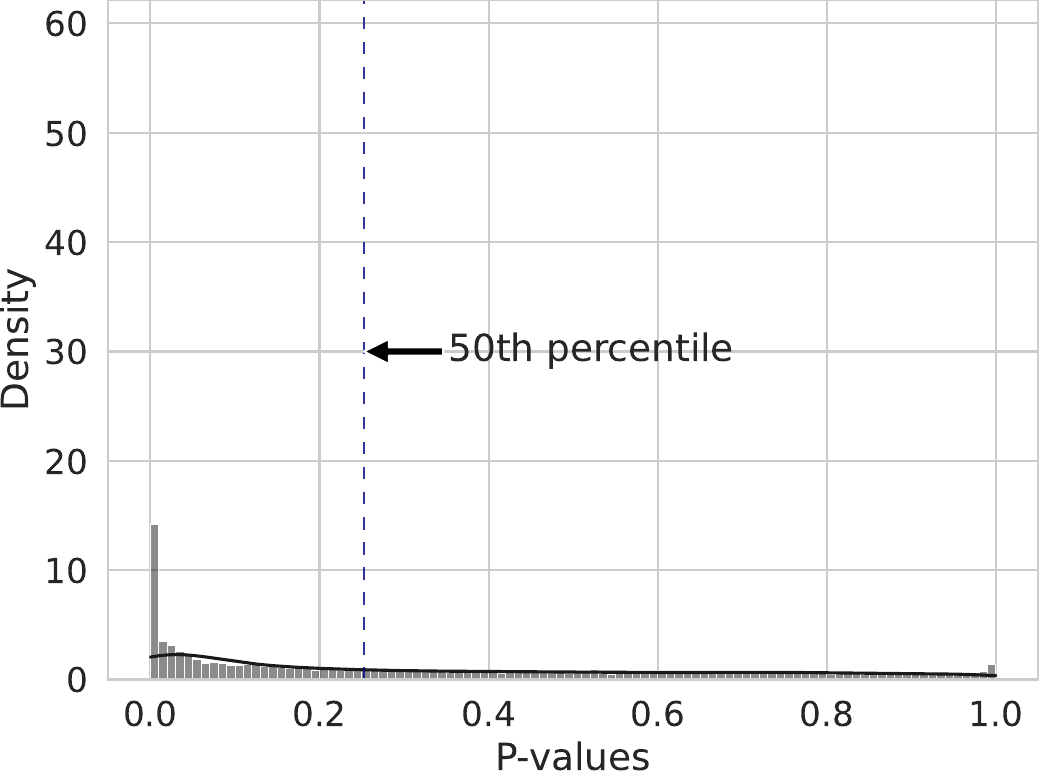}
        \caption{Bot $\rightarrow$ human}
        \label{fig:50kpypi_90days_bot2human}
    \end{subfigure}
    \hfill % add some space between images
    \begin{subfigure}[b]{0.495\textwidth}
        \centering
        \includegraphics[width=\textwidth]{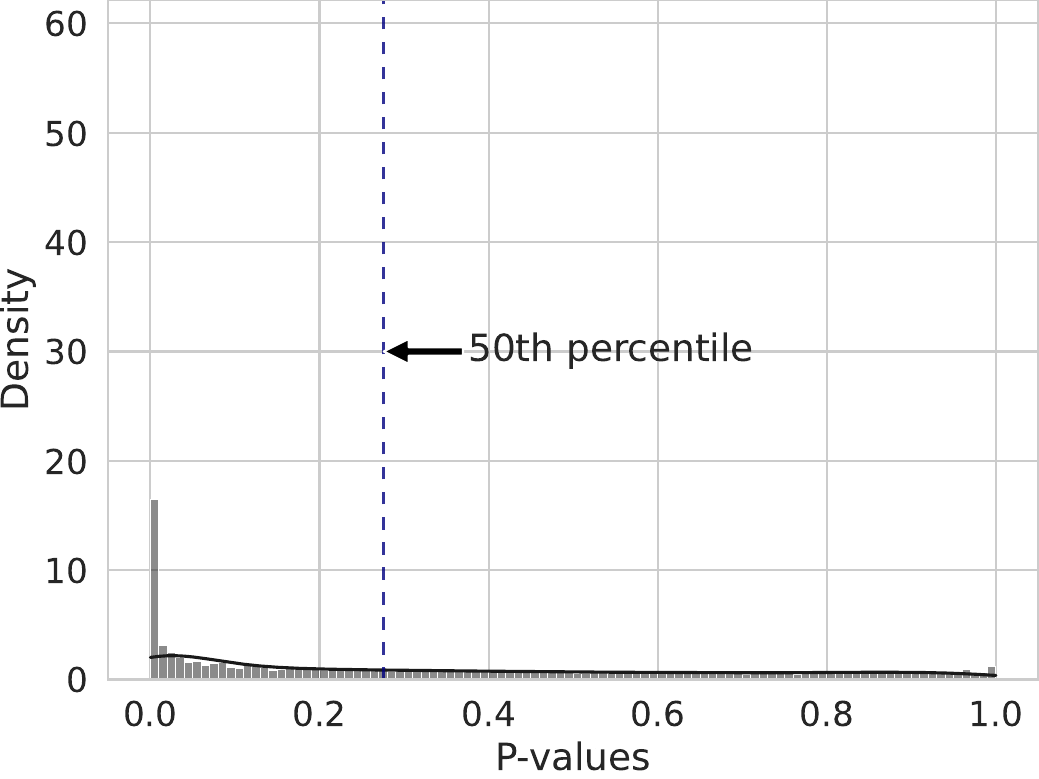}
        \caption{Bot $\rightarrow$ bot}
        \label{fig:50kpypi_90days_bot2bot}
    \end{subfigure}

    \caption{
    \textbf{Distribution of P-values From Granger Causality Tests.}
    The figure plots the distribution of p-values for each of the four versions (by panels) of the Granger tests for 6,620 packages and 602,420 package days. 
    All plots have the same scale. 
    Each package's downloads over three months are first estimated in a VAR model, with human downloads and bot downloads as the two outcomes. Estimated coefficients are then used to implement the Granger causality tests. 
    P-values are binned into 100 equal-width bins, each with a range of .01. The vertical dashed line indicates the 50th percentile value in the p-values.
    }
    \label{fig:granger_causality_pvalues}
\end{figure}

\subsubsection{Results}
\cref{fig:granger_causality_pvalues} reports the collection of p-values from all the successful VAR runs for the four versions of the Granger tests. The vertical dashed line indicates the 50th percentile in p-values. Overall, we observe that only past human downloads can predict human downloads (\cref{fig:50kpypi_90days_human2human}). Its 50th percentile in p-values is approximately .002, implying more than 50\% of the packages' VAR model have p-values small enough to reject the null hypothesis at the 1 percent level that past human downloads do not predict human downloads. 
Specifically, the percentile value for a p-value of .05 is the 69th percentile.

This same observation is not true for the three other versions (\crefrange{fig:50kpypi_90days_human2human}{fig:50kpypi_90days_bot2bot}). Some packages have Granger tests where the p-values are small enough to reject the null at conventional levels. However, they are in the minority, with the 50th percentile p-values that is much higher. The percentile values for these three groups, with a p-value of .05, are the 25th, 26th, and 26th percentiles.

In all, we conclude that past human downloads do not predict bot downloads, and past bot downloads predict neither human nor bot downloads.
%TC:endignore
\end{document}